\documentclass[prd,showpacs,nofootinbib,floatfix,preprintnumbers, twocolumn, superscriptaddress]{revtex4}%
\pdfoutput=1 

\usepackage{amssymb,amsmath}
\usepackage{hyperref}
\usepackage{color}
\usepackage{graphicx}
\usepackage{multirow}

\usepackage{latexsym} 

\begin{document}

\bibliographystyle{apsrev}

\preprint{JLAB-THY-10-1171}
\preprint{TCDMATH-10-03}

\title{Toward the excited meson spectrum of dynamical QCD}

\author{Jozef J. Dudek}
\email{dudek@jlab.org}
\affiliation{Jefferson Laboratory, 12000 Jefferson Avenue,  Newport News, VA 23606, USA}
\affiliation{Department of Physics, Old Dominion University, Norfolk, VA 23529, USA}

\author{Robert G. Edwards}
\affiliation{Jefferson Laboratory, 12000 Jefferson Avenue,  Newport News, VA 23606, USA}

\author{Michael J. Peardon}
\affiliation{School of Mathematics, Trinity College, Dublin 2, Ireland}

\author{David G. Richards}
\affiliation{Jefferson Laboratory, 12000 Jefferson Avenue,  Newport News, VA 23606, USA}

\author{Christopher E. Thomas}
\affiliation{Jefferson Laboratory, 12000 Jefferson Avenue,  Newport News, VA 23606, USA}

\collaboration{for the Hadron Spectrum Collaboration}

\begin{abstract}

We present a detailed description of the extraction of the highly excited isovector meson
spectrum on dynamical anisotropic lattices using a new quark-field construction
algorithm and a large variational basis of operators.  With careful
operator construction, the combination of these techniques is used to
identify the continuum spin of extracted states reliably, overcoming the reduced rotational symmetry of the cubic lattice. Excited states,
states with exotic quantum numbers ($0^{+-}$, $1^{-+}$ and
$2^{+-}$) and states of high spin are resolved, including, for the first time in
a lattice QCD calculation, spin-four states. The determinations of the spectrum
of isovector mesons and kaons are performed on dynamical lattices with two
volumes and with pion masses down to $\sim 400\,\mathrm{MeV}$, with statistical
precision typically at or below 1\% even for highly excited states.

\end{abstract}


\maketitle 

\section{Introduction}\label{sec:intro}

Computing the bound states of QCD is vital if we
are to claim a complete description of the strong interactions. 
Confronting high-precision calculations of the spectrum with
future experimental measurements will test the theoretical
framework for such a description rigorously.  There has been a resurgence of 
interest in 
the experimental investigations of the spectrum, notably in the charmonium
sector where a wealth of high-quality data from the $B$-factories
has challenged our understanding of spectroscopy.
A comprehensive investigation of the spectrum of mesons composed of
light quarks is the goal of the GlueX collaboration, after the 12 GeV
upgrade of Jefferson Laboratory.  Here the aim is to photoproduce
mesons, and in particular those with exotic quantum numbers, as a
means of revealing the role of gluonic degrees of freedom in the
spectrum.

Lattice calculations offer a method of performing a first-principles
computation of the spectrum of QCD and the calculation of the masses
of the lowest-lying states has been an important benchmark of lattice
studies since their inception.  However, recently there has been
considerable progress aimed at extracting the spectrum of excited
states, both for mesons and for baryons.  This has been accomplished
through the use of the variational method, employing a large basis of
interpolating operators satisfying the symmetries allowed by the cubic
lattice\cite{Basak:2005aq,Basak:2005ir,Basak:2007kj,Bulava:2009jb}.
In a series of recent papers, we have applied this methodology to the
extraction of the meson spectrum~\cite{Dudek:2007wv}, and the
radiative transitions between excited and low-lying meson
states~\cite{Dudek:2006ej,Dudek:2009kk}.  The first studies were
performed in the quenched approximation to QCD, for mesons composed of
the heavier charm quark and its antiquark, a system which is
computationally less demanding yet for which there is a wealth of
high-quality experimental data.  In this paper, we investigate the
spectrum of mesons for quark masses below the strange quark mass,
going down to pion masses of around $ 400\,\mathrm{MeV}$, and expand on our
earlier letter~\cite{Dudek:2009qf} focused on the spectrum of mesons
in full QCD with three mass-degenerate quark flavors.

Several previous studies of the spectrum have focussed on obtaining
precision results for ground state masses through controlling
systematic errors\cite{Aubin:2004wf, Burch:2009az, Jansen:2009hr,
WalkerLoud:2008bp}.  Others have aimed at extracting the masses of
some of the excited states\cite{Burch:2009wu, Petry:2008rt,
Gattringer:2008be, Burch:2006dg, Burch:2004he, Lacock:1996vy,
Yamazaki:2001er}.  The extraction of excited state masses is more
difficult owing to the decrease in the signal-to-noise ratio with
increasing time as we move progressively higher in the spectrum.  To
circumvent this difficulty, we use anisotropic lattices, with finer
temporal than spatial lattice spacing, enabling the behavior of the
Euclidean-space correlation functions to be examined at small temporal
separations.

The (hyper-) cubic lattice does not
possess the full rotational symmetry of the continuum.  Thus in a
lattice calculation, states at rest are classified not according to
the spin $(J, J_z)$, but rather according to the irreducible
representations (irreps) of a cube; for states of higher spin, the
different continuum degrees of freedom are distributed across several
lattice irreps.  In this study, we use a large basis of interpolating
operators, decomposed into their lattice irreps, that enable us to
explore all $J^{PC}$ up to spin 4, except for the exotic $4^{+-}$,
with as many as 26 operators in a
given symmetry channel. Our ability to calculate correlation functions
efficiently for such a large basis of operators relies on a new
method, ``distillation''\cite{Peardon:2009gh}, for the construction of
quark-antiquark operators, including those with non-local construction.

As the lattice spacing approaches zero, full rotational symmetry is
restored and thus in principle the spins can be identified through the
emergence of energy degeneracies between different irreps.  The
increasing density of states in each irrep as we rise in the
spectrum makes the identification of such degeneracies a challenging
task, and thus the assignment of the continuum spins has been a
formidable barrier to the interpretation of lattice calculations.  In
this paper, we adopt a very different approach, in which by
judiciously constructing operators so as to have a known continuum
behavior, the spins of the excited states are determined and thus the
barrier imposed by the reduced cubic symmetry of the lattice is overcome.

A particularly interesting class of mesons are those with exotic $J^{PC}$, such as the $1^{-+}$ channel. Extracting
clean signals for even the lightest state with these quantum numbers
has proven difficult \cite{Lacock:1996vy, Bernard:2003jd,
  Bernard:1997ib, Lacock:1998be, Lacock:1996ny, McNeile:2006bz,
  Hedditch:2005zf}, with statistical noise levels typically being
significantly higher than for other states. There is also a
requirement to determine that any such extracted state is, in fact,
the exotic spin 1 and not a non-exotic $4^{-+}$ state which would live
in the same lattice irreducible representation. We found that we
can extract information about exotic state masses at the same level of
precision as excited non-exotic states ($\lesssim 1\%$), with the
spins clearly identified.

First results, exploiting the full planopy of anisotropic lattices,
``distillation'' for efficient computation of the interpolating
operators, and the identification of the continuum spins, have been
presented for the case of three degenerate ``strange''
quarks \cite{Dudek:2009qf}; the spectrum of excited states, including
those with high spin, was extracted with
confidence.

In this paper, we expand on the earlier work (at $m_\pi \approx 700$ MeV) to include calculations
in full QCD both with three degenerate flavors of quark, and with a
strange quark and two light quarks ($N_f = 2 \oplus 1$), corresponding to pion masses $m_\pi \simeq
(520, 440, 400)~{\rm MeV}$.  Furthermore, we perform
calculations at two spatial volumes, enabling us to seek possible
finite-volume effects, and potentially the presence of multi-hadron
states in the calculated spectrum. Precise extraction of such states is required in order to carry out analyses of resonances in the manner suggested by L\"uscher \cite{Luscher:1991cf}.

The structure of the paper is as follows.  We begin in Section \ref{sec:lattice}
by presenting details of the QCD gauge-field configurations to be used. 
We describe the technology of two-point correlator measurement using 
distillation on dynamical gauge-field configurations in Section \ref{sec:distillation}. 
In Section \ref{sec:ops} the construction of a set of
composite QCD operators suitable for use in the extraction of the
meson spectrum is outlined, along with a description of the procedure
used to make these operators transform irreducibly under the limited
rotations allowed on a cubic lattice. In Section \ref{sec:fitting} we
present the methodology utilised to extract meson spectral information
from correlation functions. Section \ref{sec:spin} discusses the
possibility of using the information embedded in vacuum-operator-state
matrix elements to determine the spin of a meson, overcoming
ambiguities introduced by the reduced rotational symmetry of a cubic
lattice. The stability of the extracted spectral quantities to changes
in the details of the correlator analysis is considered in Section
\ref{sec:stability}, where it is seen that we can extract results of
considerable robustness. In Section \ref{sec:results} we report meson
spectrum results extracted from calculations at four pion
masses and two different lattice volumes. Results for isovector
mesons, kaons and the connected part of $s\bar{s}$ (``strangeonium")
are shown. Section \ref{sec:two-meson} considers the apparent absence
of multi-particle states within our extracted spectra. Finally, in Section
\ref{sec:summary} we summarise our observations and suggest future
directions.

\section{Lattice Gauge fields}\label{sec:lattice}
\begin{table}[t]
	\begin{tabular}{cccc|cccc|c}
	 & $\substack{m_\ell\\m_s}$ & $\substack{m_\pi\\/\mathrm{MeV}}$ & $m_K/m_\pi$ & volume   &$N_{\mathrm{cfgs}}$ & $N_{\mathrm{t_{srcs}}}$ & $N_{\mathrm{vecs}}$ & $\substack{N_{\mathrm{inv.}} \\ /10^6}$\\
		\hline
	\multirow{2}{*}{\emph{743}} & \multirow{2}{*}{$\substack{-0.0743\\-0.0743}$} & \multirow{2}{*}{702}& \multirow{2}{*}{1} & $16^3\times 128$ & 536 & 9 & 64 & 1.2 \\
				&&&& $20^3\times 128$ & 198 & 6 & 128 & 0.6\\
		\hline
	\multirow{2}{*}{\emph{808}} & \multirow{2}{*}{$\substack{-0.0808\\-0.0743}$} & \multirow{2}{*}{524} &\multirow{2}{*}{1.15}& $16^3\times 128$ & 500 & 7 & 64 & 0.9\\
				&&&& $20^3\times 128$ & 382 & 4 & 96 & 0.6	\\
		\hline
	\emph{830} & $\substack{-0.0830\\-0.0743}$ & 444 & 1.29 & $16^3\times 128$ & 601 & 10 & 64 & 1.5 \\				
		\hline
	\multirow{2}{*}{\emph{840}} & \multirow{2}{*}{$\substack{-0.0840\\-0.0743}$} & \multirow{2}{*}{396} &\multirow{2}{*}{1.39}& $16^3\times 128$ & 479 & 32 & 64 & 3.9\\
				&&&& $20^3\times 128$ & 600 & 6 & 128 & 1.8	
	\end{tabular}
\caption{The lattice data sets and propagators used in this paper.  The lattice size and number of configurations are listed, as well as the number of time-sources and the number of distillation vectors $N_{\mathrm{vecs}}$. The total number of inversions for each quark mass = $4\times N_{\mathrm{t_{srcs}}} \times N_\mathrm{vecs} \times N_\mathrm{cfgs}$ is shown.}
\label{tab:lattices}
\end{table}

In Euclidean space, excited state correlation functions decay faster than the ground state, and at large times are swamped by the signals of lower states, thus complicating the resolution of excited states. To ameliorate this problem we have adopted a dynamical anisotropic lattice formulation whereby the temporal extent is discretized with a finer lattice spacing than in the spatial directions~\cite{Edwards:2008ja,Lin:2008pr}; this has proven crucial to obtain the results shown in this paper.
This method avoids the computational cost that would come from reducing the spacing in all directions. Improved gauge and fermion actions are used, corresponding to two light dynamical quarks and one strange dynamical quark. Details describing the formulation of the actions as well as the techniques used to determine the anisotropy parameters can be found in Refs.~\cite{Edwards:2008ja,Lin:2008pr}.  The lattices have a spatial lattice spacing $a_s\sim 0.12$ fm with a temporal lattice spacing $3.5$ times smaller corresponding to a temporal scale $a_t^{-1}\sim 5.6$ GeV.

Previous work~\cite{Dudek:2009qf} showed results using the three-flavor degenerate quark-mass dataset corresponding to bare light and strange quark masses $a_t m_l=a_t m_s = -0.0743$ and lattice size $16^3\times 128$. The pion mass (degenerate with the kaon and $\eta$ masses) is roughly $700$ MeV.  In this work, results are extended to lighter masses $a_t m_l=(-0.0808, -0.0830, -0.0840)$ and $a_t m_s=-0.0743$ corresponding to $2+1$ flavors of dynamical quarks, and lattice sizes $16^3\times 128$ as well as $20^3\times 128$. These fully dynamical datasets, described in more detail in Table \ref{tab:lattices}, allow for some investigations of the quark mass dependence as well as finite volume dependence of the spectrum.

The lattice scale, quoted above, is determined in the physical quark mass limit using the $\Omega$ baryon mass (denoted by $a_t m_\Omega$). As noted, these calculations are carried out away from this limit. To facilitate comparisons of the spectrum at different quark masses, the ratio of hadron masses with the $\Omega$ baryon mass is used to remove the explicit scale dependence~\cite{Lin:2008pr}.

\section{Correlator construction}\label{sec:distillation}

The determination of the excited spectrum proceeds from the calculation
of correlation functions between a basis of hermitian creation and annihilation operators $\cal O$
at Euclidean times $0$ and $t$,
\begin{equation}
   C_{ij}(t)  = \Big\langle 0 \Big|{\cal O}_i(t)\,  {\cal O}_j(0) \Big|0\Big\rangle. \nonumber
\end{equation}
Inserting a complete set of eigenstates of the Hamiltonian, such that 
$\hat{H} | \mathfrak{n} \rangle = E_\mathfrak{n}  | \mathfrak{n}\rangle$, this correlation function
decomposes into a sum of contributions from all states in the spectrum
with the same quantum numbers as the source operators,
\begin{equation}
   C_{ij}(t)  = \sum_\mathfrak{n} \frac{1}{2 E_\mathfrak{n}} \langle 0|{\cal O}_i | \mathfrak{n} \rangle 
  \langle \mathfrak{n}| {\cal O}_j|0 \rangle  \, e^{-{E_\mathfrak{n}} t},\nonumber
\end{equation}
where the discrete character of the spectrum follows because the calculation is performed in finite-volume. In order to measure energies of low-lying states, it is crucial to construct operators that overlap predominantly with light modes. 

Smearing is a well-established means to improve operator overlap, whereby a smoothing function is applied to the quark fields used in the creation operators.  This smoothing function should effectively remove noisy short-range modes which should not make a significant contribution to the low-energy correlation functions. The Jacobi method \cite{Allton:1993wc} uses the gauge-covariant second-order three-dimensional lattice Laplacian operator
\begin{equation}
   -\nabla^2_{xy}(t) = 6 \delta_{xy} - \sum_{j=1}^3
      \left(
	  \tilde{U}_j(x,t)         \delta_{x+\hat\jmath,y}
	 +\tilde{U}^\dagger_j(x-\hat\jmath,t) \delta_{x-\hat\jmath,y}
	  \right), \nonumber
\end{equation}
where the gauge fields, $\tilde{U}$ may be constructed from an appropriate covariant gauge-field-smearing algorithm \cite{Morningstar:2003gk}. To suppress high-energy modes of $\nabla^2$, this operator is exponentiated, $\exp{(\sigma\nabla^2)}$, with some smearing weight $\sigma$. The resulting smoothed operator is then applied to the quark fields $\psi$.

The suppression of the high energy modes of the Jacobi smearing operator $\exp{(\sigma\nabla^2)}$ means that only a small number of modes contribute significantly to the construction of the smeared quark fields, $\widetilde{\psi}$. As suggested in Ref.~\cite{Peardon:2009gh}, this smearing function can be replaced with a low-rank approximation. The ``distillation'' operator defines a smearing function
\begin{equation}
\Box_{xy}(t) = \sum_{k=1}^N \xi_x^{(k)} (t) \xi_y^{(k)\dag} (t),
    \label{eqn:box}
\end{equation}
where $\xi^{(k)}_x$ are a finite number, $N$, of eigenvectors of $\nabla^2$ evaluated on the background of the spatial gauge-fields of time-slice $t$, once the eigenvectors have been sorted by eigenvalue. This is the projection operator into the subspace spanned by these eigenmodes, so $\Box^2 = \Box$. 

This smearing function is used in the construction of isovector meson operators
of the form $\widetilde{\bar{\psi}} \boldsymbol{\Gamma} \widetilde{\psi}$, where
$\boldsymbol{\Gamma}$ acts in spin and color as well as coordinate space.
Applying the distillation operator $\Box$ onto each quark field, $\widetilde{\psi} \equiv \Box \psi$, the creation operators at zero three-momentum are written as
\begin{equation}
{\cal O}_i(t) 
  = \bar{\psi}_x(t) \Box_{xy}(t) \cdot \, \boldsymbol{\Gamma}^i_{yz}(t) \cdot \Box_{zw}(t) \psi_w(t), \nonumber
\end{equation}
where there is an implied volume summation over repeated spatial indices. In a shorthand notation the correlation function can be written as
\begin{equation}
  C_{ij}(t) = \Big\langle \bar{\psi}_t\Box_t \boldsymbol{\Gamma}^{i}_t\Box_t\psi_t\,\cdot\,
                      \bar{\psi}_0\Box_0 \boldsymbol{\Gamma}^{j}_0\Box_0 \psi_0
    \Big\rangle. \nonumber
\end{equation}
After evaluating the quark-field path-integral and inserting the outer-product 
definition of the distillation operator $\Box$ from Eq.~\ref{eqn:box}, the 
correlator can be written
\begin{equation}
  C_{ij}(t) = \mbox{Tr} \Bigl[
      \Phi^{j}(t) \,
      \tau(t,0) \,\Phi^{i}(0) \, \tau(0,t) \Bigr],\nonumber
\end{equation}
where
\begin{equation}
  \Phi^{i}_{\alpha\beta}(0) = \xi^\dagger(0) \left[ \boldsymbol{\Gamma}^{i}(0) \right]_{\alpha\beta} \xi(0) ,\nonumber
\end{equation}
encodes the structure of the operator and
\begin{equation}
  \tau_{\alpha\beta }(t,0) = \xi^\dagger(t) M^{-1}_{\alpha\beta}(t,0) \xi(0), \nonumber
\end{equation}
is the ``permabulator", with $M$ the lattice representation of the Dirac operator 
and where the quark spin indices, $\alpha,\beta$ of $\Phi$ and $\tau$ have been
explicitly written. $\Phi$ has a well-defined momentum, while there is no
explicit momentum projection in the definition of $\tau$. The $\Phi$ and $\tau$ are
square matrices of dimension $N  N_\sigma$ where $N_\sigma = 4$ are the number
of spin components in a lattice Dirac spinor. Construction of the $\tau$ require $N N_\sigma$ inversions
of the fermion matrix to compute all elements. These matrices are small compared to the dimension of the space of quark fields. Once the $\tau$ have been computed and stored, the correlation of any source and sink operators can be computed {\it a posteriori}. 
The method straightforwardly extends to the determination of multi-hadron
two-point correlation functions~\cite{Foley:2010te} 
as well as three-point functions~\cite{Peardon:2009gh}.

\section{Construction of meson operators}\label{sec:ops}

Meson spectral information will follow from analysis of two-point correlators featuring a large basis of composite QCD operators having mesonic quantum numbers. The simplest such operators are color-singlet local fermion bilinears, $\bar{\psi}_{i\alpha}(\vec{x},t)\Gamma_{\alpha\beta} \psi_{i\beta}(\vec{x},t)$ where the quantum numbers are determined by the choice of gamma matrix, $\Gamma$. In distillation \cite{Peardon:2009gh} the quark fields, $\psi$ are replaced by the smeared quark fields $\tilde{\psi}$ but the rotationally symmetric nature of the smearing does not change the quantum numbers of the bilinear operators.
These simple local operators are extremely limited in that they allow access only to the set $J^{PC} = 
0^{-+}, 0^{++}, 1^{--}, 1^{++}, 1^{+-}$ and they do not offer significant redundancy within any $J^{PC}$. In order to consider higher spins, exotic $J^{PC}$ and to produce multiple operators within a given symmetry channel, one must consider extending to the use of non-local operators\cite{Burch:2009wu, Petry:2008rt, Gattringer:2008be, Burch:2006dg, Burch:2004he, Lacock:1996vy}. Our approach is to use spatially-directed gauge-covariant derivatives within a fermion bilinear, that is to construct operators of essential structure
\begin{equation}
	\sum_{\vec{x}}  \bar{\psi}(\vec{x},t) \Gamma \overleftrightarrow{D}_i \overleftrightarrow{D}_j \ldots \psi(\vec{x},t) \nonumber
\end{equation}
where $\overleftrightarrow{D} \equiv \overleftarrow{D} - \overrightarrow{D}$ and where spin and color indices are suppressed for clarity. The use of the ``forward-backward" derivative, $\overleftrightarrow{D}$ is not strictly necessary at zero momentum (projected by the sum over spatial sites), the only case we consider here, but it does somewhat simplify the construction of eigen-operators of charge-conjugation as will be discussed below.

With the continuum $SO(3)$ rotational symmetry it is straightforward to produce operators of this type that are of definite spin, parity and charge-conjugation at zero momentum. This follows from forming a circular basis of the cartesian-vector-like derivatives and gamma matrices, $\overleftrightarrow{D}_i, \gamma_i, \gamma_5 \gamma_i, \gamma_0 \gamma_i, \epsilon_{ijk} \gamma_j \gamma_k$, e.g.
\begin{align}
  \overleftrightarrow{D}_{m=-1} &= \tfrac{i}{\sqrt{2}} \left(
    \overleftrightarrow{D}_x - i \overleftrightarrow{D}_y     \right)
  \nonumber\\
 \overleftrightarrow{D}_{m=0} &= 
    i \overleftrightarrow{D}_z
  \nonumber\\
  \overleftrightarrow{D}_{m=+1} &= -\tfrac{i}{\sqrt{2}} \left(
    \overleftrightarrow{D}_x + i \overleftrightarrow{D}_y     \right).\nonumber
\end{align} 
Once expressed in this basis, which transforms like spin-1, operators of definite spin can be constructed using the standard $SO(3)$ Clebsch-Gordan coefficients. For example, with a vector-like gamma matrix and one covariant derivative, operators of $J=0,1,2$ can be formed
\begin{equation}
 (\Gamma \times D^{[1]}_{J=1} )^{J, M} = \sum_{m_1, m_2}\big\langle 1, m_1 ; 1, m_2 \big| J, M
  \big\rangle\,  \bar{\psi} \Gamma_{m_1}
  \overleftrightarrow{D}_{m_2} \psi. \nonumber
\end{equation}
The choice of $\Gamma$ plays a role in setting the parity and charge-conjugation quantum numbers of the operator - our naming scheme for these matrices is given in Table \ref{table:gamma}. 

\begin{table}
	\begin{tabular}{r|cccccccc}
		 & 		$a_0$ & $\pi$ & 		$\pi_2$ & 		$b_0$ &		$\rho$ & 	$\rho_2$	&			$a_1$ 			& $b_1$\\ 
		\hline
		$\Gamma$ & 	$1$ & 	$\gamma_5$ & 	$\gamma_0\gamma_5$ & 	$\gamma_0$ &$\gamma_i$& $\gamma_i \gamma_0$ &	$\gamma_5 \gamma_i$ & $\gamma_i \gamma_j$
	\end{tabular}
\caption{Gamma matrix naming scheme. \label{table:gamma}}
\end{table}

At the two-derivative level we adopt the convention of first coupling the two derivatives to a definite spin, $J_D$, then coupling with the vector-like gamma matrix (if any) as
\begin{multline}
(\Gamma \times D^{[2]}_{J_D})^{J,M} = \sum_{\substack{m_1, m_2,\\ m_3, m_D}}\big\langle 1, m_3 ; J_D, m_D \big| J, M
  \big\rangle \nonumber\\  \times \big\langle 1, m_1 ; 1, m_2 \big| J_D, m_D
  \big\rangle\, \bar{\psi} \Gamma_{m_3}
  \overleftrightarrow{D}_{m_1} \overleftrightarrow{D}_{m_2} \psi.\nonumber
\end{multline}
It is worth noting here that while the Clebsch-Gordan for $1 \otimes 1 \to 1$ is antisymmetric and $\overleftrightarrow{D}_{m_1}\overleftrightarrow{D}_{m_2}$ appears to be symmetric, there are non-zero operators with $J_D = 1$ because the gauge-covariant derivatives do not commute with each other in QCD. Rather these ``commutator" operators are proportional to the gluonic field-strength tensor which does not vanish on non-trivial gluonic field configurations. 

At the three-derivative level we need to choose a convention for the ordering in which we couple the derivatives. A natural choice comes from insisting the operators have definite charge-conjugation symmetry. By exchanging the quark and anti-quark fields, the charge-conjugation operation effectively acts as a transpose of the operators between the quark
fields - for three derivatives then, one ensures definite
charge conjugation by coupling the outermost derivatives together first
since this gives them a definite exchange symmetry (even for $J_{13}=0,2$,
odd for $J_{13}=1$). This simple formulation is possible because we have used the ``forward-backward" derivatives, $\overleftrightarrow{D}$. 
\begin{multline}
  (\Gamma \times D^{[3]}_{J_{13}, J_D})^{J,M} = \\ \sum_{\substack{m_1,m_2,\\m_3,
    m_4, \\m_{13}, m_D}}\big\langle 1, m_4 ; J_D, m_D \big| J, M
  \big\rangle    \big\langle 1, m_2 ; J_{13}, m_{13} \big| J_D, m_D
  \big\rangle \\
  \times \big\langle 1, m_1 ; 1, m_3 \big| J_{13}, m_{13}
  \big\rangle  \bar{\psi} \Gamma_{m_4}
  \overleftrightarrow{D}_{m_1} \overleftrightarrow{D}_{m_2}  \overleftrightarrow{D}_{m_3}\psi. \nonumber
\end{multline}

Clearly this procedure can be extended to as many covariant derivatives as one wishes. In this paper we will use operators with up to three derivatives providing access to all $J^{PC}$ with $J \le 4$.\footnote{except the exotic $4^{+-}$ which requires a minimum of four derivatives.} 

The operators as formed are eigenstates of charge-conjugation in the case that the $\psi$ and $\bar{\psi}$ fields are of the same flavour. In the case that the fields are degenerate but not identical (e.g. the $u$ and $d$ quarks in our calculation), the $C$-parity is trivially generalised to $G$-parity. For kaons, where the light and strange quarks are not degenerate, there is no $C$-parity or any generalisation of it. In this case the symmetry channels are labelled by $J^P$ and operators of both $C$ can be used.

\subsection{Subduction into lattice irreps.}\label{Sec:Subduction}

In lattice QCD calculations the theory is discretized on a four-dimensional
hypercubic Euclidean grid.  The full three-dimensional rotational symmetry that classifies energy eigenstates in the continuum is hence reduced to the symmetry group of a cube (the cubic symmetry group, or equivalently the octahedral group).  Instead of the infinite number of irreducible representations labelled by spin $J$, the single-cover cubic group relevant for integer spin has only five irreducible representations (\emph{irreps}): $A_1$, $T_1$, $T_2$, $E$, $A_2$.  The distribution of the various $M$ components of a spin-$J$ meson into the lattice irreps is known as \emph{subduction}, the result of which is shown in Table \ref{Table:Subduce}.

\begin{table}
\begin{tabular}{ccl}
$J$ & & irreps \\
\hline
$0$ & & $A_1(1)$ \\
$1$ & & $T_1(3)$ \\
$2$ & & $T_2(3) \oplus E(2)$\\
$3$ & & $T_1(3) \oplus T_2(3) \oplus A_2(1)$\\
$4$ & & $A_1(1) \oplus T_1(3) \oplus T_2(3) \oplus E(2)$
\end{tabular}  
\caption{Continuum spins subduced into lattice irreps $\Lambda(\mathrm{dim})$.}
\label{Table:Subduce}
\end{table}

To be of any use in lattice computations, the continuum operators described above must be subduced into lattice irreps.  Noting that each class of operator is closed under rotations, the subductions can be performed using known linear combinations of the $M$ components for each $J$:
\begin{eqnarray}
\lefteqn{{\cal O}^{[J]}_{\Lambda,\lambda} \equiv (\Gamma \times D^{[n_D]}_{\ldots})^J_{\Lambda, \lambda} = } \nonumber\\
& & \sum_M {\cal
     S}^{J,M}_{\Lambda, \lambda}   \; (\Gamma \times
   D^{[n_D]}_{\ldots})^{J,M} \equiv \sum_M {\cal S}^{J,M}_{\Lambda,\lambda} {\cal O}^{J,M}, \nonumber
\end{eqnarray}
where $\lambda$ is the ``row'' of the irrep ($1\ldots\mathrm{dim}(\Lambda)$).  Note that, although ${\cal O}^{[J]}_{\Lambda, \lambda}$ can have an overlap with all spins contained within $\Lambda$ (as listed in Table \ref{Table:Subduce} for $J\le 4$) it still carries the memory of the $J$ from which it was subduced, a feature we exploit in Section \ref{sec:spin}.  
The subduction coefficients, ${\cal S}^{J,M}_{\Lambda, \lambda}$, form an orthogonal matrix, $\sum_M {\cal S}^{J,M}_{\Lambda, \lambda} {\cal S}^{J,M*}_{\Lambda', \lambda'} = \delta_{\Lambda, \Lambda'} \delta_{\lambda, \lambda'}$, and this fixes their normalisation.  

The subduction coefficients can be constructed in a number of different ways and here we give a simple derivation.  More details and an alternative method using a group-theoretic projection formula can be found in Appendix\ \ref{App:SubductionCoeffs}.

The simplest case is the subduction of the $J=0$ operator; from Table \ref{Table:Subduce} this only subduces into the $A_1$ irrep and so trivially we have $\mathcal{S}^{0,0}_{A_1, 1} = 1$.  The $J=1$ operator is also relatively straightforward, only subducing into the the $T_1$ irrep with subduction coefficients shown in Appendix \ref{App:SubductionCoeffs}.  
Note that $\mathcal{S}^{1,M}_{T_1, \lambda} = \delta_{\lambda, 2-M}$, where the shift by 2 places $\lambda$ in the range $1\ldots \mathrm{dim}(\Lambda)$.

Subduction coefficients for all higher spins can be constructed by iteration, starting from the $J=0$ and $J=1$ coefficients and using
\begin{eqnarray}
\mathcal{S}^{J,M}_{\Lambda, \lambda} &=& N \sum_{\lambda_1, \lambda_2} \sum_{M_1, M_2} 
\mathcal{S}^{J_1,M_1}_{\Lambda_1, \lambda_1} \mathcal{S}^{J_2,M_2}_{\Lambda_2, \lambda_2} \nonumber \\
&& C\Bigl( \begin{array}{ccc}\Lambda & \Lambda_1 & \Lambda_2 \\ \lambda & \lambda_1 & \lambda_2\end{array} \Bigr) 
\left<J_1, M_1; J_2, M_2 | J, M \right>  ~ . \nonumber
\label{Equ:SubductionIteration}
\end{eqnarray}
Here $\langle J_1, M_1; J_2, M_2 | J, M \rangle$ is the usual $SO(3)$ Clebsch-Gordan coefficient for $J_1 \otimes J_2 \rightarrow J$ and $C\Bigl( \begin{array}{ccc}\Lambda & \Lambda_1 & \Lambda_2 \\ \lambda & \lambda_1 & \lambda_2\end{array} \Bigr)$ is the octahedral group Clebsch-Gordan coefficient for $\Lambda_1 \otimes \Lambda_2 \rightarrow \Lambda$.  $N$ is a normalisation factor, fixed by the requirement that the subduction coefficients form an orthogonal matrix as discussed above.  We give explicit values for the subduction coefficients up to $J=4$ in Appendix \ref{App:SubductionCoeffs}.

\begin{table}
\begin{tabular}{cc|cc|cc|cc}
$A_1^{++}$ & 13  & $A_1^{+-}$ & 5 & $A_1^{-+}$ & 12 & $A_1^{--}$ & 6 \\
$T_1^{++}$ & 22 & $T_1^{+-}$ & 22 & $T_1^{-+}$ & 18 & $T_1^{--}$ & 26
\\
$T_2^{++}$ & 22 & $T_2^{+-}$ & 14 & $T_2^{-+}$ & 18 & $T_2^{--}$ & 18
\\
$E^{++}$ & 17 & $E^{+-}$ & 9 & $E^{-+}$ & 14 & $E^{--}$ & 12
\\
$A_2^{++}$ & 5 & $A_2^{+-}$ & 5 & $A_2^{-+}$ & 4 & $A_2^{--}$ & 6 
\end{tabular}  
\caption{Number of operators in each lattice irrep $\Lambda^{PC}$, using all operators with up to three derivatives.}
\label{table:opnumbers}
\end{table}

In Table \ref{table:opnumbers} we show the number of operators we have in each lattice irrep, i.e.\ using all operators with up to three derivatives.  We have performed extensive tests of this operator set to check that two-point correlators having operators in differing irreps at source and sink are consistent with zero and that similarly within an irrep, correlators of differing rows at source and sink are consistent with zero. All such ``orthogonality tests'' are passed in explicit calculation.

For our final spectral extractions, we form a correlator matrix in a given irrep $\Lambda$ and average over equivalent rows, $\lambda$, 
\begin{equation}
 C^{\Lambda}_{ij} \equiv \tfrac{1}{\mathrm{dim}(\Lambda)} \sum_{\lambda=1}^{\mathrm{dim}(\Lambda)}
  C^{\Lambda}_{i\lambda, j\lambda} \equiv  \tfrac{1}{\mathrm{dim}(\Lambda)} \sum_\lambda
  \langle 0 | {\cal O}^{[J]}_{i(\Lambda) \lambda}  {\cal
    O}^{[J]}_{j(\Lambda) \lambda} | 0 \rangle,  \nonumber
\end{equation}
where $i,j$ labels the different operator constructions within the irrep $\Lambda$.

\pagebreak
\section{Correlator analysis}\label{sec:fitting}

The variational method for spectral extraction \cite{Michael:1985ne,Luscher:1990ck}, which takes advantage of a redundancy of operators within a given symmetry channel, is now in common usage \cite{Burch:2009wu, Gattringer:2008be, Burch:2006dg, Burch:2004he, Dudek:2007wv}. This method finds the best (in a variational sense) linear combination of operators within a finite basis for each state in the spectrum. Mathematically it boils down to the solution of a linear system of generalised eigenvalue type:
\begin{equation}
	C(t) v^\mathfrak{n}(t) = \lambda_\mathfrak{n}(t) C(t_0)  v^\mathfrak{n}(t) \label{var}
\end{equation}
where $\lambda_\mathfrak{n}(t_0) = 1$ and where there is an orthogonality
condition on the eigenvectors of different states ($\mathfrak{n},
    \,\mathfrak{n}'$), $v^{\mathfrak{n}'\dag} C(t_0) v^\mathfrak{n} =
\delta_{\mathfrak{n}, \mathfrak{n}'}$. As discussed in Ref.~\cite{Dudek:2007wv} this orthogonality condition is very powerful in extracting near degenerate states which would be difficult to distinguish by their time dependence alone.

In our particular implementation of this method, equation \ref{var} is solved for eigenvalues $\lambda_\mathfrak{n}$ and eigenvectors $v^\mathfrak{n}$, independently on each timeslice, $t$. Ensuring the same ordering of states between timeslices requires some care owing to the high degree of mass degeneracy in the meson spectrum. Rather than the obvious ordering by size of eigenvalue which might fluctuate timeslice-by-timeslice for nearby masses, we associate states between timeslices using the similarity of their eigenvectors. We choose a reference timeslice on which reference eigenvectors are defined, $v^\mathfrak{n}_{\mathrm{ref}} \equiv v^{\mathfrak{n}}(t_\mathrm{ref})$, and compare eigenvectors on other timeslices by finding the maximum value of $v^{\mathfrak{n}'\dag}_\mathrm{ref} C(t_0) v^{\mathfrak{n}}$ which associates a state $\mathfrak{n}$ with a reference state $\mathfrak{n}'$. Using this procedure we observe essentially no ``flipping" between states in either the principal correlators, $\lambda_\mathfrak{n}(t)$ or the eigenvectors $v^\mathfrak{n}(t)$, as functions of $t$.

\begin{figure}
 \centering
\includegraphics[width=0.5\textwidth, bb=0 0 514 661]{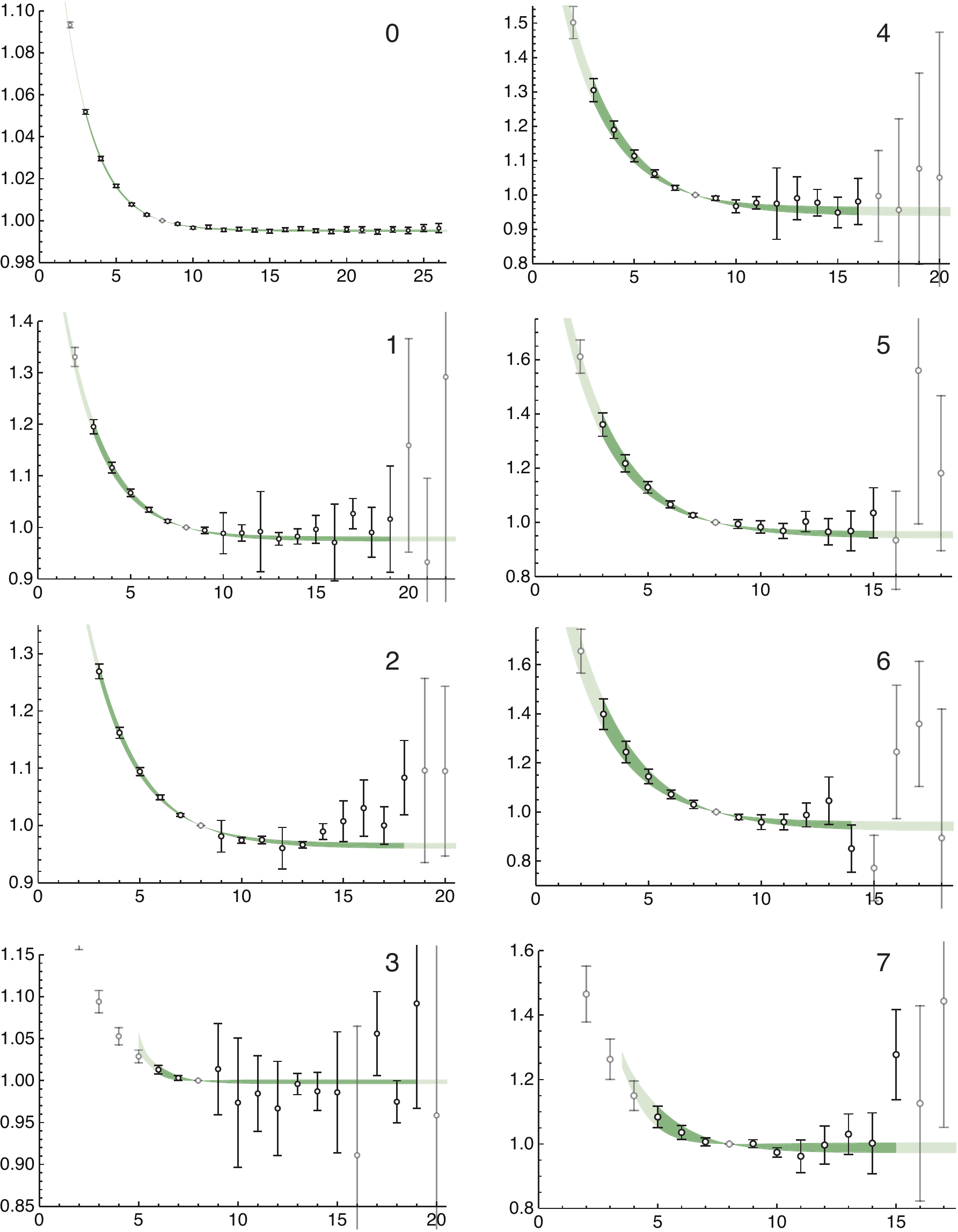}
\caption{Principal correlator fits according to eqn(\ref{eqn:fitprincorr}).  Eight states from the $T_1^{--}$ irrep (\emph{743}, $16^3$). Plotted are $\lambda^{\mathfrak{n}}(t)~\!\cdot~\!e^{m_\mathfrak{n}(t-t_0)}$ data and the fit for $t_0 = 8$. Data used in the fit are shown in black, while points excluded from the fit are in grey.  
\label{fig:princorrfits}}
\end{figure}

Within finite-volume field theory, which has only discrete eigenstates, any two-point correlator can be expressed as a spectral decomposition, 
\begin{equation}
	C_{ij}(t) = \sum_\mathfrak{n} \frac{Z_i^{\mathfrak{n}*} Z_j^{\mathfrak{n}}}{2m_\mathfrak{n}} e^{-m_\mathfrak{n} t}
	\label{spectro_decomp}
\end{equation}
where this is an approximation valid providing $t \ll L_t$, the temporal length of the box. The ``overlap factors", $Z^\mathfrak{n}_i \equiv \langle \mathfrak{n} | {\cal O}_i | 0 \rangle$ are related to the eigenvectors by $Z^\mathfrak{n}_i = \sqrt{2 m_\mathfrak{n}} e^{m_\mathfrak{n} t_0/2}\, v^{\mathfrak{n}*}_j C_{ji}(t_0)$. The state masses follow from fitting the principal correlators, $\lambda_\mathfrak{n}(t)$, which for large times should tend to $e^{- m_\mathfrak{n} (t - t_0)}$. In practice we allow a second exponential in the fit form and use even relatively low timeslices in order to stabilise the fit. The fit function is
\begin{equation}
	\lambda_\mathfrak{n}(t) = (1 - A_\mathfrak{n}) e^{-m_\mathfrak{n}(t-t_0)} + A_\mathfrak{n} e^{-m_\mathfrak{n}' (t-t_0)},\label{eqn:fitprincorr}
\end{equation}
where the fit parameters are $m_\mathfrak{n}, m_\mathfrak{n}'$ and $A_\mathfrak{n}$. Typical fits for a set of excited states within an irrep are shown in figure \ref{fig:princorrfits} where we plot the principal correlator with the dominant time-dependence due to state $\mathfrak{n}$ divided out. In such a plot one would see a horizontal line of value 1.0 in the case that a single exponential dominates the fit and clearly the data shows flat behaviour for $t>t_0$. 

Empirically we find that the size of the second exponential term decreases rapidly as one increases $t_0$. Further we find, in agreement with the perturbative analysis of Ref.~\cite{Blossier:2009kd}, that for large $t_0$ values the $m_\mathfrak{n}'$ extracted are larger than the value of $m_{\mathfrak{n}=\mathrm{dim(C)}}$, the largest ``first" exponential mass extracted. At smaller $t_0$ values this is not necessarily true and is indicative of forcing an incorrect orthogonality as discussed below. The values of $A_\mathfrak{n}$ and $m_\mathfrak{n}'$ are not used elsewhere in the analysis.

From the spectral decomposition of the correlator, equation \ref{spectro_decomp}, it is clear that there should in fact be no time dependence in the eigenvectors, while our independent solution of the generalised eigenvalue problem as a function of $t$ has allowed there to be. The time-independent overlap factors, $Z^\mathfrak{n}_i$, which will be used later to identify the spin of extracted states, follow from fitting the $Z^\mathfrak{n}_i(t)$, obtained from the eigenvectors, with a constant or a constant plus an exponential (in the spirit of the perturbative corrections outlined in \cite{Blossier:2009kd}).

The importance of choosing an appropriately large value of $t_0$ was emphasised
in Ref.~\cite{Dudek:2007wv}. In this paper we will follow the ``reconstruction"
scheme outlined therein in the selection of $t_0$. In short, the masses, $m_\mathfrak{n}$, extracted from fits to the principal correlators and the $Z^\mathfrak{n}_i$ extracted from the eigenvectors on a single timeslice are used in equation \ref{spectro_decomp} to ``reconstruct" the correlator matrix. This reconstructed matrix is compared to the original data for all $t>t_0$ with the degree of agreement indicating the acceptability of the spectral description. The description generally improves as one increases $t_0$ until at some point the increase in statistical noise prevents further improvement. In particular see figure 6 in Ref.~\cite{Dudek:2007wv} where the effect of choosing $t_0$ too small is clearly seen. Forcing the $\mathrm{dim}(C)$-state orthogonality, $v^{\mathfrak{m}\dag}\, C(t_0) \, v^\mathfrak{n} = \delta_{\mathfrak{n}, \mathfrak{m}}$, in a situation where accurate description of $C(t_0)$ requires more than $\mathrm{dim}(C)$ states leads to a poor description of the correlator matrix at times $t > t_0$. The reconstruction procedure gives a guide to the minimal $t_0$ for which the correlator matrix is well described by the variational solution. The sensitivity of extracted spectral quantities to the value of $t_0$ used will be discussed in detail in section \ref{sec:t0}, but in short it is usually necessary for us to use $t_0 \gtrsim 7$. 

The {\tt reconfit2} code used for variational analysis is available within the {\tt adat} suite \cite{adat}.

\section{Determining the spin of a state}\label{sec:spin}

\begin{figure}
 \centering
\includegraphics[width=0.35\textwidth,bb= 0 0 264 257]{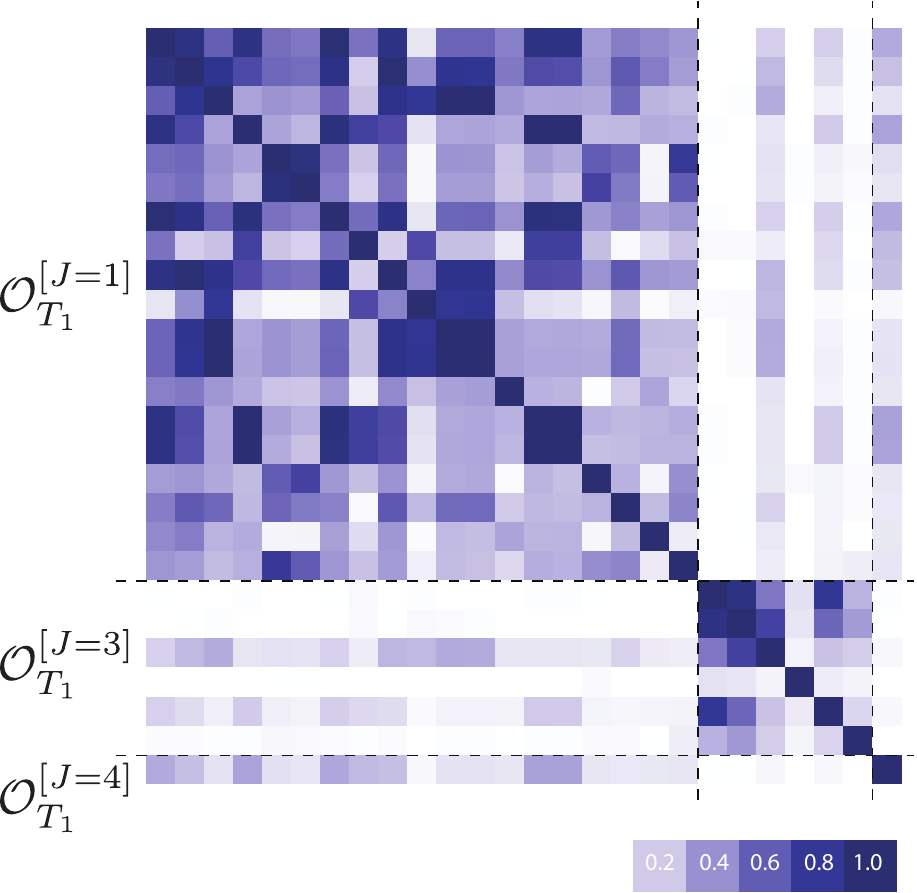}        
\caption{Normalised corrrelation matrix ($C_{ij}/\sqrt{C_{ii} C_{jj}}$) on timeslice 5 in the $T_1^{--}$ irrep (\emph{743}, $16^3$). Operators are ordered such that those subduced from spin 1 appear first followed by spin 3 then spin 4. \label{matrixplot}}
\end{figure}

\begin{figure}
 \centering
\includegraphics[width=0.5\textwidth,bb= 0 0 578 754]{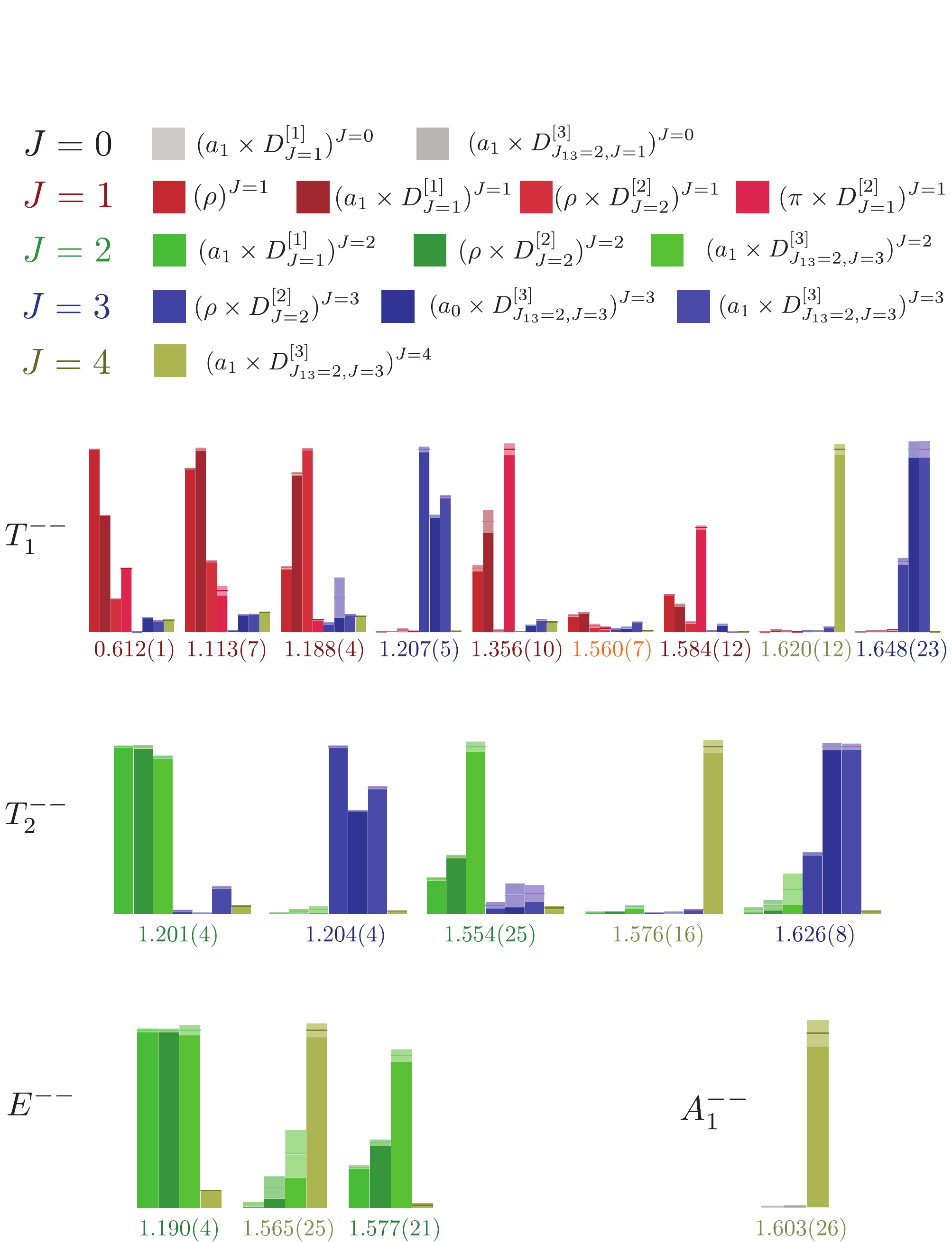}        
\caption{Overlaps, $Z$, of a selection of operators onto states labelled by $m/m_\Omega$ in each lattice
   irrep, $\Lambda^{--}$ (\emph{743}, $16^3$). $Z$'s are normalised so that the largest value across all
   states is equal to $1$. Lighter area at the head of each bar
   represents the one sigma statistical uncertainly.\label{histogram}}
\end{figure}

\begin{figure*}
 \centering
\includegraphics[width=0.95\textwidth,bb= 0 0 1393 387]{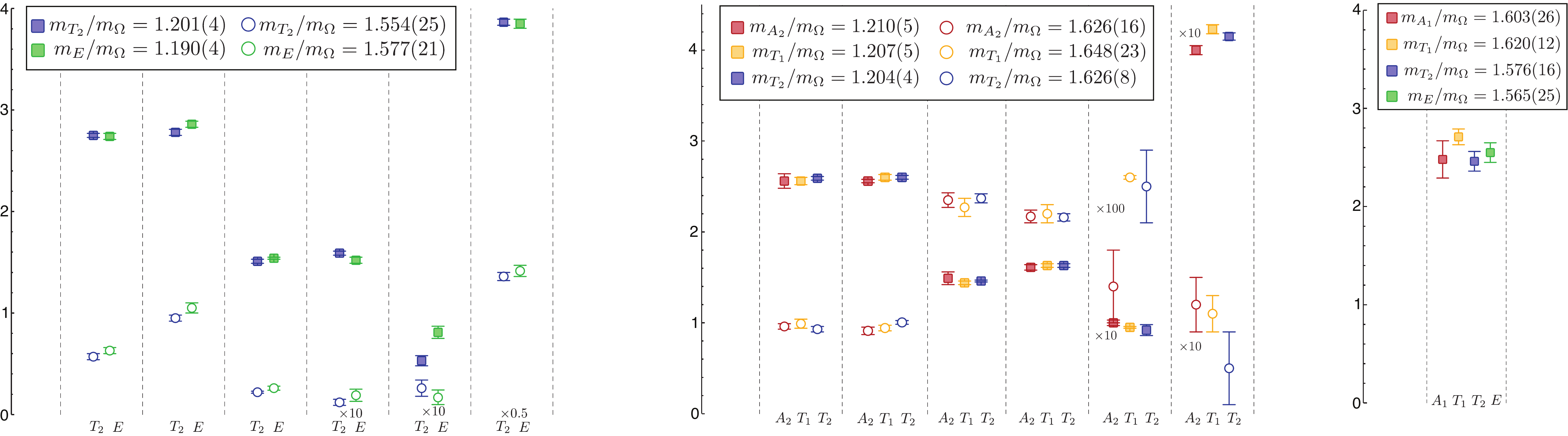}  
 \caption{A selection of $Z$ values across irreps $\Lambda^{--}$ for states suspected of being
   $J=2,3,4$ (\emph{743}, $16^3$).  Left to right the operators are $(a_1\times D^{[1]}_{J=1})^{J=2}, (\rho\times D^{[2]}_{J=2})^{J=2}, (\rho_2 \times D^{[2]}_{J=2})^{J=2}, (a_0\times D^{[3]}_{J_{13}=2, J=2})^{J=2}, (b_0\times D^{[3]}_{J_{13}=1, J=2})^{J=2}, (a_1\times D^{[3]}_{J_{13}=0, J=1})^{J=2},      (\rho\times D^{[2]}_{J=2})^{J=3}, (\rho_2 \times D^{[2]}_{J=2})^{J=3}, (a_0 \times D^{[3]}_{J_{13}=2, J=3})^{J=3}, (a_1 \times D^{[3]}_{J_{13}=2, J=3})^{J=3},  (a_1 \times D^{[3]}_{J_{13}=2, J=2})^{J=3}, (b_1 \times D^{[3]}_{J_{13}=1, J=2})^{J=3}$ and $ (a_1 \times D^{[3]}_{J_{13}=2, J=3})^{J=4}  $. }
 \label{Zvalues}
\end{figure*}

In principle the most rigourous method to determine the spin of a
state is to perform the extraction of the spectrum for each lattice
irrep at successively finer lattice spacings, and then to
extrapolate the energies in each irrep to the continuum limit. There
one expects to see degeneracies emerge according to the pattern of
subduction, free of splittings arising from the discretisation 
effects.  Thus, for example, a spin-3 state would appear as
degenerate energies within the $A_2, T_1$ and $T_2$ irreps. This
procedure has been successfully applied to identify a number of
low-lying states in the calculation of the glueball spectrum within
pure $SU(3)$ Yang-Mills theory\cite{Morningstar:1999rf}.

There are two reasons why this technique is not currently practical
for the QCD meson spectrum. Firstly, the procedure relies on a series
of calculations on progressively finer lattices, and hence at
increasing computational cost.  
Secondly, the
continuum spectrum, classified according to the continuum quantum
numbers, exhibits a high degree of degeneracy; when classified
according to the symmetries of the cube, the degree of degeneracy is
vastly magnified.  Identification of degeneracies between irreps 
would require a statistical precision far beyond even that of the
high-quality data presented here, as seen in Figure~\ref{743_16_irrep}
and subsequent figures.

To alleviate these difficulties it would be useful to have a
spin-identification method that is effective when using data obtained
at only a single lattice spacing. Obviously this lattice spacing
should be fine enough that rotation symmetry has been restored to a
sufficient degree in order that it be describing QCD. The mass
degeneracy complications outlined above suggest that any alternative
method needs to use state information beyond just the mass. Our
suggestion is to consider the values of the vacuum-to-state matrix
elements, or ``overlaps" ($\big\langle \mathfrak{n}\big| {\cal O}
\big|0\big\rangle$) of our carefully constructed subduced operators.

The operators constructed in Section \ref{sec:ops} transform irreducibly under the
allowed cubic rotations, that is they faithfully respect the symmetries of the
lattice. However it is also clear from the method of construction that each
operator ${\cal O}^{[J]}_{\Lambda}$ carries a ``memory" of the continuum spin,
$J$, from which it was subduced. If our lattice is reasonably close to
restoring rotational symmetry we would expect an operator subduced from spin $J$ to overlap strongly only onto states of continuum spin $J$. In fact this is clearly apparent even at the level of the correlator matrix as seen in figure \ref{matrixplot}. Here the correlator matrix for $T_1^{--}$ is observed to be approximately block diagonal when the operators are ordered according to the spin from which they were subduced. The elements outside the diagonal blocks are smallest when the operators feature zero, one or two derivatives and are somewhat larger for three-derivative operators, possible reasons for this will be discussed later in this section.

The effect is seen even more strongly at the level of individual states, where the ``overlaps", $Z^\mathfrak{n}_i = \langle \mathfrak{n} | {\cal O}_i |0\rangle$ for a given state show a clear preference for overlap onto only operators of a single spin. 
In figure \ref{histogram} we show the overlaps for a set of low-lying states in the $\Lambda^{--}$ irreps of the \emph{743} $16^3$ calculation, the mass spectrum being shown in the second pane of figure \ref{743_16_irrep}. While we show only a subset of the operators for clarity, the same pattern is observed for the full operator set. 

\begin{figure*}
 \centering
\includegraphics[width=0.95\textwidth,bb= 0 0 1012 155]{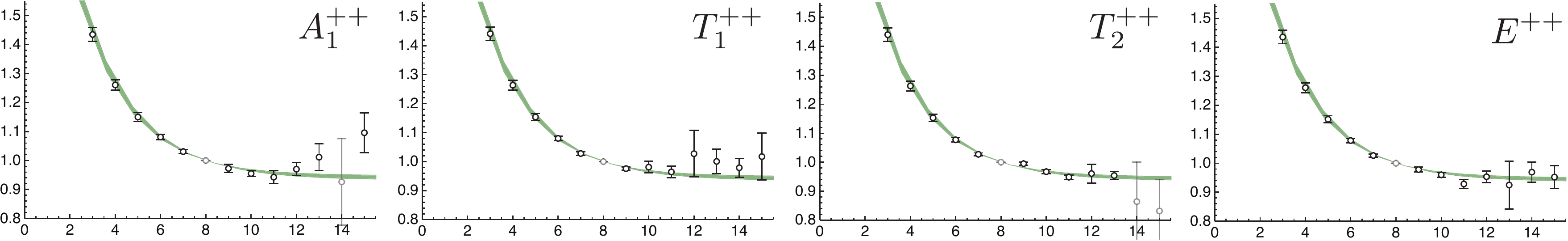}  
 \caption{Fit to the four subduced principal correlators of a $4^{++}$ meson using a common mass (\emph{743}, $16^3$). Plotted is $\lambda(t) \cdot e^{m(t-t_0)}$. Grey points not included in the fit.}
 \label{fit_across_irreps}
\end{figure*}

In fact we can be more quantitative in our analysis and compare the
overlaps obtained in different irreps.  In the continuum our operators
are of definite spin such that $\langle 0 | {\cal O}^{J,M} |J',
M'\rangle = Z^{[J]} \delta_{J,J'} \delta_{M,M'}$ and therefore
$\langle 0 | {\cal O}^{[J]}_{\Lambda, \lambda} | J', M\rangle = {\cal
  S}^{J,M}_{\Lambda, \lambda} Z^{[J]} \delta_{J,J'}$ so that only the
spin $J$ states will contribute, and not any of the other spins
present in the irrep $\Lambda$.  Inserting a complete set of meson
states between the operators in the correlator and using the fact that
the subduction coefficients form an orthogonal matrix, $\sum_M {\cal
  S}^{J,M}_{\Lambda, \lambda} {\cal S}^{J,M*}_{\Lambda', \lambda'} =
\delta_{\Lambda, \Lambda'} \delta_{\lambda, \lambda'}$, we obtain
terms in the correlator spectral decomposition proportional to
$Z^{[J]*} Z^{[J]}$; these terms do not depend upon which $\Lambda$ we
have subduced into, up to discretisation uncertainties as described
below. Hence, for example, a $J=3$ meson created by a $[J=3]$ operator
will have the same $Z$ value in each of the $A_2, T_1, T_2$
irreps. This suggests that we compare the independently obtained
$Z$-values in each irrep.  In figure \ref{Zvalues} we show the
extracted $Z$ values for states suspected of being spin 2,3 and 4
across the $\Lambda^{--}$ irreps.

As seen in figure \ref{Zvalues}, at finite lattice spacing there are deviations
from exact equality. Some discretisation effects scale with positive powers of
the lattice spacing, such as the effect of using finite differences to represent
derivatives. 
There are no dimension-five operators made of quark bilinears that respect the
symmetries of lattice actions based on the Wilson formalism and that do not also
transform trivially under the continuum group of spatial rotations. Thus, 
rotational symmetry breaking terms do not appear until ${\cal O}(a^2)$. This
argument holds even though the action used in this work describes an anisotropic
lattice. 
As a result, we expect the rotational breaking between lattice irreps to be suppressed in both the spectrum as well as for the wave-function overlaps. 
On the other hand, renormalisation mixing of high mass-dimension operators with lower mass dimension operators can give rise to effects scaling with negative powers of the lattice spacing and these we would expect to cause more trouble when $Z$ values are to be used to determine spin. In practice we do see the largest discrepancies for operators with three derivatives, but even here the effects are not so large as to prevent use of the method. We suggest that it is our use of smoothed fields\footnote{All gauge links are stout-smeared and the distilled quark fields are effectively low-momentum filtered} that has rendered these mixings relatively small, sensitive as they are to high-energy physics which has been filtered out. 

In summary, we have demonstrated that the $Z$ values of carefully constructed subduced operators can be used to identify the continuum spin of states extracted in explicit computation, as least on the lattices we have used.

Given that this is possible, suppose we confidently identify the components of a spin-$J$ meson subduced across multiple irreps; what then should we quote as our best estimate of the mass of the state? The mass determined from fits to principal correlators in each irrep can differ slightly due to unavoidable discretisation effects and avoidable fitting fluctuations (such as variations in fitting time-region). In practice we have found that variations due to changes in the fitting of principal correlators are typically much larger than any discretisation differences and we propose a simple scheme to minimise these in a final quoted mass. Rather than averaging the masses from independent fits to multiple principal correlators, we perform a joint fit to the principal correlators with the mass being common. We allow a differing second exponential in each principal correlator so that the fit parameters are $m_\mathfrak{n}, \{ m^{'\Lambda}_\mathfrak{n}\} $ and $\{A_\mathfrak{n}^\Lambda \}$. These fits are typically very successful with correlated $\chi^2/N_\mathrm{dof}$ close to 1. An example for the case of $4^{++}$ components identified in $A_1^{++}, T_1^{++}, T_2^{++}, E^{++}$ is shown in figure \ref{fit_across_irreps}. When we present our final, spin-assigned spectra it is the results of such fits that we show.

\section{Stability of spectrum extraction}\label{sec:stability}

In this section we consider to what extent the extracted spectrum changes as we vary details of the calculation, ``keeping the physics constant".  Variations to be considered are the specific reference timeslice, $t_0$, used in the variational analysis, the set of meson operators used and the number of distillation vectors.
We will use the $T_1^{--}$ irrep in the \emph{743}, $16^3$ dataset to demonstrate our findings.
\subsection{Variational analysis and $t_0$}\label{sec:t0}
\begin{figure}
 \centering
\includegraphics[width=0.5\textwidth,bb= 0 0 723 419]{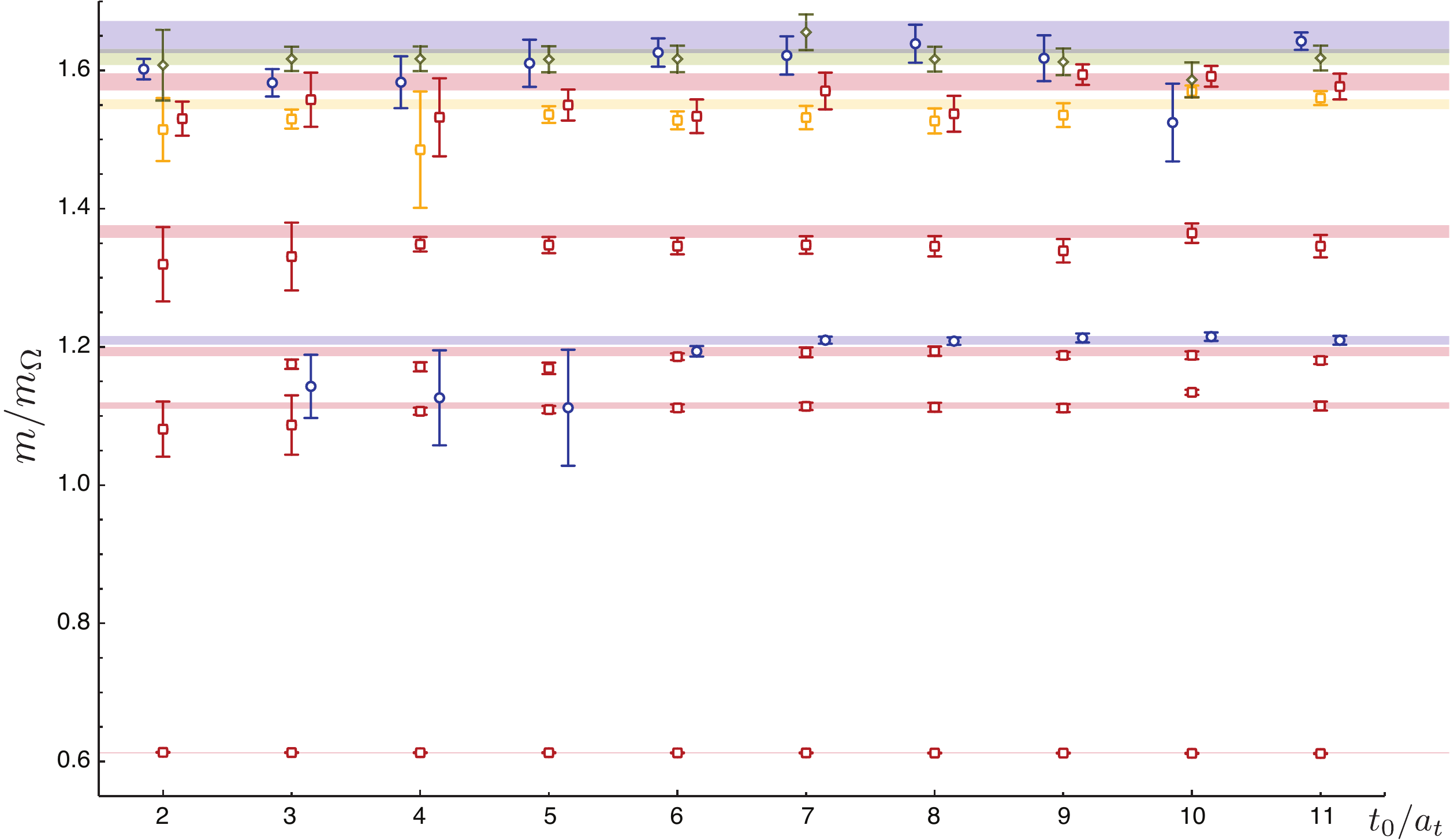}        
\caption{Extracted $T_1^{--}$ mass spectrum as a function of $t_0$. Horizontal bands to guide the eye. \label{masses_vs_t0}}
\end{figure}

Our fitting methodology was described in Section \ref{sec:fitting} where reconstruction of the correlator was used to guide us to an appropriate value of $t_0$. As seen in Figure \ref{masses_vs_t0}, for $t_0 \gtrsim 6$, the low-lying mass spectrum is rather stable under variations of $t_0$. This appears to be mostly due to the inclusion of a second exponential term in equation \ref{eqn:fitprincorr} which is able to absorb much of the effect of other states ``leaking" into this principal correlator through use of an inaccurate orthogonality. The contribution of this second exponential typically falls rapidly with increasing $t_0$ both by having a smaller $A$ and a larger $m'$.

\begin{figure}[b]
 \centering
\includegraphics[width=0.45\textwidth,bb= 0 0 560 343]{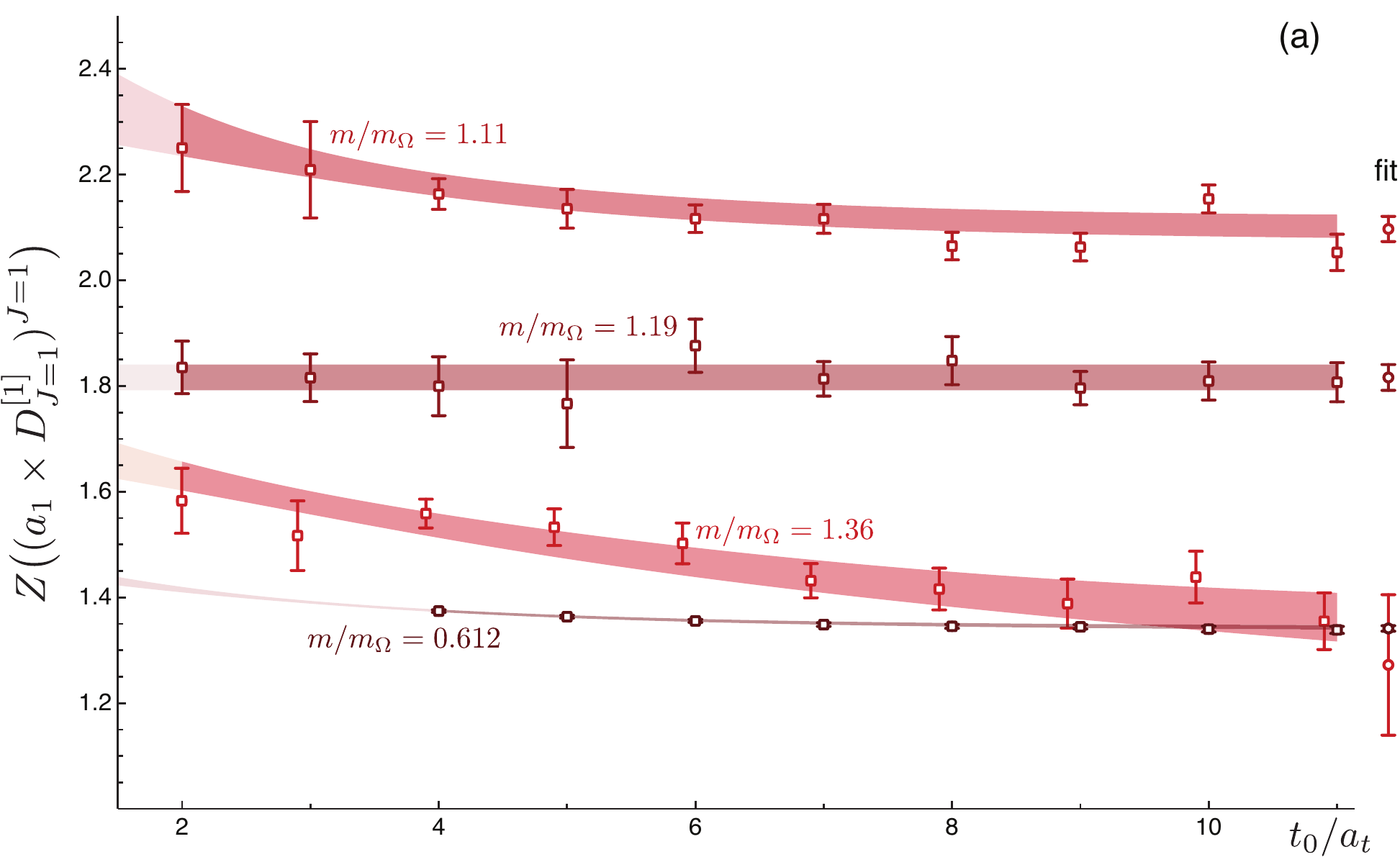}  
\includegraphics[width=0.45\textwidth,bb= 0 0 627 384]{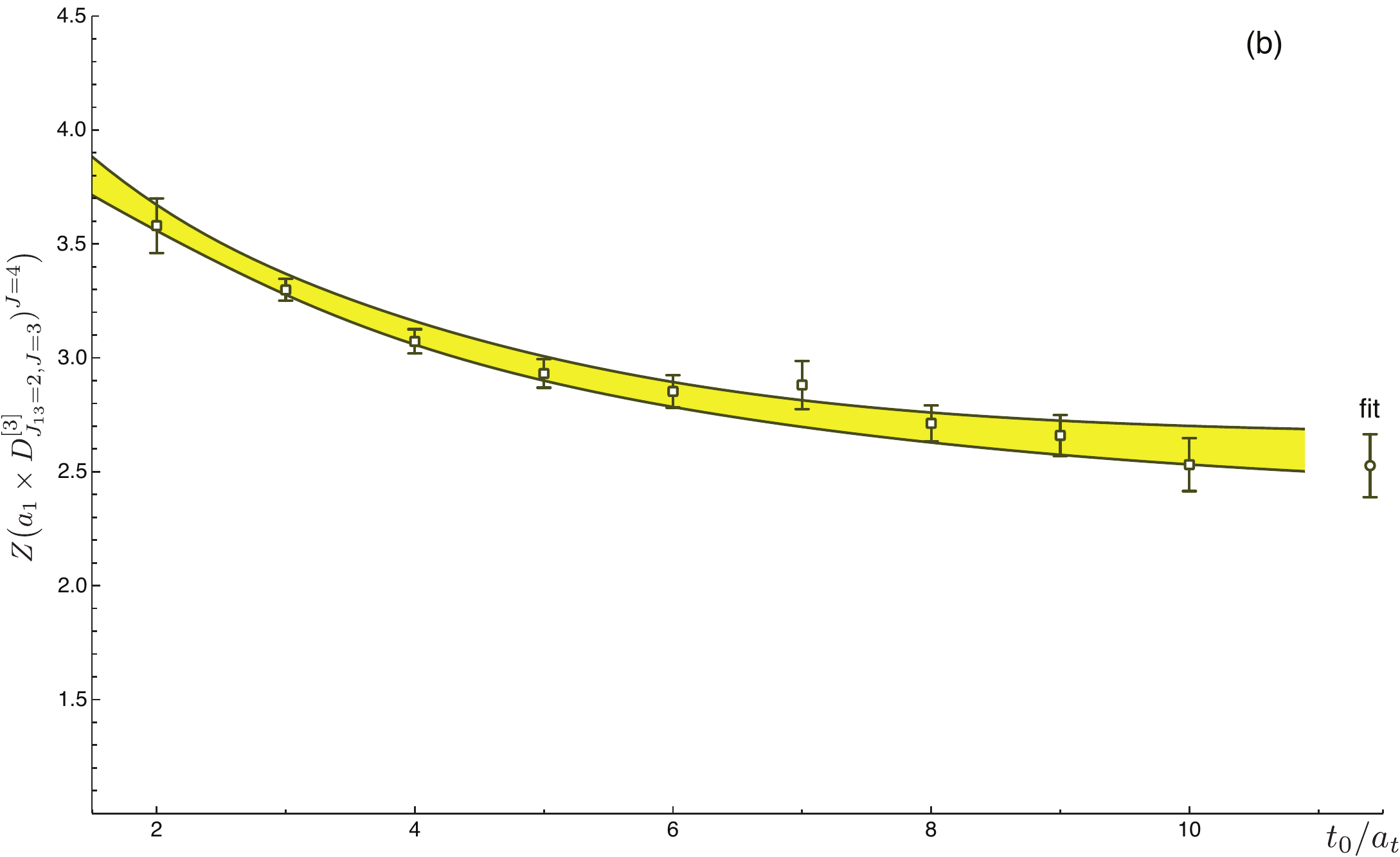}        
\caption{Extracted overlaps as a function of $t_0$. Fits to a constant or to a constant plus a decaying exponential shown by the coloured regions. \\
(a) $J^{PC}=1^{--}$ overlaps onto $(a_1 \times D^{[1]}_{J=1})^{J=1}$ \\(b) $J^{PC}=4^{--}$ overlap onto $(a_1 \times D^{[3]}_{J_{13}=2,J=3})^{J=4}$. \label{Z_vs_t0}}
 
\end{figure}

Overlaps, $Z^i_\mathfrak{n} = \big\langle \mathfrak{n} \big| {\cal O}_i \big| 0 \big\rangle$, can show more of a sensitivity to $t_0$ values being too low as one might expect given the argument of an incorrect orthogonality in the generalised eigenvector space at small $t_0$. In Figure \ref{Z_vs_t0} we present overlaps of various $J=1$ states onto an operator subduced from $J=1$ and the overlap of an extracted $J=4$ state onto the only $J^{PC}=4^{--}$ operator in our basis. Clearly in the $J=4$ case one only extracts a stable $Z$ for large $t_0$, which is likely due to heavier $J=4$ states only here becoming negligible contributions to $C(t_0)$. 

We note that one may fit the extracted $Z(t_0)$ with either a constant or a constant plus an exponential as shown in figure \ref{Z_vs_t0}. Since the data between $t_0$ values are strongly correlated, the statistical uncertainty is not significantly decreased by this procedure, but it does seem to ``average out" some of the fluctuations in fitting at each $t_0$.

In summary it appears that variational fitting is reliable provided $t_0$ is ``large enough". Using two-exponential fits in principal correlators we observe relatively small $t_0$ dependence of masses, but more significant dependence for the $Z$ values which we require for spin-identification.

\subsection{Changing the operator basis}\label{sec:basis}
Our approach, as described in Section \ref{sec:ops}, is to construct a variational correlator matrix featuring all operators available to us in a given irrep at up to three derivatives. Here we consider the effect on the extracted spectrum of reducing the size of this operator basis. In the plots that follow we will use the color-coding described in Table \ref{table:colours} to indicate spin-assignments.
\begin{figure}[b]
 \centering
\includegraphics[width=0.35\textwidth,bb= 0 0 371 646]{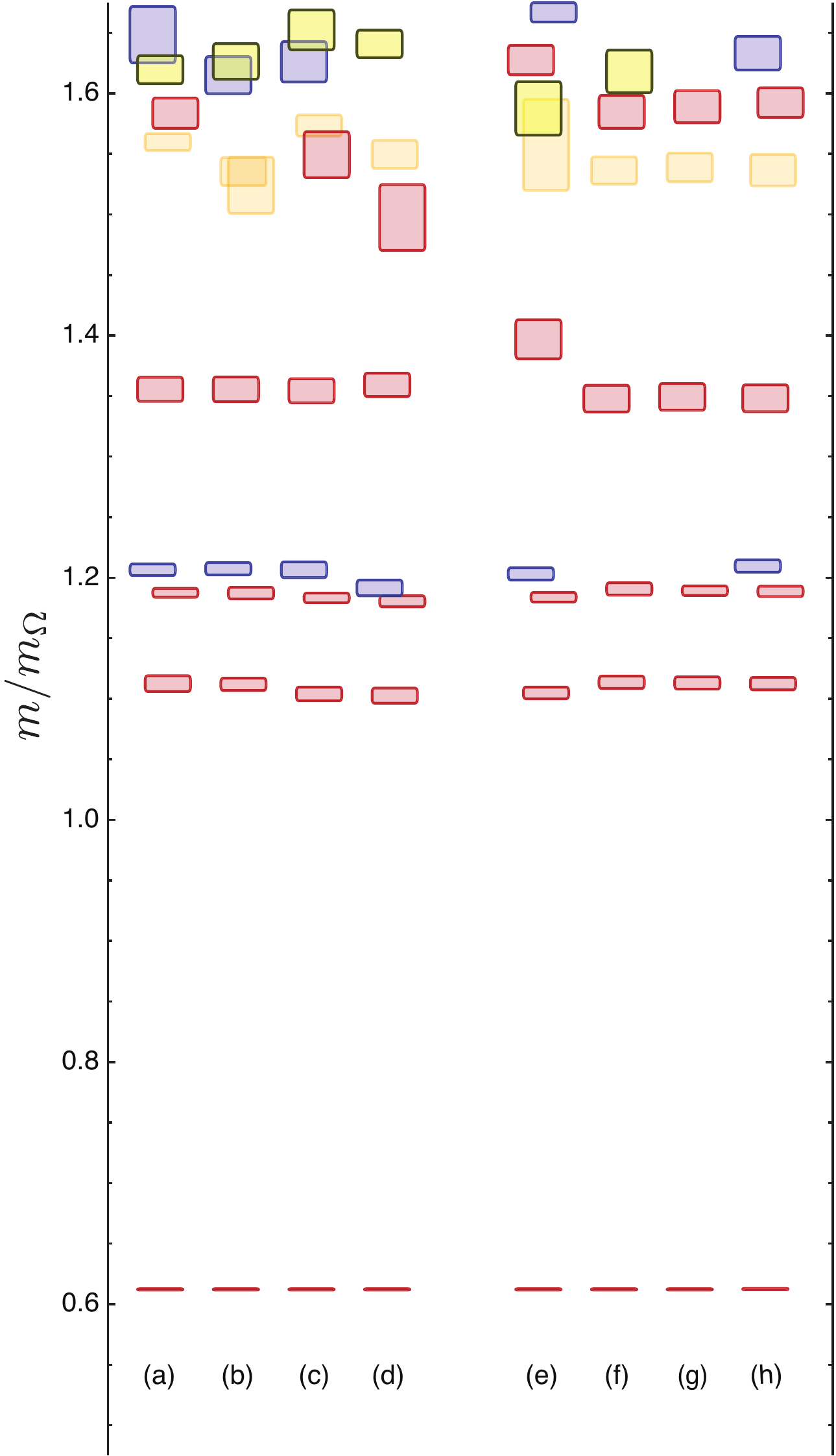}        
\caption{Extracted $T_1^{--}$ mass spectrum for various operator bases. (a)-(d) are ``reasonable" operator bases, (e)-(h) discard important operators.\\
(a) Full basis (dim = 26), (b) Full basis less two noisiest ${\cal O}^{J=1}$ and noisiest ${\cal O}^{J=3}$ (dim = 23), (c) Full less four noisiest ${\cal O}^{J=1}$ and two noisiest ${\cal O}^{J=3}$ (dim = 20), (d) No three-derivative operators except ${\cal O}^{J=4}$ (dim = 13), (e) No operators with commutators of derivatives (dim = 15), (f) No ${\cal O}^{J=3}$ (dim = 20), (g) No ${\cal O}^{J=3,4}$ (dim = 19), (h) No ${\cal O}^{J=4}$ (dim = 25).  \label{vary_ops}}
\end{figure}

\begin{figure*}
 \centering
\includegraphics[width=0.65\textwidth,bb= 0 0 708 427]{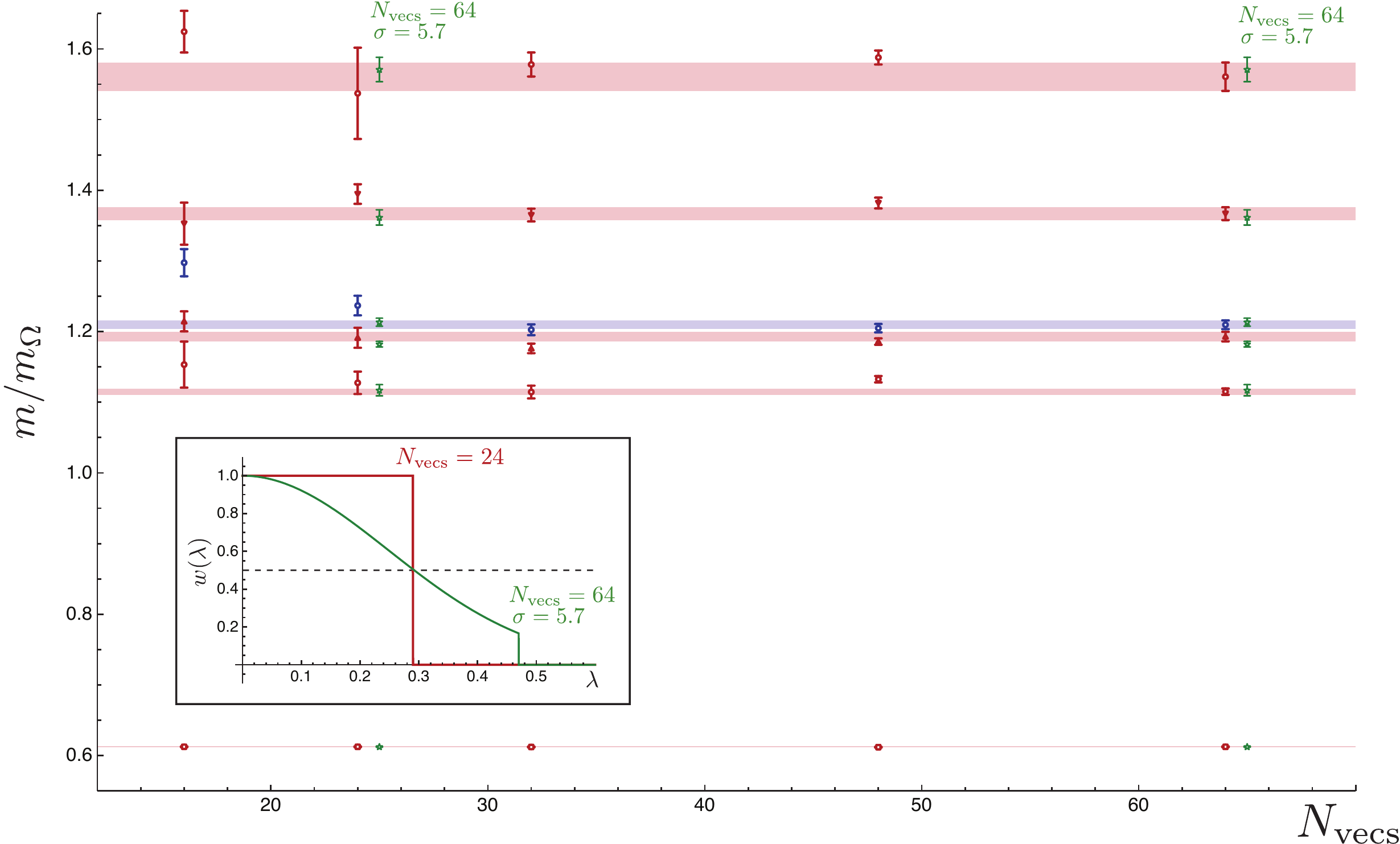}        
\caption{Extracted $T_1^{--}$ mass spectrum as a function of number of distillation vectors. Green points and inset show the effect of using a modified smearing operator $\Box_w = \sum_{n=1}^N w(\lambda_n) \xi^n \xi^{n\dag}$ with $w(\lambda_n)= e^{-\sigma^2 \lambda_n^2 / 4}$.
 \label{masses_vs_vecs}}
\end{figure*}

In the first four columns of Figure \ref{vary_ops}, the operator basis is reduced by discarding operators, first discarding those whose diagonal correlators are noisiest and then, in column (d), discarding all three-derivative operators except the one $J=4$ operator. We observe that it is only high in the spectrum that any change takes place and that there it is only at the level of statistical fluctuations. In (d) the first excited $J=3$ state is no longer extracted - this is most likely due to the basis only retaining two $J=3$ operators.

In the next four columns we deliberately discard operators we believe are important. In column (e), all operators featuring commutators of derivatives are discarded. As discussed later in this manuscript these operators may have good overlap onto states containing an excited gluonic field. The state at $m/m_\Omega \sim 1.35$ is observed, using the full basis of operators, to have large overlap onto a commutator operator, $(\pi \times D^{[2]}_{J=1})^{J=1}$ along with smaller overlaps onto non-commutator operators such as $\rho$. In column (e), the commutator operators being discarded, we find that this state is less cleanly extracted when it can only be produced through its suppressed non-commutator overlaps. Otherwise the spectrum in (e) is rather similar to that obtained with the full operator basis.

In column (f) all operators subduced from spin $J=3$ are discarded. The observed spectrum is almost identical to that with the full basis with the expected exception of the previously identified $J=3$ states. This would appear to suggest that one cannot rely upon discretisation corrections to continuum $J=1$ operators to reliably produce $J=3$ states. In column (g) the continuum $J=4$ operator is also discarded and the $J=4$ state vanishes. Finally, in column (h), the full operator basis, less the continuum $J=4$ operator is used yielding a spectrum that lacks only the $J=4$ state. 

In summary one should be sure to have operators with continuum overlap on to all the spins you expect to see. There is apparently little benefit, in terms of improving the precision of determination of the low-lying spectrum, of reducing the operator basis size by discarding operators.

\subsection{Number of distillation vectors}\label{sec:nvecs}
The results presented so far came from analysis of correlators computed on
$16^3$ lattices using 64 distillation vectors. We might wonder how the
determination of the spectrum varies if one reduces the number of distillation
vectors and thus reduces the computational cost of the calculation. This is
particularly important given that, as shown in \cite{Peardon:2009gh}, to get the
same smearing operator on larger volumes one must scale up the number of
distillation vectors by a factor equal to the ratio of spatial volumes. To scale
up to a $32^3$ lattice this would require $64 \times
\left(\tfrac{32}{16}\right)^3 = 512$ vectors which is not currently a realizable
number without using stochastic estimation \cite{Morningstar:2010ae}.

In figure \ref{masses_vs_vecs} we show the low-lying part of the extracted $T_1^{--}$ spectrum on the \emph{743} $16^3$ lattice as a function of the number of distillation vectors used in the correlator construction. It is clear that the spectrum is reasonably stable for $N \gtrsim 32$ but that the spectrum quality degrades rapidly for fewer vectors.

It is also possible within distillation to implement a smearing operator other than $\Box = \sum_{n=1}^N \xi_n \xi_n^\dag$. This is particularly relevant for large $N$ where this smearing choice tends toward the identity and hence does not actually filter out high-energy modes. An alternative smearing includes a weight function as $\Box_w = \sum_{n=1}^N w(\lambda_n) \xi_n \xi_n^\dag$ where $w(\lambda_n)$ might, for example, be a gaussian damping $e^{-\sigma^2 \lambda_n^2 / 4}$. In figure \ref{masses_vs_vecs} we also show the spectrum obtained using $N=64$ and $\sigma = 5.7$ which, as shown in the inset, is a smearing radius that crudely approximates using only 24 distillation vectors. The thus extracted spectrum differs very little from the $N=64$ spectrum, suggesting that with 64 vectors we are still far away from the ``unsmeared" limit.

One place (not shown in figure \ref{masses_vs_vecs}) where the effect of reduction of the number of distillation vectors is seen clearly is for high-spin states. In particular the $4^{--}$ state seen at $m/m_\Omega \sim 1.6$ in $T_1^{--}$ is not reliably extracted for $N < 48$. 
The need to have large numbers of distillation vectors to reliably extract high spin mesons can be described in a simple free-field picture: In the continuum without gauge-fields, the eigenvectors of the laplacian ($-\nabla^2 \xi = k^2 \xi$) can be expressed as $\xi(\vec{r}) = e^{i\vec{k}\cdot\vec{r}} = 4\pi \sum_\ell i^\ell j_\ell(k r)  \sum_m Y_\ell^{m*}(\theta, \phi) Y_\ell^m(\hat{k})$. From this expression it would appear that all $\ell$ values should contribute for any value of the eigenvalue $k^2$ and that high-spin states should be excited even by low eigenvectors. But this argument does not take account of the radial behaviour which must be compared to the typical size of hadrons. The spherical Bessel function $j_\ell(kr)$ is peaked at low values of $kr$ with the peak position moving out to larger $kr$ for larger $\ell$. Thus, since hadrons are only of finite size, $R$, in order that the peak of $j_\ell(kr)$ remain within $r<R$ as $\ell$ increases, one must also increase the value of $k$. Thus, to have considerable amplitude multiplying $Y_\ell$ and hence large overlap onto high-spin mesons, as $\ell$ increases one must include eigenvectors of higher eigenvalue, $k$.\footnote{this analysis is somewhat oversimplified since in fact the size of hadrons is likely to increase with increasing spin}

In summary one is limited as to how few distillation vectors can be used if one requires reliable extraction of high-spin states. The results shown here suggest 48 distillation vectors on a $16^3$ lattice is the minimum, so 96 distillation vectors on a $20^3$ lattice are likely to be required.


\section{Results}\label{sec:results}

As outlined in Section \ref{sec:lattice}, our first results are obtained on lattices with pion masses between 400 and 700 MeV.  The heaviest pion mass corresponds to the three-flavour symmetric point, while the lower masses have lower light-quark masses but the same bare strange quark mass; these are found to have only mild $SU(3)_F$ breaking, as indicated by $m_K/m_\pi$ which reaches only $1.4$ on the lightest lattice, well below the physical value of $3.5$. For each lattice (except one) we compute correlators on two volumes, $16^3 \times 128$ and $20^3 \times 128$. More complete details of the number of configurations, time sources and distillation vectors used are given in Table \ref{tab:lattices}. 

\subsection{$m_\pi \sim 700 $ MeV results (\emph{743})}

This is an example of an exact $SU(3)$-flavour symmetric calculation. Since we are computing only the connected two-point correlators we will obtain the mass spectrum of degenerate octets (e.g.\ degenerate pions, kaons and $\eta_{\mathbf{8}}$). Results on $16^3$ lattices have already been reported in Ref.~\cite{Dudek:2009qf}. To obtain singlet states like the $\eta_{\mathbf{1}}$, which with exact flavour symmetry cannot mix with the $\eta_{\mathbf{8}}$, we would need disconnected two-point correlators. 

\begin{figure}
 \centering
\includegraphics[width=0.22\textwidth,bb= 0 0 272 437]{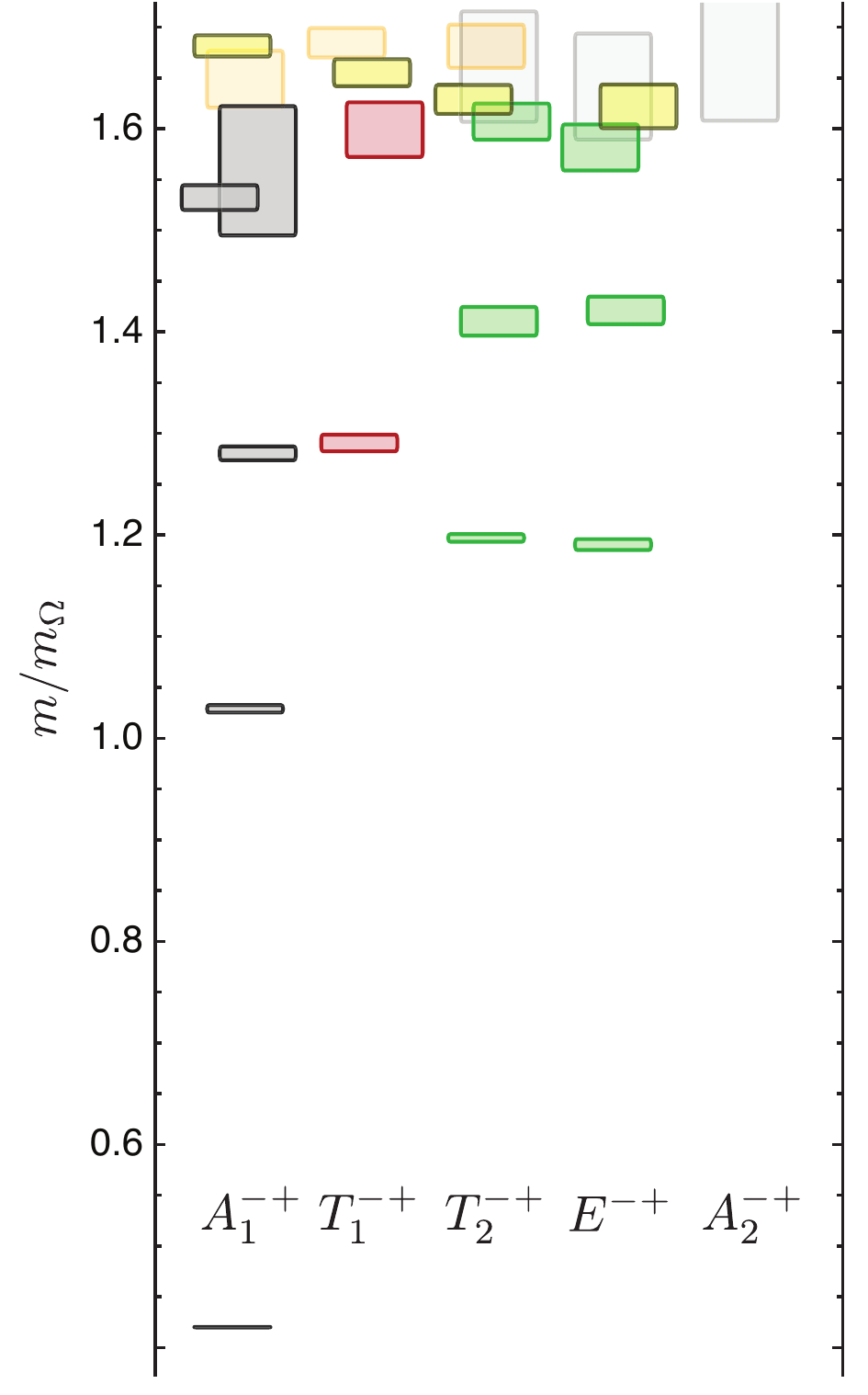} 
\includegraphics[width=0.22\textwidth,bb= 0 0 271 437]{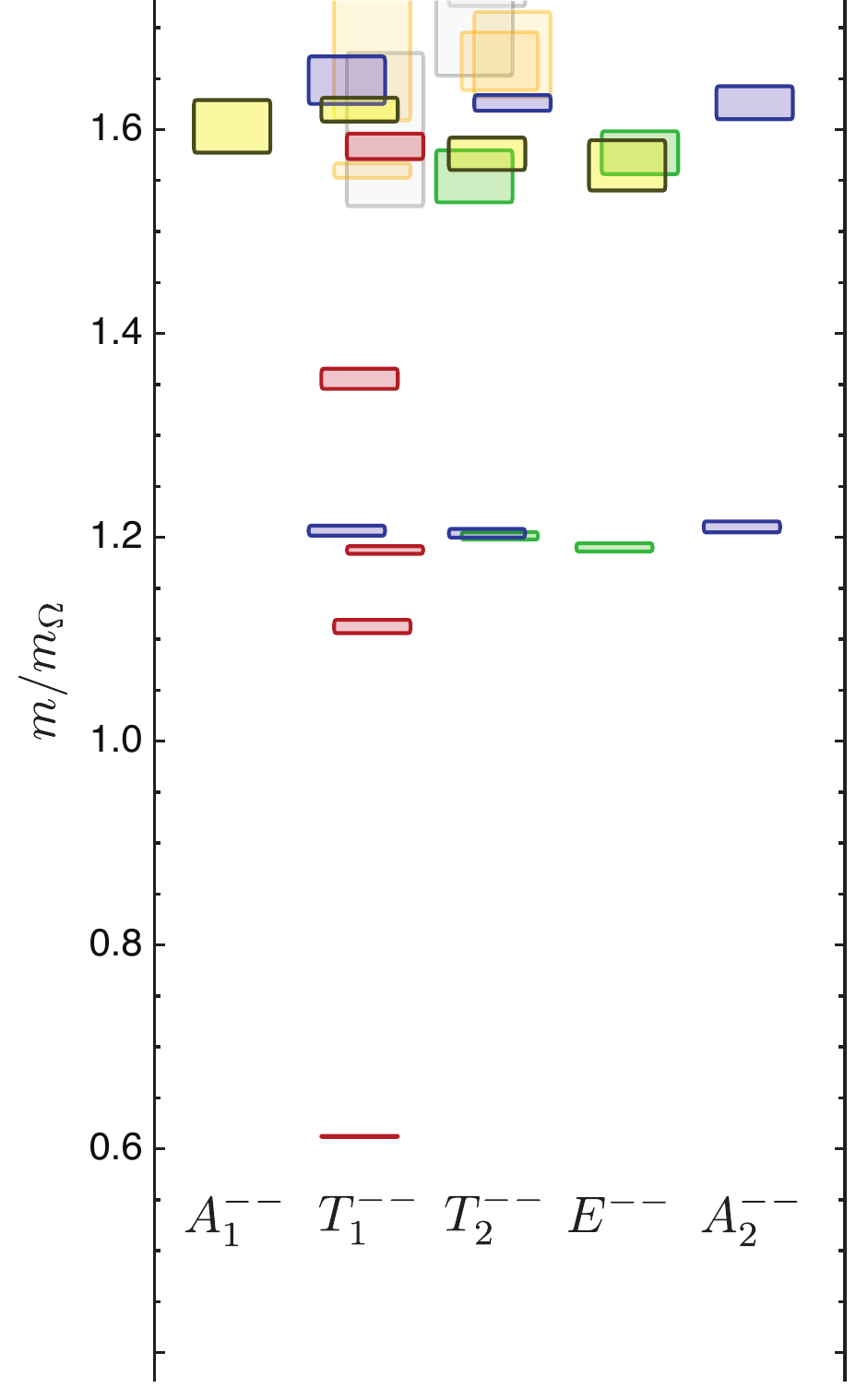}    
\includegraphics[width=0.22\textwidth,bb= 0 0 272 434]{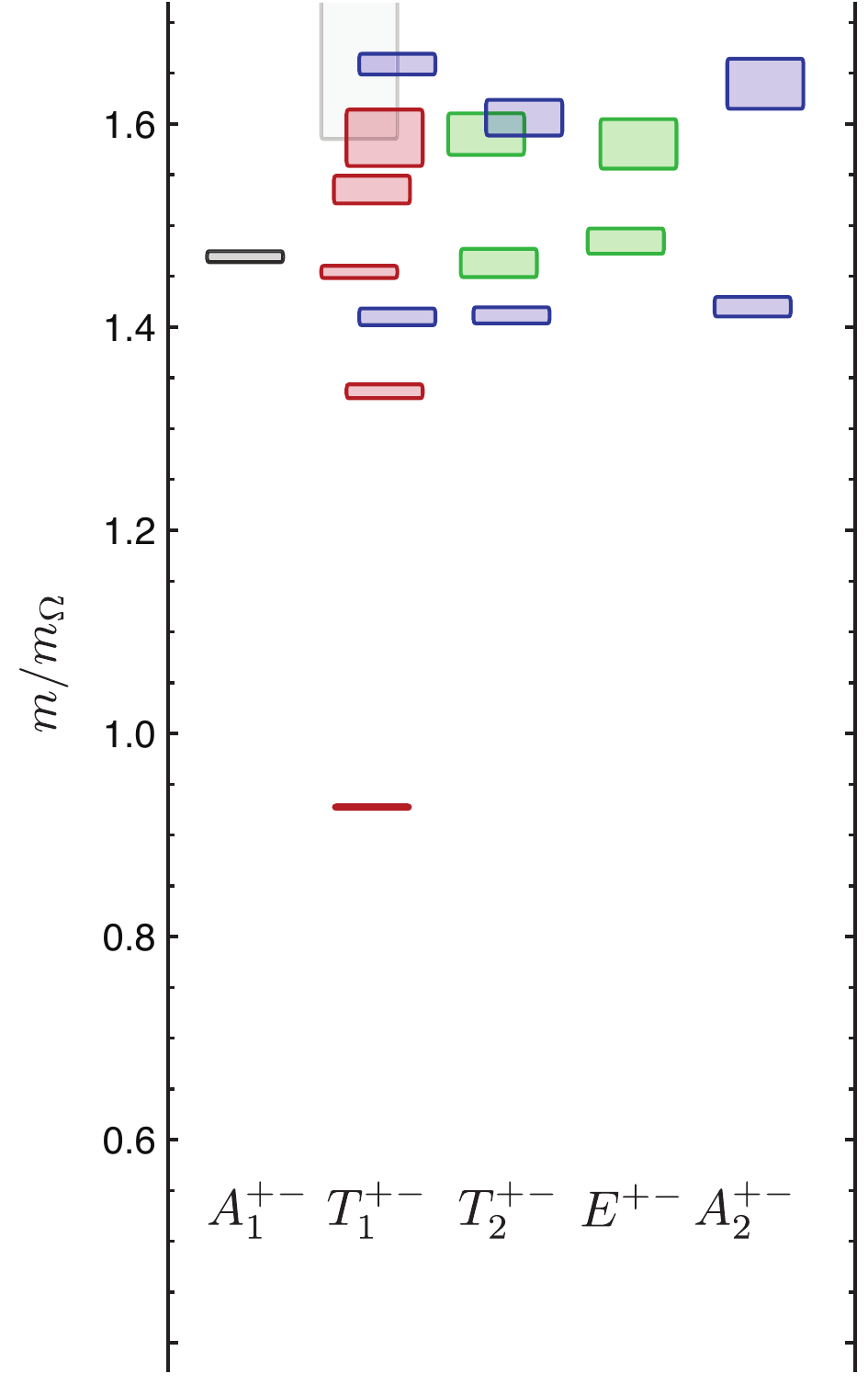} 
\includegraphics[width=0.22\textwidth,bb= 0 0 261 433]{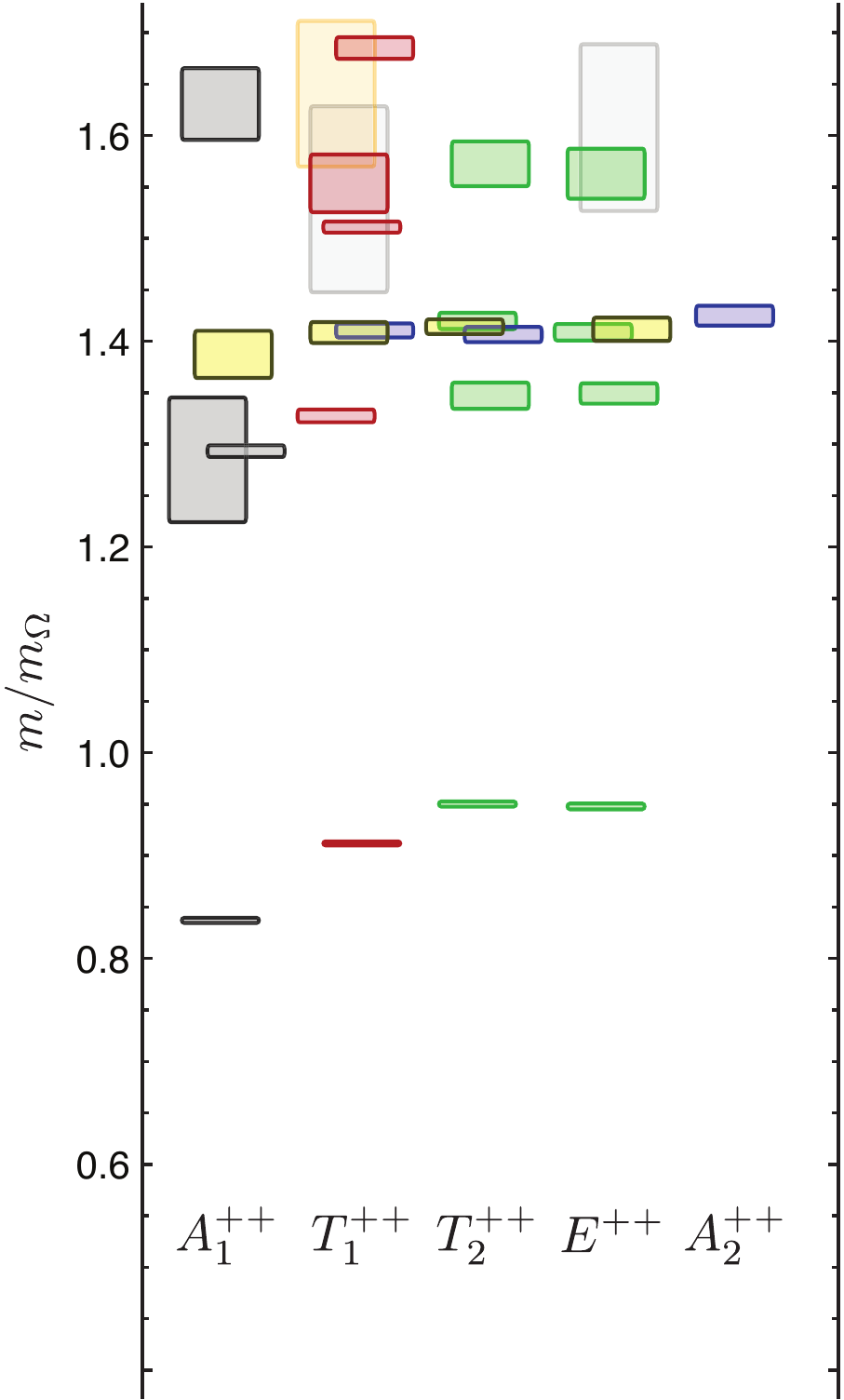}        
\caption{Extracted spectra by irrep from \emph{743} $16^3$ lattices.\label{743_16_irrep}}
\end{figure}

Variational analysis as described in previous sections leads to the irrep spectra from $16^3$ lattices shown in figure \ref{743_16_irrep}. The colour-coding, tabulated in table \ref{table:colours}, indicates the continuum spin as determined by methods described in section \ref{sec:spin}. The $\Lambda^{--}$ overlaps are shown in figures \ref{histogram} and \ref{Zvalues}. If the spin of a state is not unambiguously determined by the methods of section \ref{sec:spin}, it is represented by an orange box.  On the other hand, if a state is extracted in the variational analysis but its mass cannot be accurately determined from fits to the principal correlator, it is represented by a grey box. 

\begin{table}[h]
\begin{tabular}{r|l}
black & $J=0$\\
red & $J=1$\\
green & $J=2$\\
blue & $J=3$\\
yellow & $J=4$\\
\hline
orange & undetermined $J$ \\
grey & badly determined mass
\end{tabular}
\caption{Colour-coding used in spectrum irrep plots.\label{table:colours}}
\end{table}

\begin{figure*}
 \centering
\includegraphics[width=0.95\textwidth,bb= 0 0 1035 439]{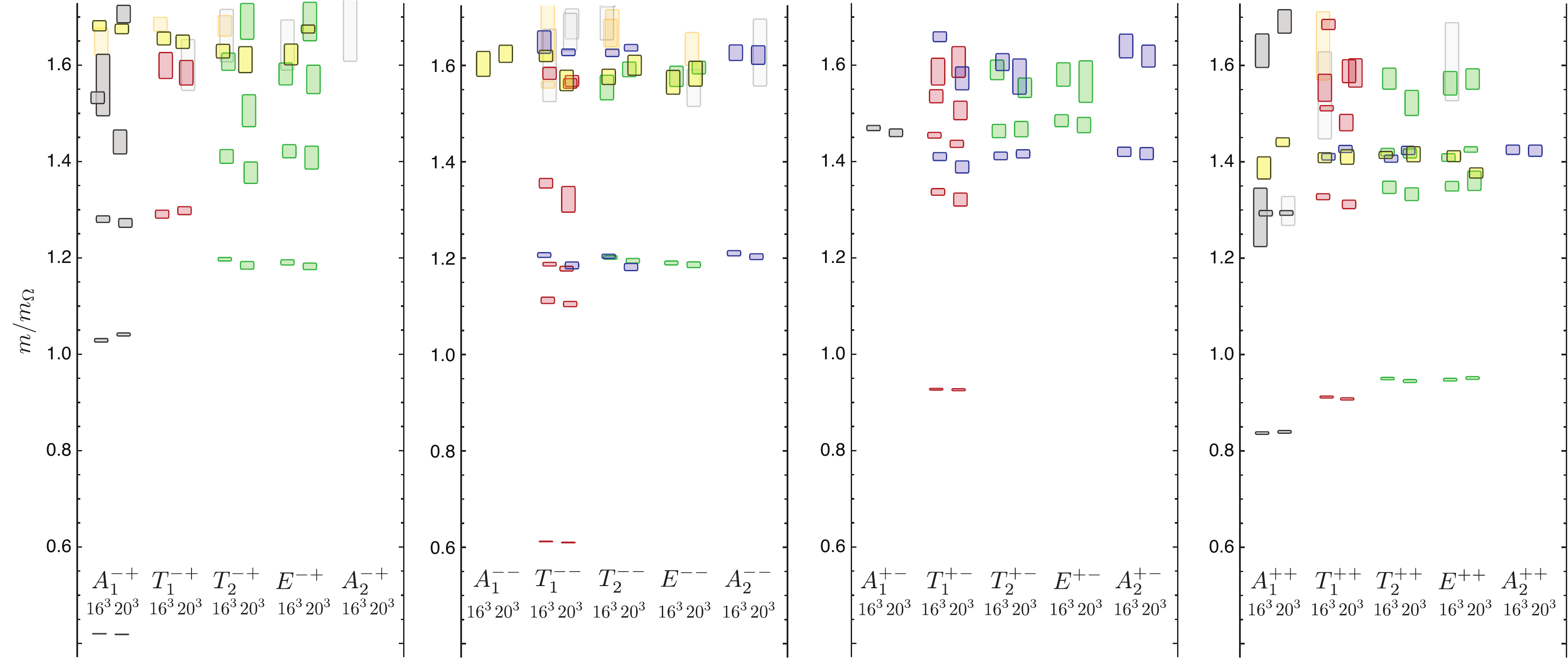} 
\caption{Extracted spectra by irrep from \emph{743} $16^3$ and $20^3$ lattices.\label{743_16_20_irrep}}
\end{figure*}

In figure \ref{743_16_20_irrep} we show side-by-side the spectra obtained from $16^3$ and $20^3$ lattices. We note that there is really no change significantly outside statistical fluctuations between the two volumes. We also note in passing that the 128 distillation vectors used on the $20^3$ lattices give a smearing that is essentially equivalent to that obtained by using 64 distillation vectors on the $16^3$ lattices.  In fact we observe that the independently extracted $Z$ values between the two volumes scale rather accurately as $\sqrt{\frac{16^3}{20^3}}$.

\begin{figure*}
 \centering
\includegraphics[width=0.75\textwidth,bb= 0 0 648 438]{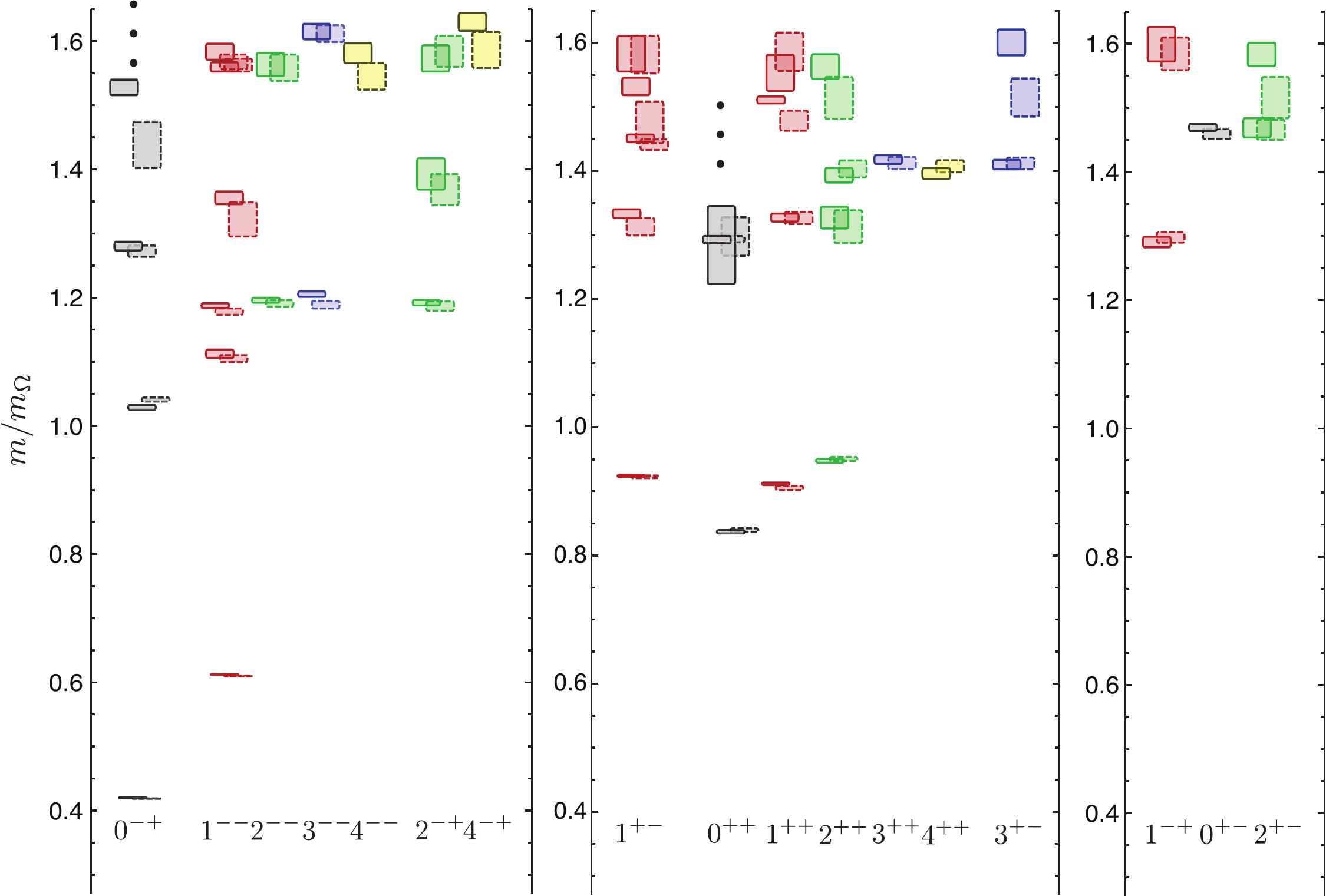}        
\caption{Spin-identified spectrum of isovector (octet) mesons from the \emph{743} lattices. $16^3$(solid) and $20^3$(dashed) spectra agree well. Ellipses indicate that there are heavier states with a given $J^{PC}$ but that they are not well determined in this calculation.\label{743_spin}}
\end{figure*}

Given the success of spin-identification we can summarise the results in a spectrum labelled by continuum $J^{PC}$ quantum numbers, figure \ref{743_spin}. Here we show only well-determined low-lying states. 

There are a number of notable features within this spectrum. Firstly there appears to be much of the $n^{2S+1}L_J$ distribution of non-exotic states predicted by the $q\bar{q}$ quark model (e.g.\ \cite{Godfrey:1985xj}). The left-hand pane of figure \ref{743_spin} has candidates for a ground-state $S$-wave pair ($0^{-+}, 1^{--}$) and a radial excitation at around $m/m_\Omega \sim 1.1$. There is a complete $(1,2,3)^{--}, 2^{-+}$ $D$-wave set around $m/m_\Omega \sim 1.2$. The region around $m/m_\Omega \sim 1.6$ appears to contain parts of a $D$-wave radial excitation and a $G$-wave ($(3,4,5)^{--}, 4^{-+}$)\footnote{we have no operators capable of producing a spin-5 meson in the continuum.}.
In the middle pane there is a clear $P$-wave, $((0,1,2)^{++}, 1^{+-})$ around $m/m_\Omega \sim 0.9$ and a probable radial excitation near $m/m_\Omega \sim 1.3$. At $m/m_\Omega \sim 1.4$ there is a complete $F$-wave, $(2,3,4)^{++}, 3^{+-}$.

In the right pane we see clearly a set of exotic $J^{PC}$ states, not accessible
to a simple $q\bar{q}$ pair. Such states can be described in terms of
constituents if additional degrees-of-freedom, either gluonic or extra 
$q\bar{q}$ pairs are included. If the extra component is gluonic the states are known as hybrid mesons and within models of gluonic excitation, such as the flux-tube model \cite{Isgur:1984bm}, there is usually a roughly degenerate set of $1^{-+}$, $0^{+-}$ and $2^{+-}$ states. 
A hybrid nature for these states is suggested by the large overlap onto operators with essential non-trivial gluonic structure as described below. In this calculation we observe a $1^{-+}$ lightest, with a pair of $0^{+-}$ and $2^{+-}$ states nearly degenerate a little higher in mass. A second $2^{+-}$ is then close to a second $1^{-+}$. States with exotic $0^{--}$ and $3^{-+}$ quantum numbers are found to be considerably heavier, well above $m/m_\Omega = 1.6$.

There are also a number of \emph{non-exotic} quantum numbered states which do not appear to fit into the $q\bar{q}$ $n^{2S+1}L_J$ classification. The $0^{-+}, 1^{--}$ pair at $m/m_\Omega \sim 1.3$ is probably too light to be the second radial excitation of the $S$-wave (most likely the pair near $m/m_\Omega \sim 1.55$) and is partnered with a totally unexpected $2^{-+}$ state. There may be excess states too in the positive parity sector but the situation is not totally clear above $m/m_\Omega \sim 1.4$. We note that the mass scale of these first excess states is comparable to that of the lowest lying exotic states shown in the right-hand pane. Furthermore, these states have characteristically different overlap behaviour compared to most other extracted states: they all overlap considerably onto operators featuring the commutator of two derivatives, that is the gluonic field-strength tensor.

Given the large overlap onto operators requiring a non-trivial gluonic field distribution, we identify these states as hybrid mesons with non-exotic quantum numbers.  Such non-exotic hybrids are predicted in models that assume non-trivial gluonic field configurations like the flux-tube model or constituent gluon models. In principle, such states can mix strongly with regular quark-model $q\bar{q}$ states leaving a spectrum which is not simple to interpret. In our results such large mixing may be present for the ``excess" $0^{-+}$ state which has a large overlap also onto operators like $\bar{\psi} \gamma_5 \psi$; however, this mixing does not appear to be present to the same degree for the $1^{--}$ state.

Detailed model-dependent interpretation of the spectrum, comparing the overlap values with the expectations of a bound-state quark model (analogous to that done for charmonium in \cite{Dudek:2008sz}) and considering the degree of mixing of non-exotic hybrids and quark model states will follow in a subsequent publication.

\subsection{Quark mass dependence}

Here we move away from the $SU(3)$ flavour point by lowering the mass of two
degenerate ``light" flavours and keeping one remaining strange flavour heavy. We
have access to isovector mesons from the connected correlators with a light quark
and a light antiquark, and kaons from the connected correlators with a light quark and
a strange antiquark. It is also possible for us to compute the connected part of
correlators with both quark and antiquark being strange, the so-called
``strangeonium". We recognise that neglecting the disconnected contributions to
these diagrams leads to a non-unitary description of this particular element of
our calculation. 
Of course it is also true that ``strangeonium" states are not necessarily QCD
eigenstates since being isoscalars they can mix with light-light isoscalars and
pure-glue states through disconnected diagrams\footnote{work is ongoing within
  the Hadron Spectrum Collaboration \cite{Peardon:2009gh} to utilise distillation methods to efficiently compute disconnected two-point functions, allowing extraction of the true QCD eigenstates}. The classic extreme examples are the $\eta,\eta'$ system which is mixed almost as $SU(3)_F$ octet-singlet and the $\omega, \phi$ system which is mixed almost as $\ell \bar{\ell}, s\bar{s}$. 

In the figures we show extracted state masses as a function of $\ell_\Omega
\equiv \frac{9}{4} \frac{(a_t m_\pi)^2}{(a_t m_\Omega)^2}$ which we use as a proxy for the quark mass \cite{Lin:2008pr}. The state masses are presented via $\frac{ a_t m_H }{a_t m_\Omega} m_\Omega^\mathrm{phys.}$.  The ratio of the state mass ($m_H$) to the $\Omega$-baryon mass computed on the same lattice removes the explicit scale dependence and multiplying by the physical $\Omega$-baryon mass conveniently expresses the result in MeV units. This is clearly not a unique scale-setting prescription, but it serves to display the data in a relatively straightforward way. We remind the reader that the data between different volumes and quark masses are uncorrelated since they follow from computations on independently generated dynamical gauge-fields.

\subsubsection{Isovector mesons}

\begin{figure*}
 \centering   
\includegraphics[width=0.45\textwidth,bb= 0 0 591 382]{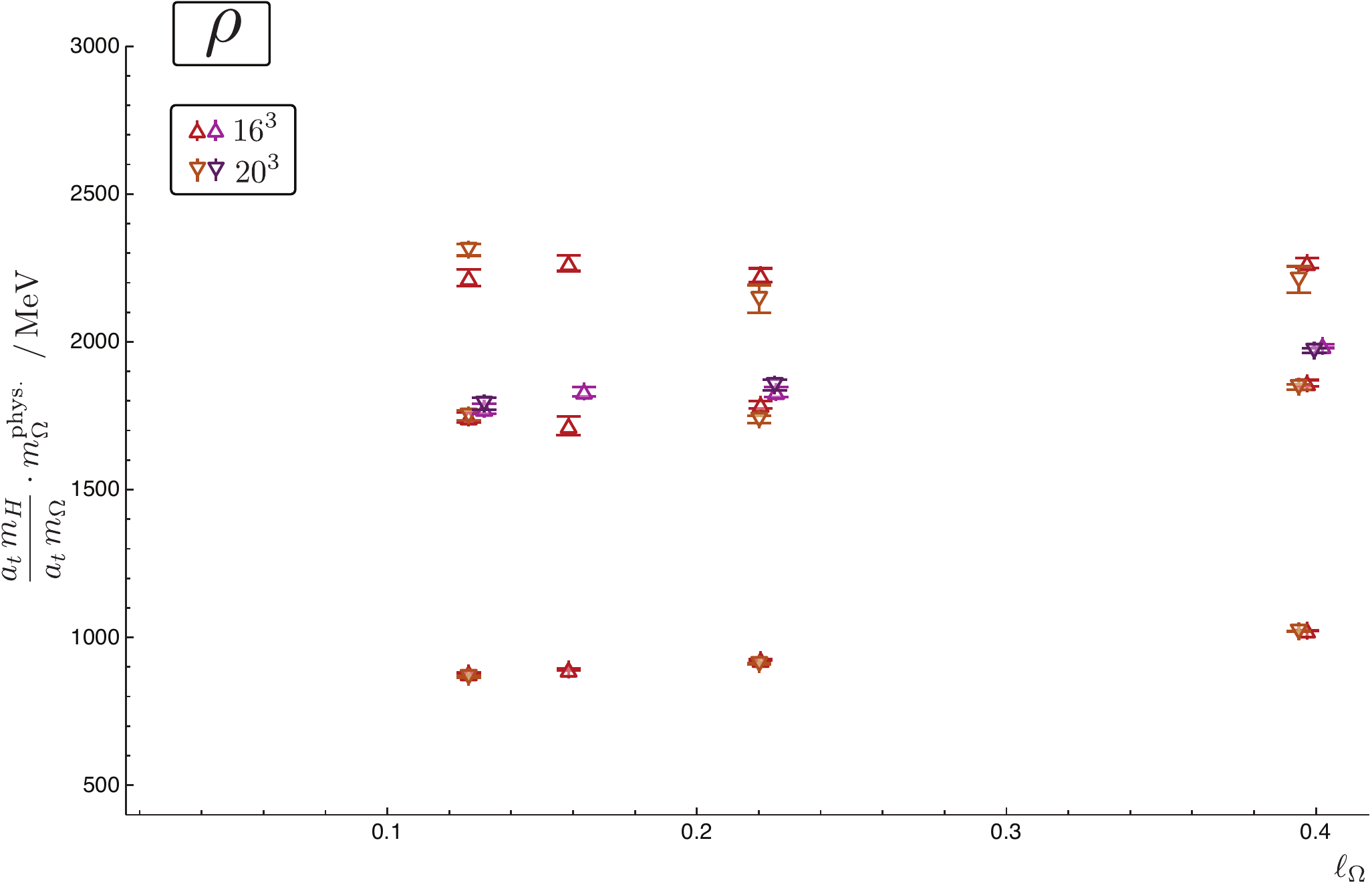}     
\includegraphics[width=0.45\textwidth,bb= 0 0 595 382]{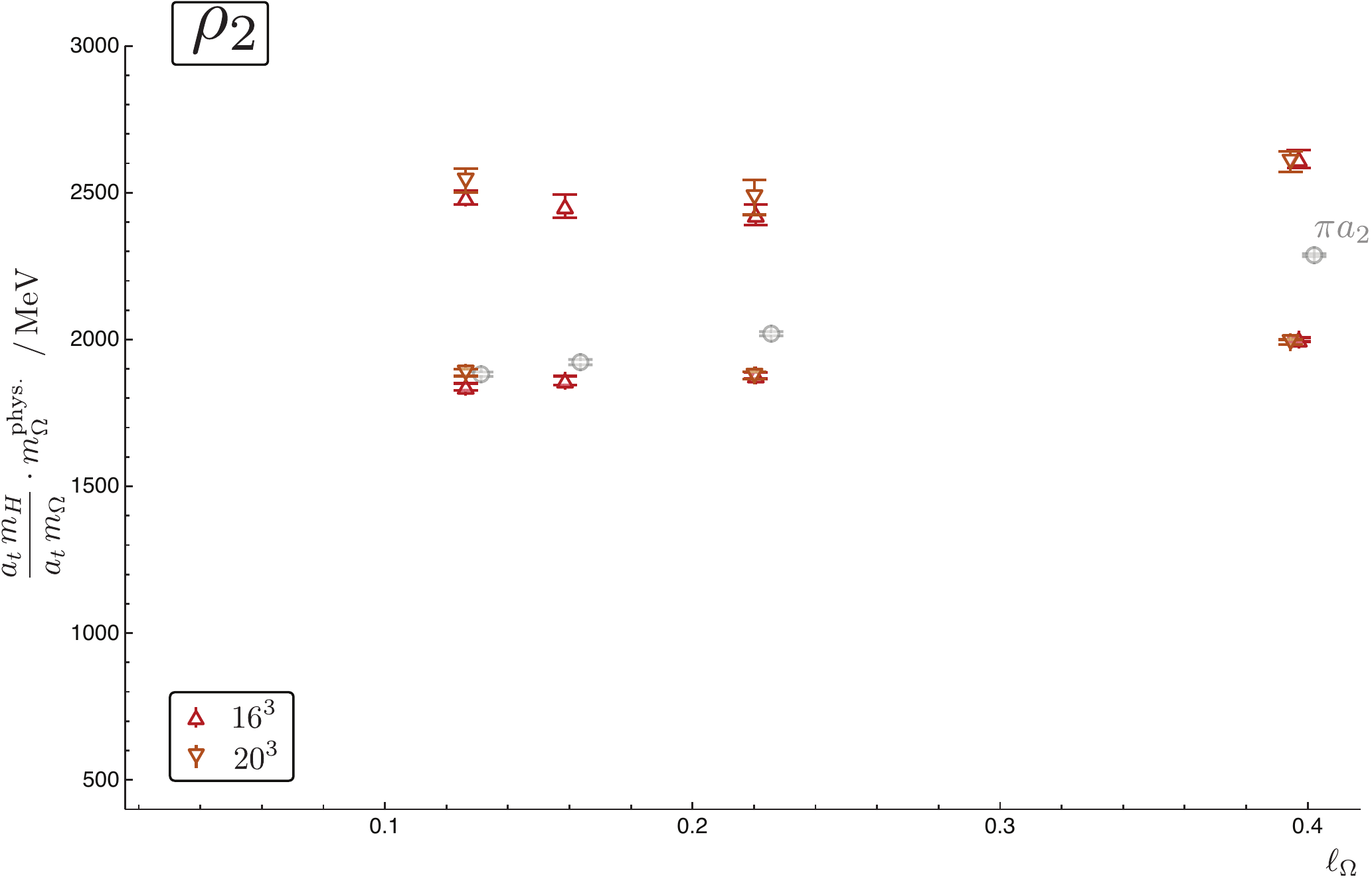}     
\includegraphics[width=0.45\textwidth,bb= 0 0 593 382]{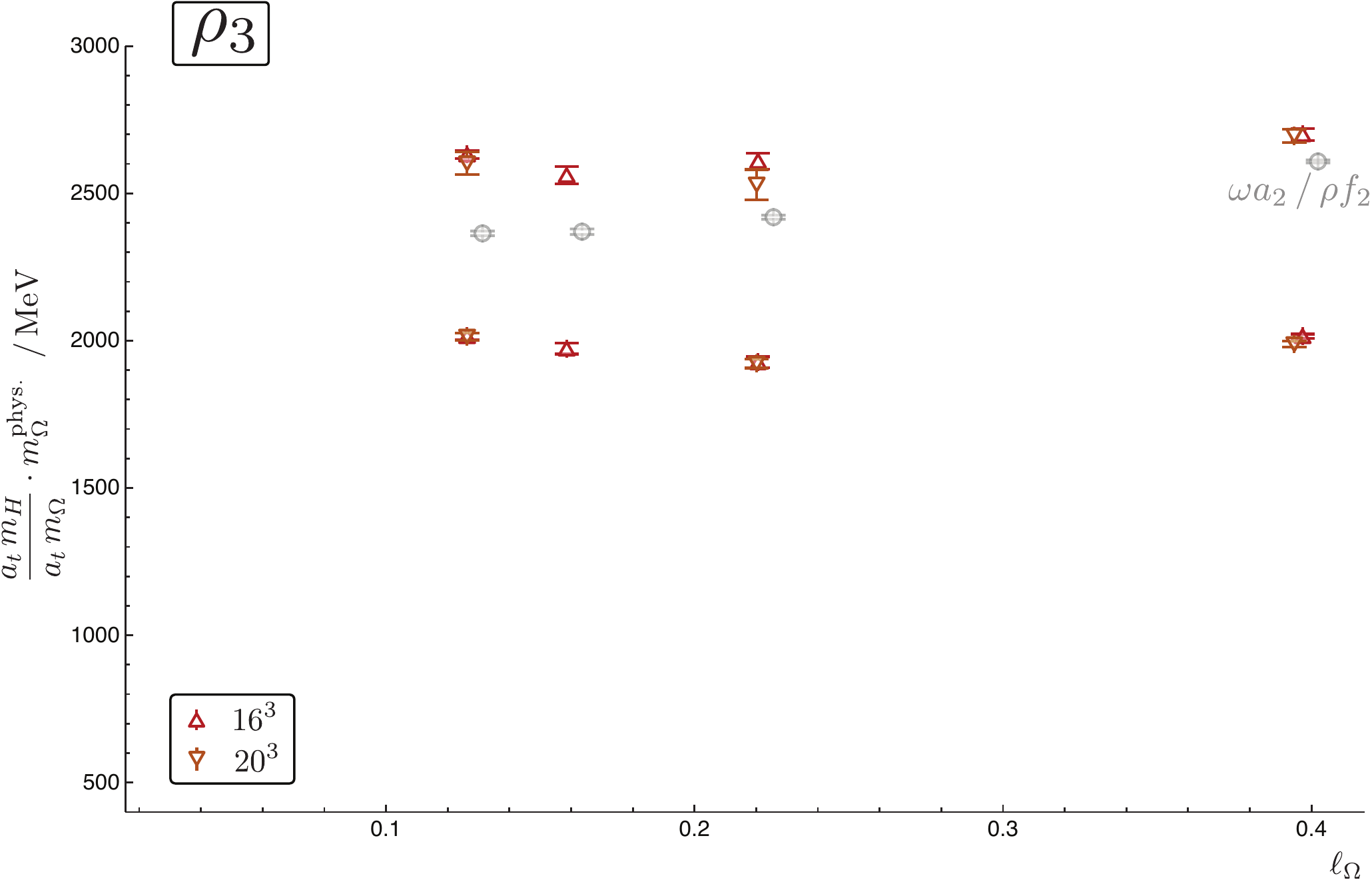}    
\includegraphics[width=0.45\textwidth,bb= 0 0 591 382]{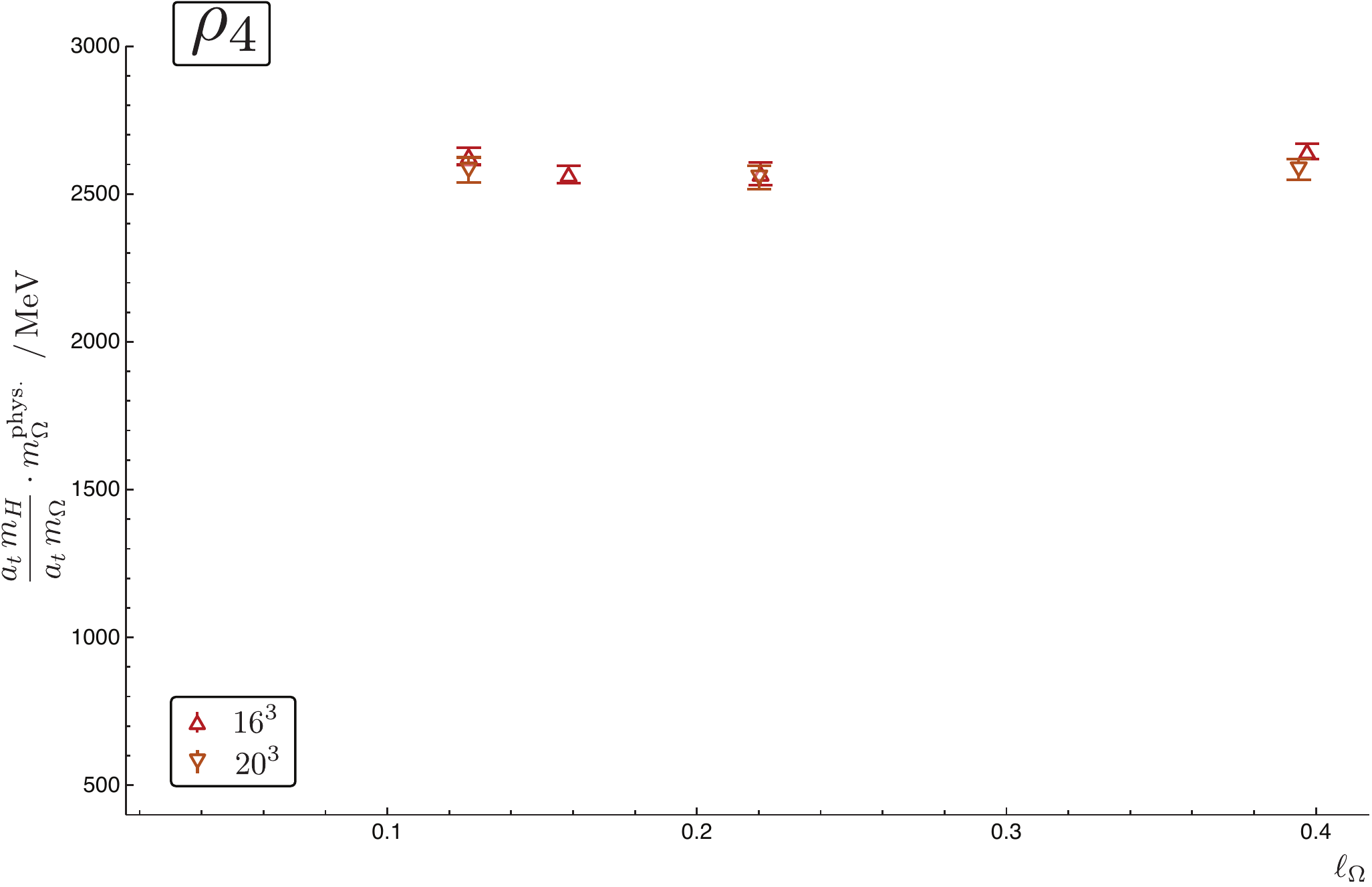}        
\caption{Lightest isovector states, $\rho_J$; neutral members with quantum numbers $J^{--}$.\label{rhos}}
\end{figure*}

\begin{figure*}
 \centering  
\includegraphics[width=0.45\textwidth,bb= 0 0 593 382]{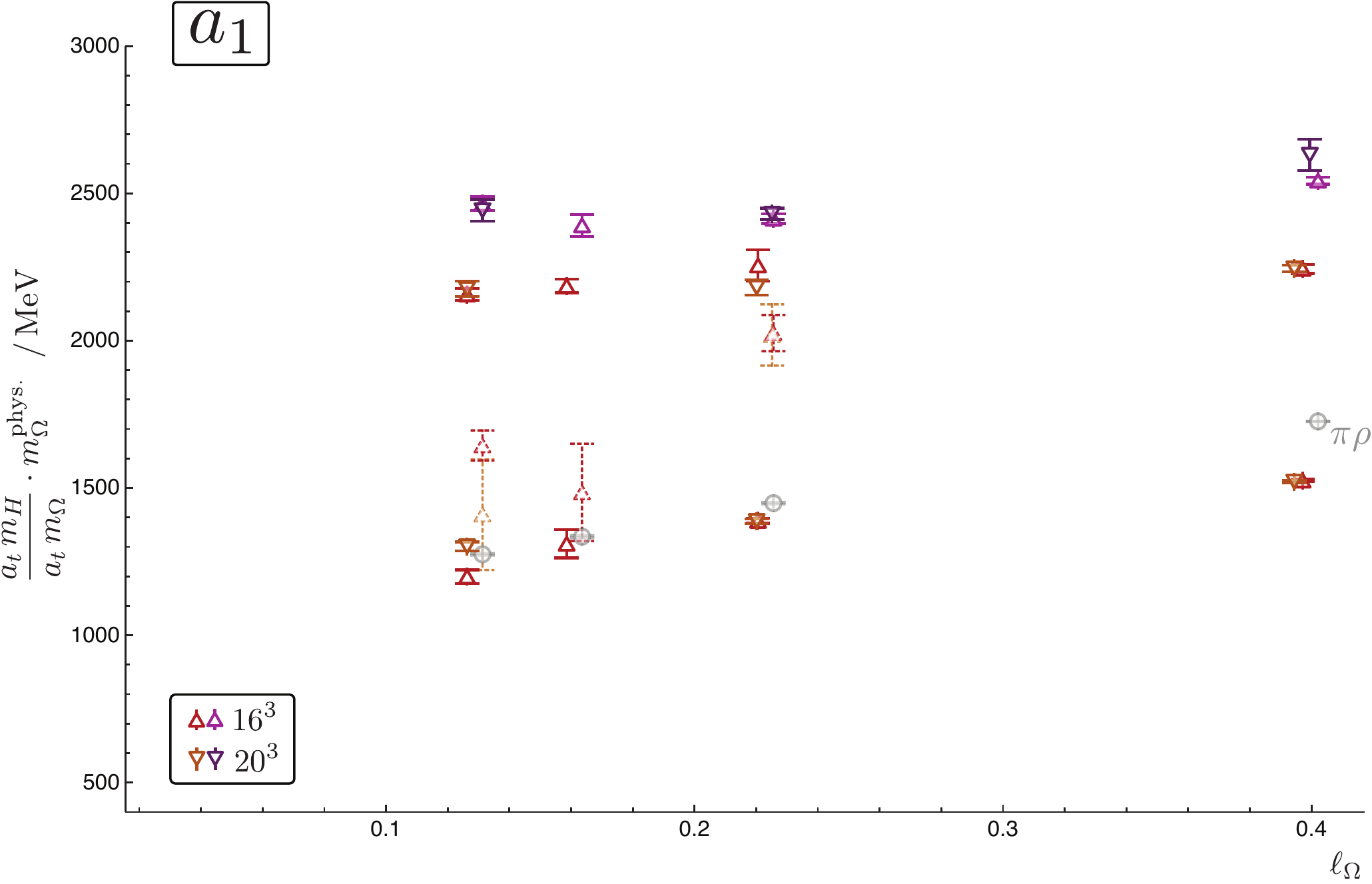}     
\includegraphics[width=0.45\textwidth,bb= 0 0 593 382]{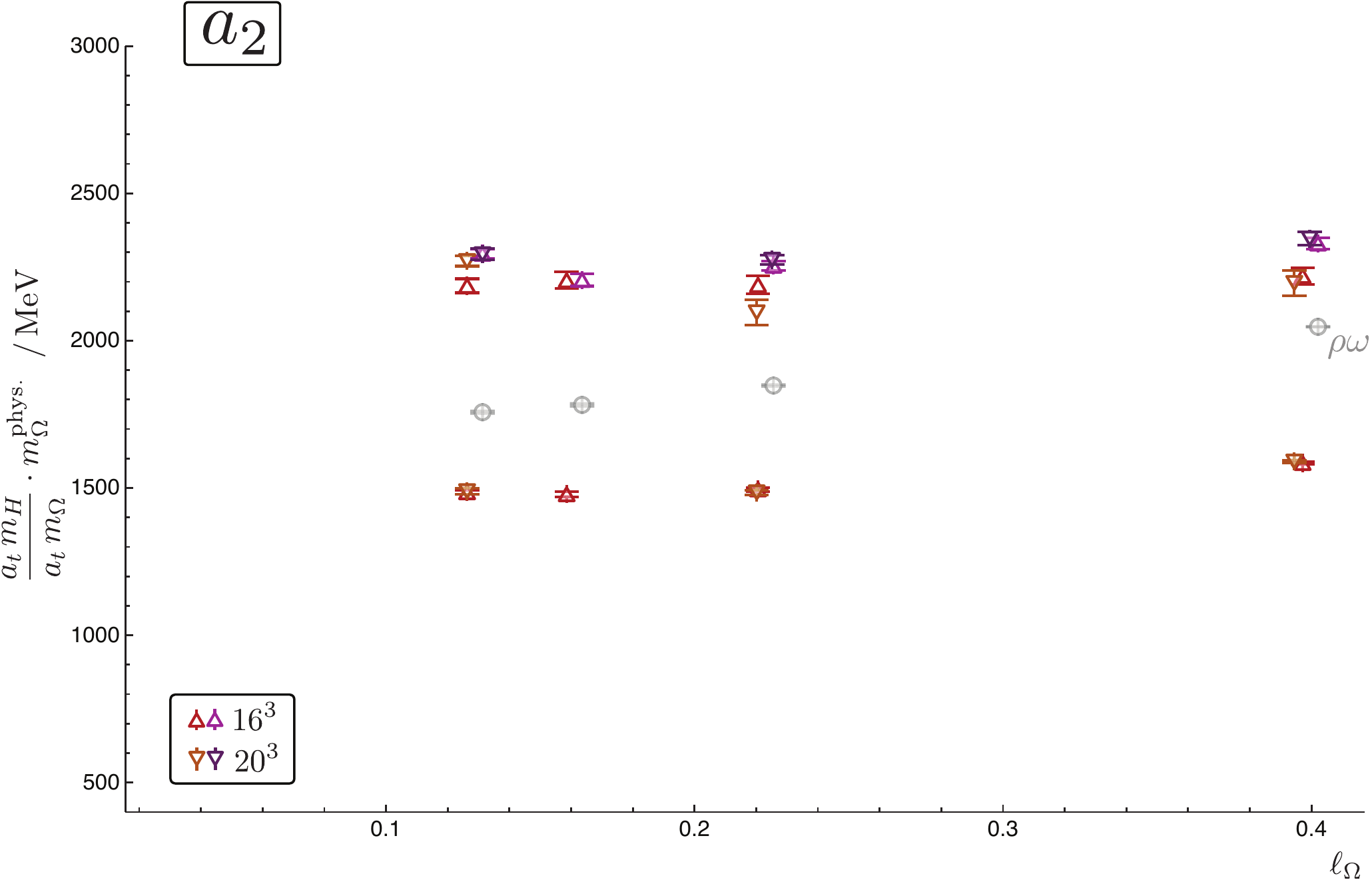}     
\includegraphics[width=0.45\textwidth,bb= 0 0 593 382]{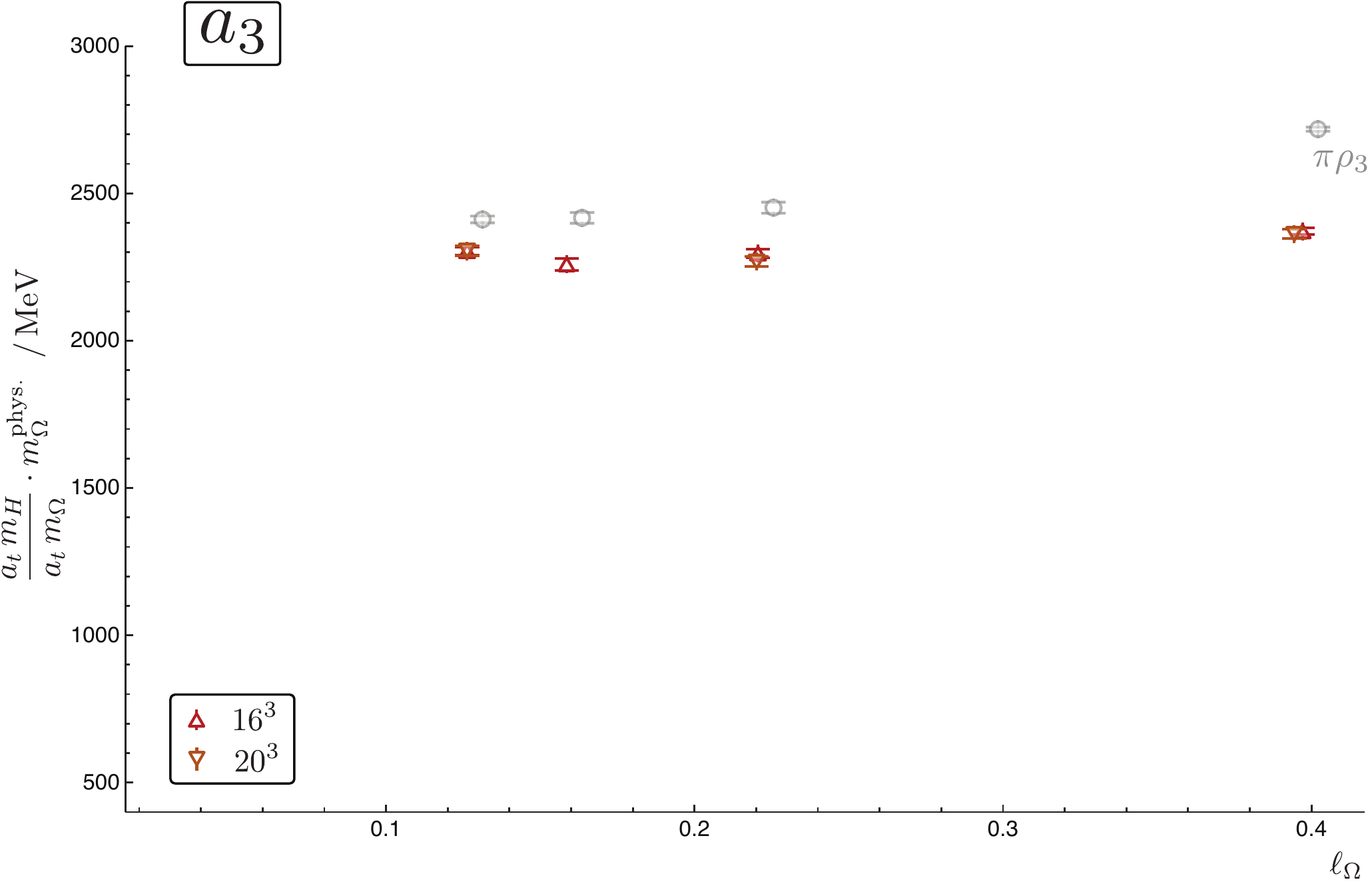}    
\includegraphics[width=0.45\textwidth,bb= 0 0 591 382]{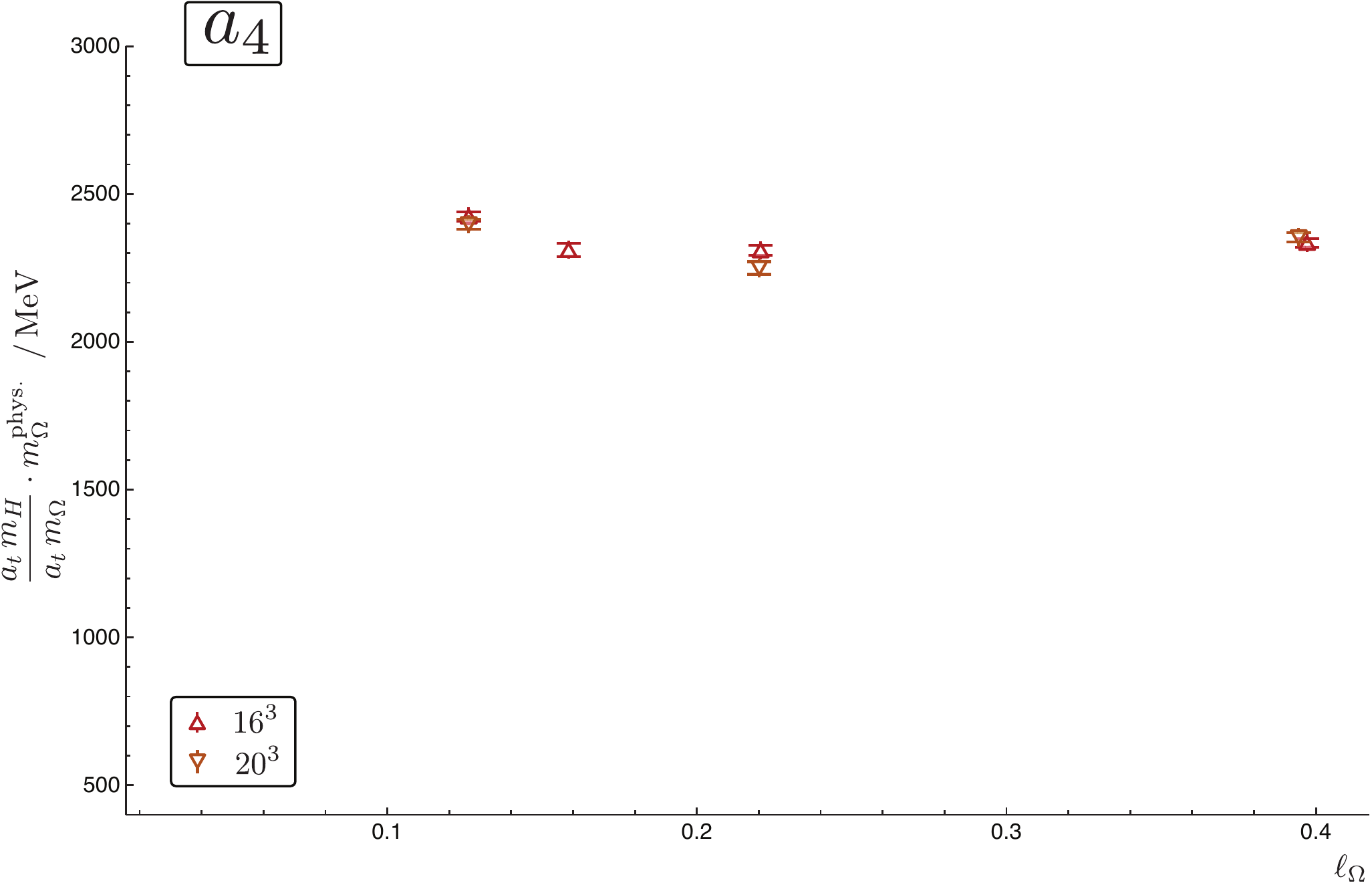}        
\caption{Lightest isovector states, $a_J$; neutral members with quantum numbers $J^{++}$ (the $a_0$ states are shown in Figure\ \ref{a0ex}).\label{as}}
\end{figure*}

\begin{figure*}
 \centering  
\includegraphics[width=0.45\textwidth,bb= 0 0 591 382]{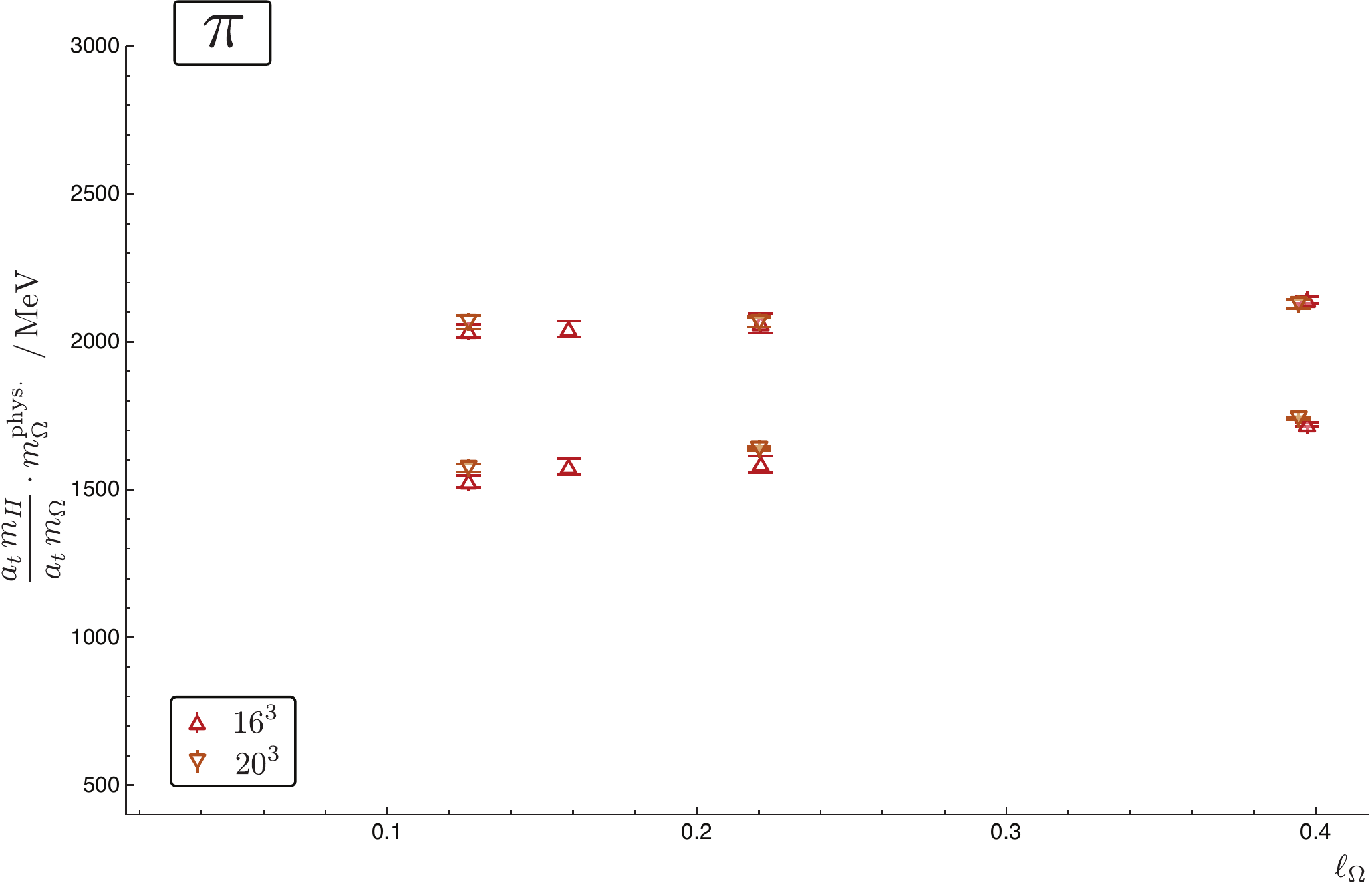}     
\includegraphics[width=0.45\textwidth,bb= 0 0 593 382]{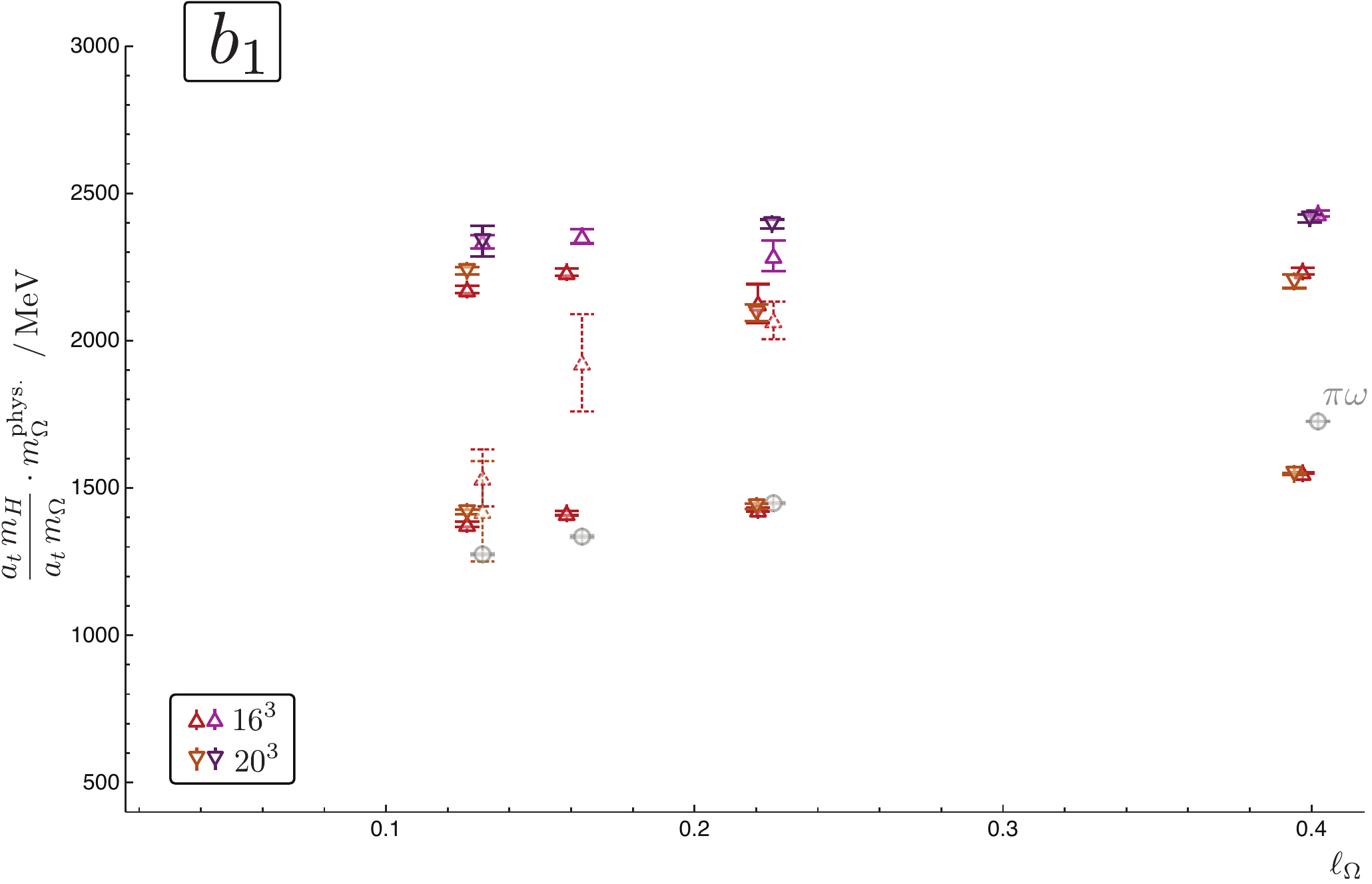}     
\includegraphics[width=0.45\textwidth,bb= 0 0 593 382]{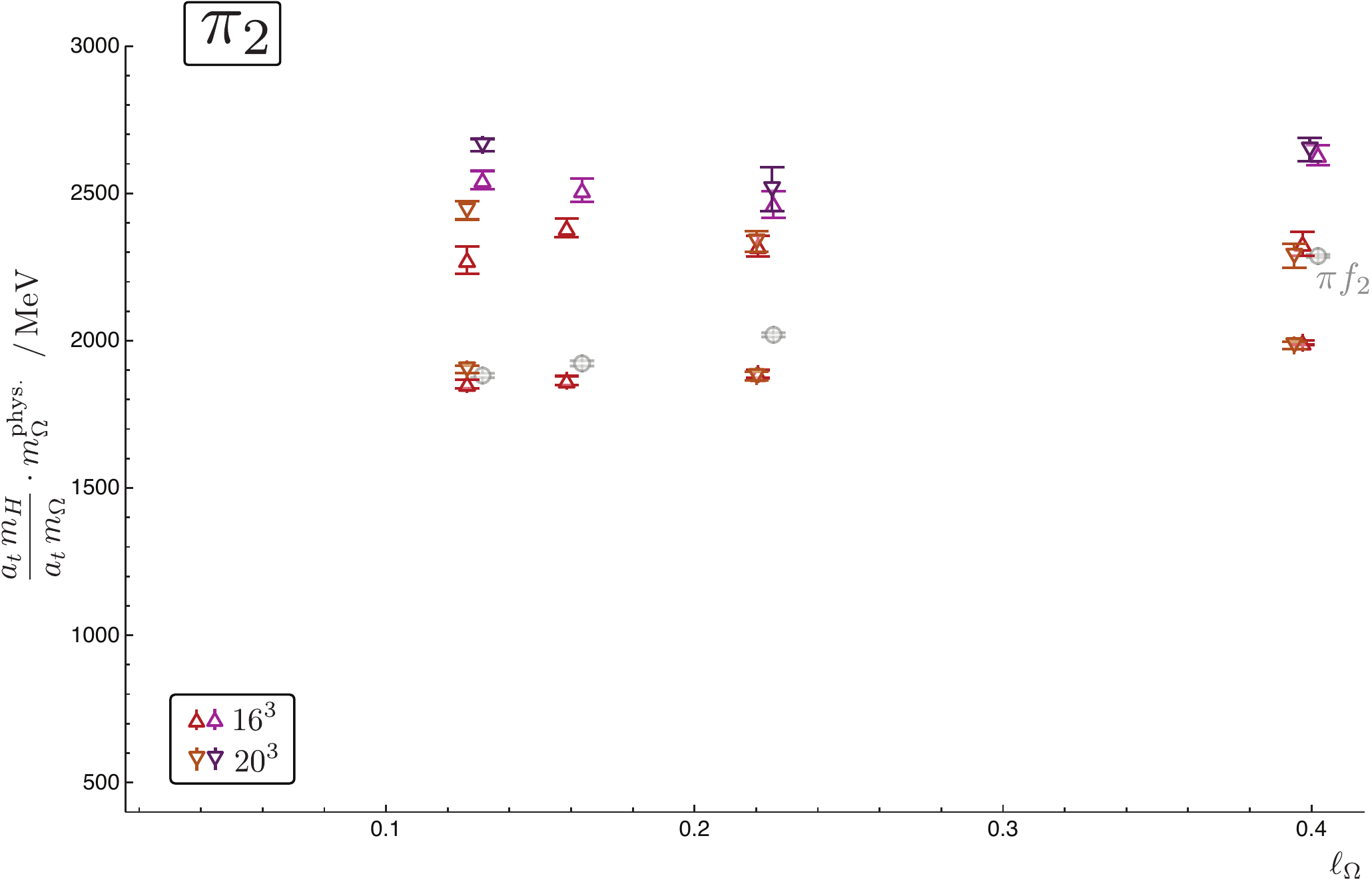}    
\includegraphics[width=0.45\textwidth,bb= 0 0 594 386]{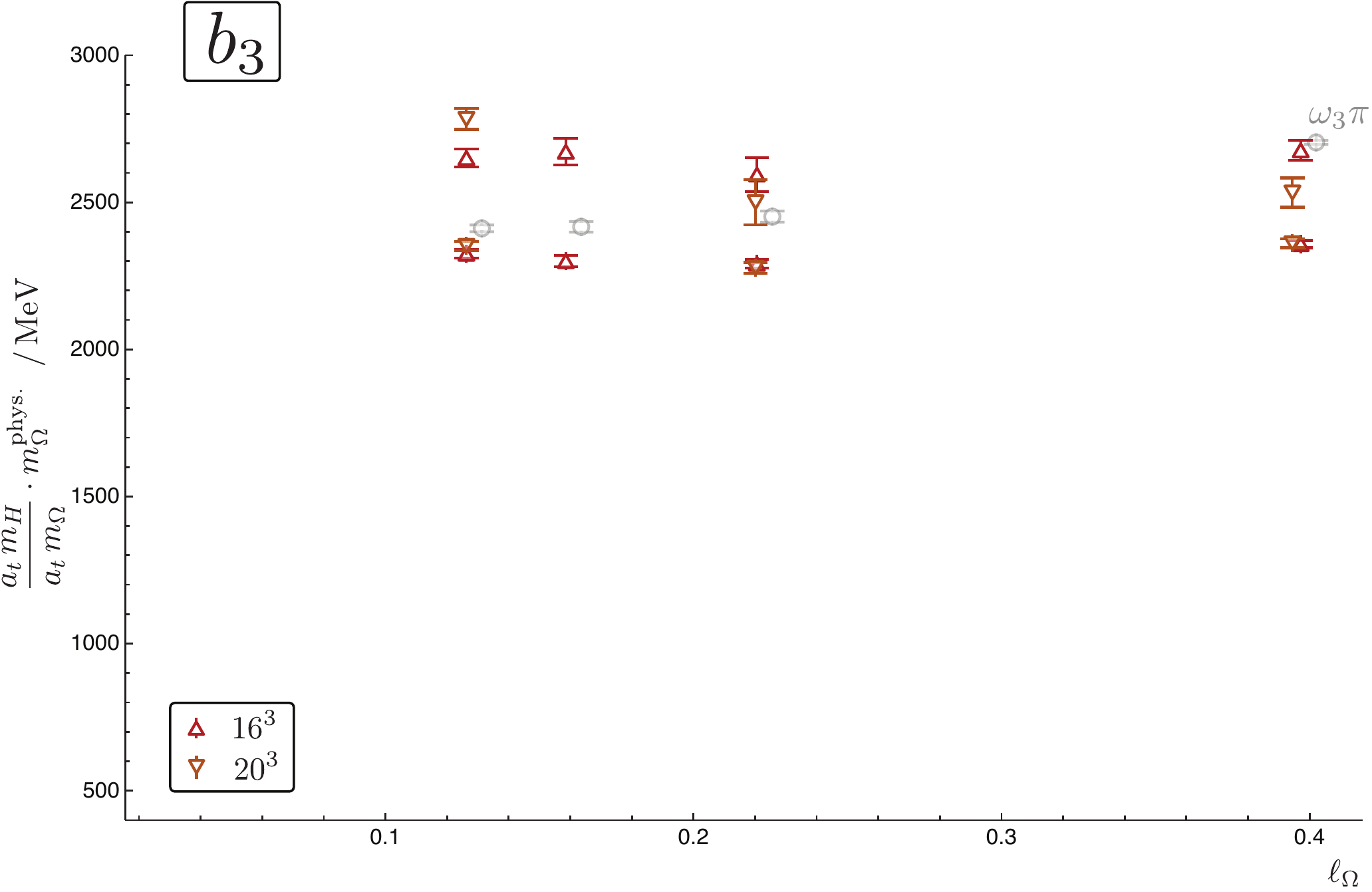}        
\caption{Lightest non-exotic isovector states, $\pi_J(J^{-+})$ and $b_J(J^{+-})$; the ground state $\pi$ is not shown. \label{pib}}
\end{figure*}

\begin{figure*}
 \centering  
\includegraphics[width=0.45\textwidth,bb= 0 0 593 382]{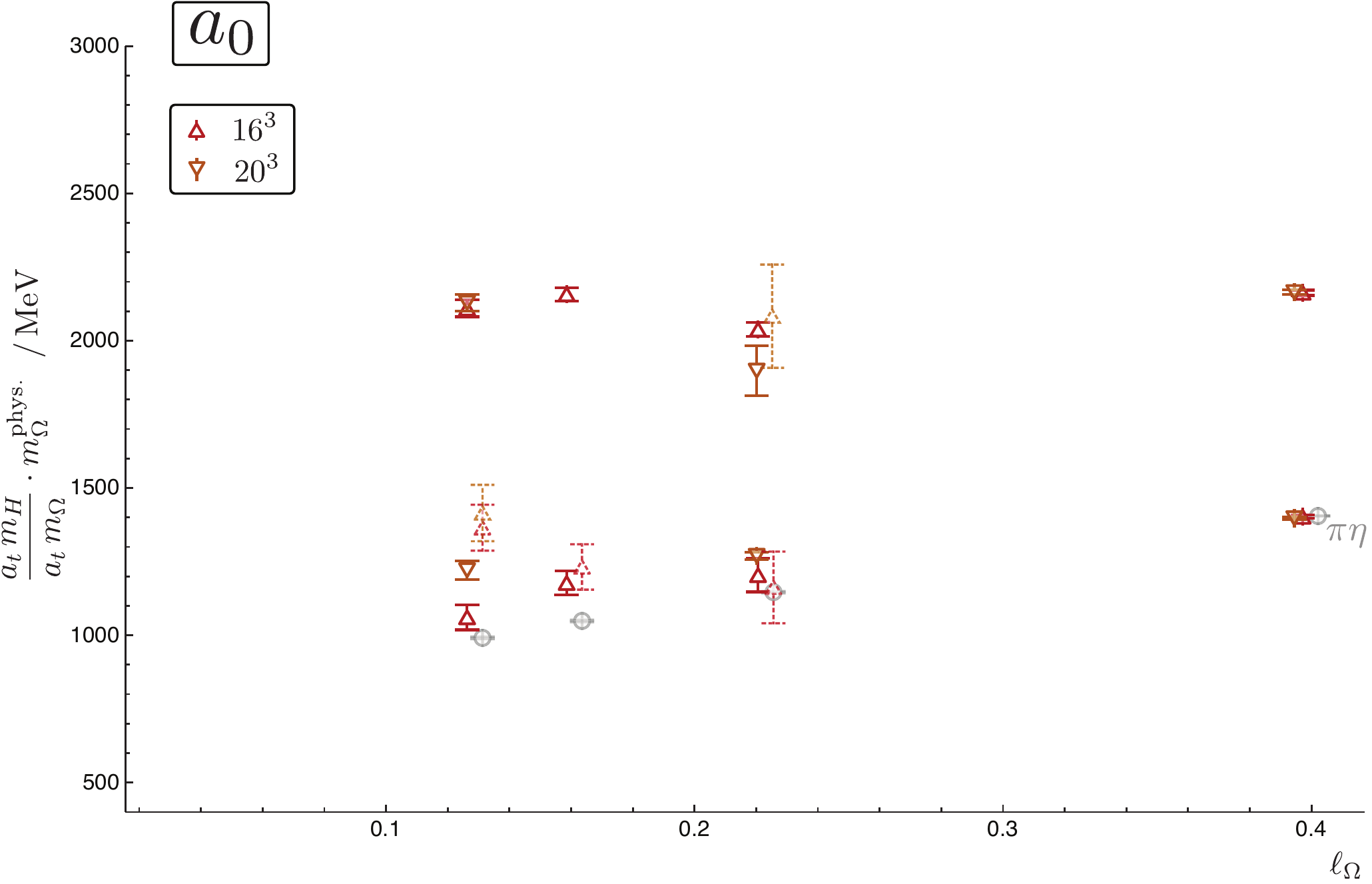}     
\includegraphics[width=0.45\textwidth,bb= 0 0 591 382]{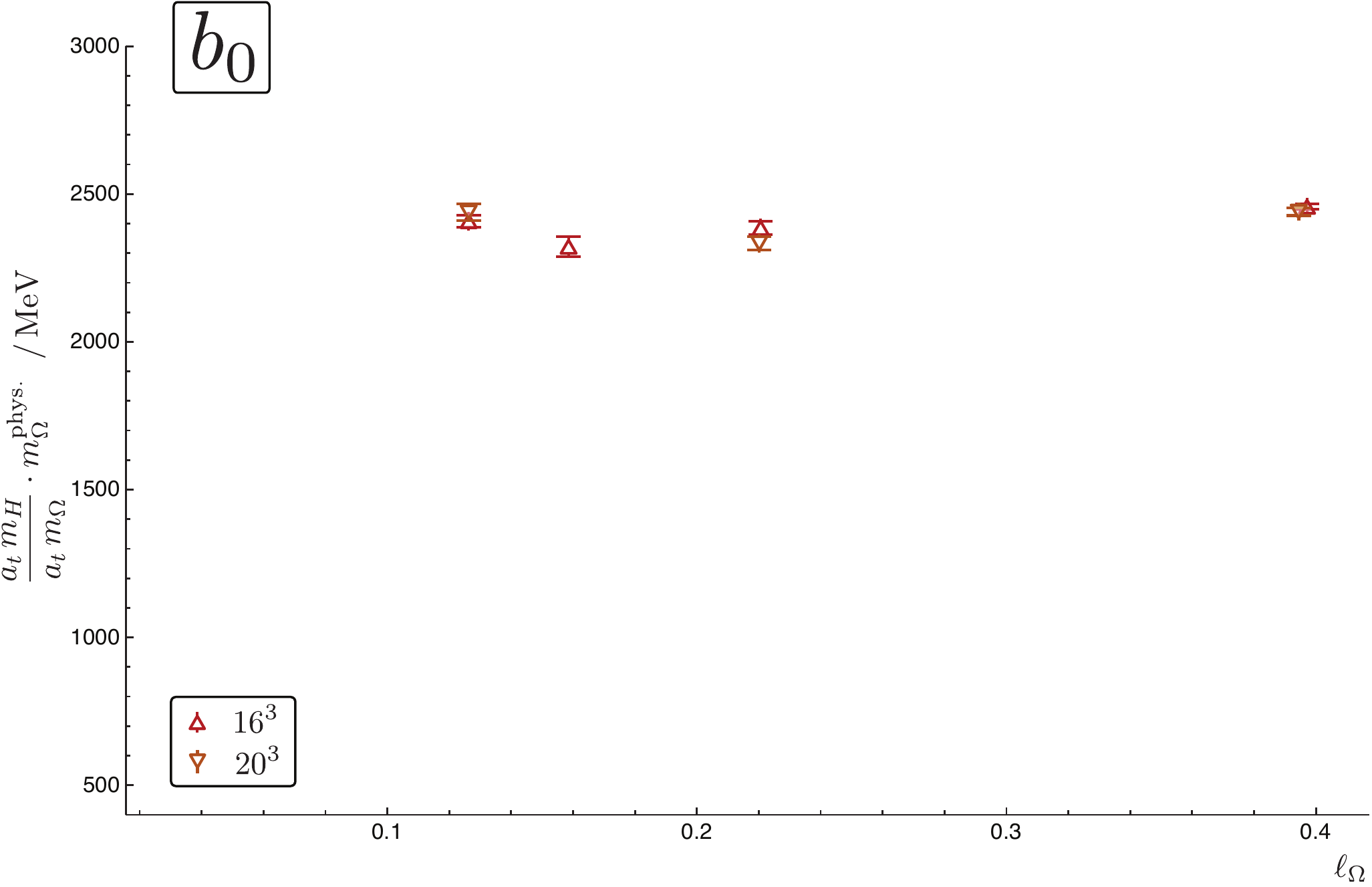}     
\includegraphics[width=0.45\textwidth,bb= 0 0 593 382]{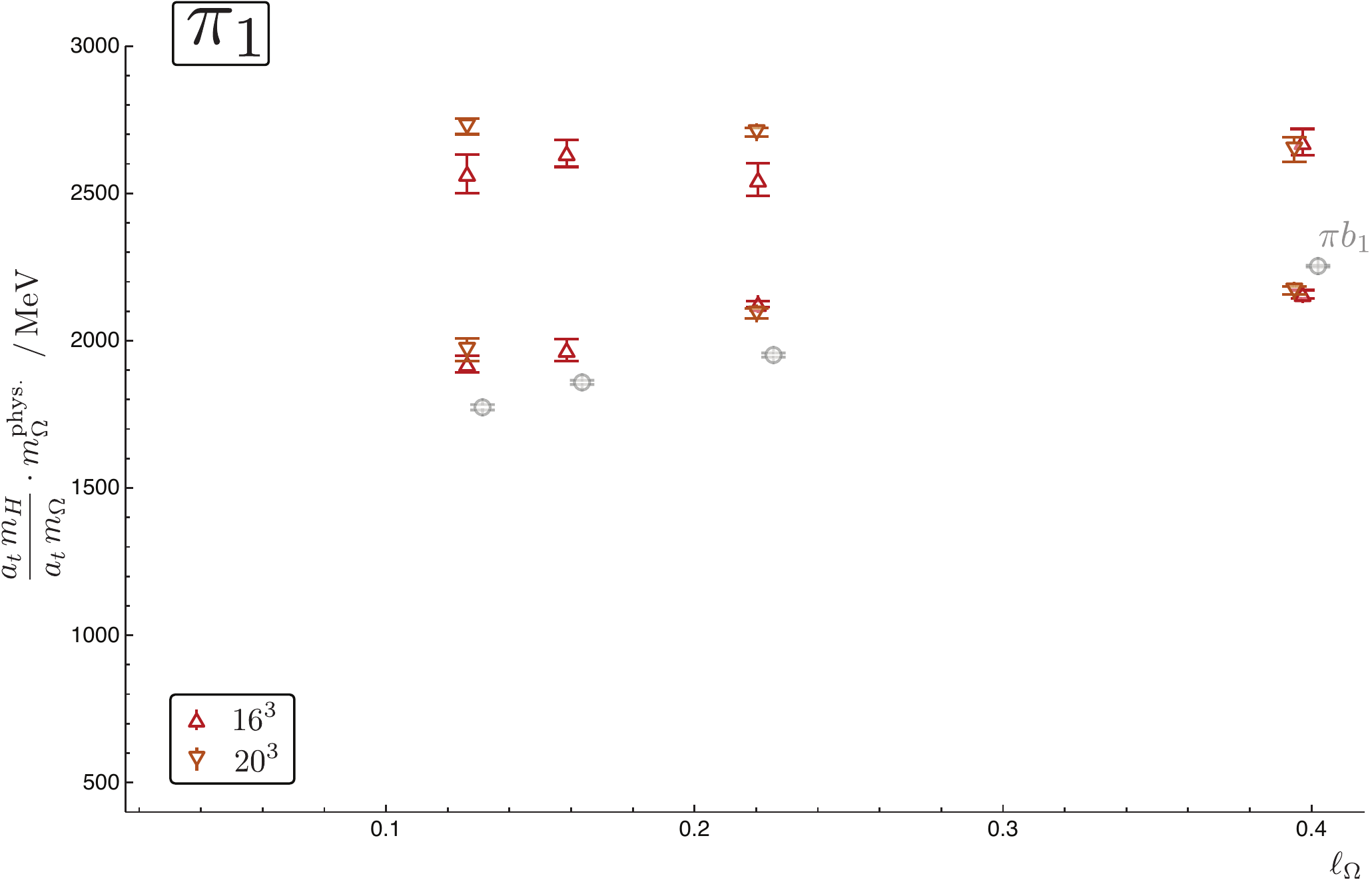}    
\includegraphics[width=0.45\textwidth,bb= 0 0 593 382]{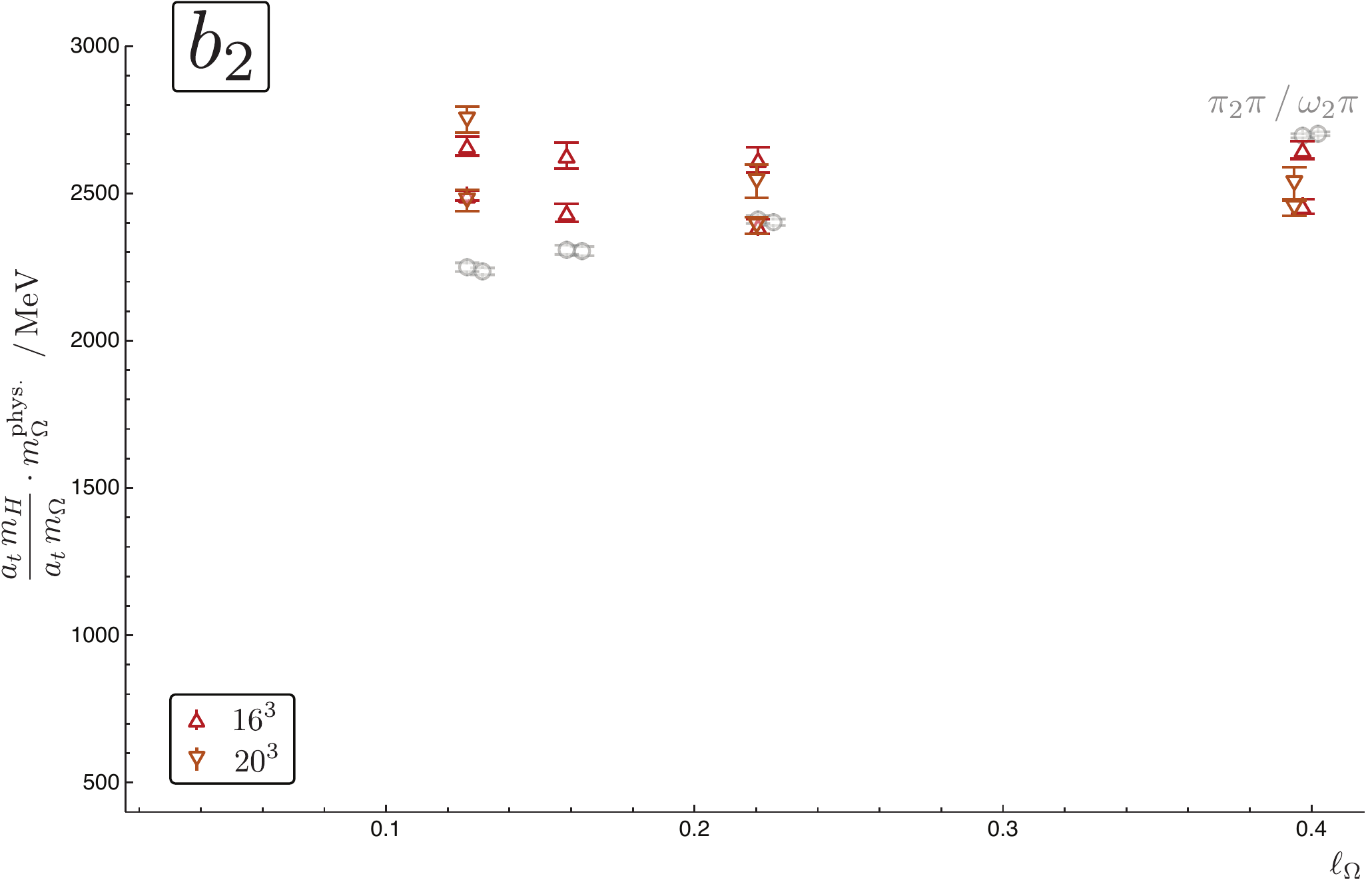}        
\caption{Lightest isovector states: scalar and exotics, $a_0(0^{++})$, $\pi_1(1^{-+})$, $b_J(J^{+-})$.\label{a0ex}}
\end{figure*}

Figures \ref{rhos}, \ref{as}, \ref{pib}, \ref{a0ex} show the extracted spin-assigned spectra for mesons of isospin-1 having a range of $J^{PC}$ quantum numbers (the neutral members of the $I=1$ triplet are eigenstates of $C$, the charged members are eigenstates of $G$-parity with $G=-C$). We use the PDG \cite{Amsler:2008zzb} nomenclature for meson states throughout. In those cases where two states are almost degenerate we shift one in the horizontal direction by an amount $\delta \ell_\Omega = 0.005$ for clarity. In some cases, for comparison, we plot the mass of the lightest meson-meson pair which in $S$-wave would have the appropriate quantum numbers - the mass follows from the simple sum of the extracted masses on these lattices. This may involve a so-far undetermined isoscalar mass and in these cases we use the approximations $m_\omega \approx m_\rho$, $m_{f_2} \approx m_{a_2}$ and the crudely estimated $\eta$ mass from \cite{Lin:2008pr}. Occasionally we extract a low-lying state that is reasonably robust against the changes in analysis detailed in section \ref{sec:stability}, but whose principal correlator is quite noisy leading to a relatively poorly determined mass - we show these states with dashed symbols. 

Not shown are results for $0^{--}$ isovectors, the $\rho_0$, which are exotic. The lightest such state we extract is at least 2 GeV heavier than the $\rho$ at all our quark masses. The exotic $3^{-+}$, the $\pi_3$, is found to be similarly heavy.

\subsubsection{Kaons}

In the kaon sector, we no longer have charge-conjugation as a good quantum number, with only $J^P$ remaining in the continuum which is then subduced into $\Lambda^P$ on a cubic lattice. We compute a correlator matrix for a given $\Lambda^P$ using the concatenated list of all $\Lambda^{P+}$ and $\Lambda^{P-}$ operators. 

\begin{figure}
 \centering
\includegraphics[width=0.4\textwidth,bb= 0 0 355 332]{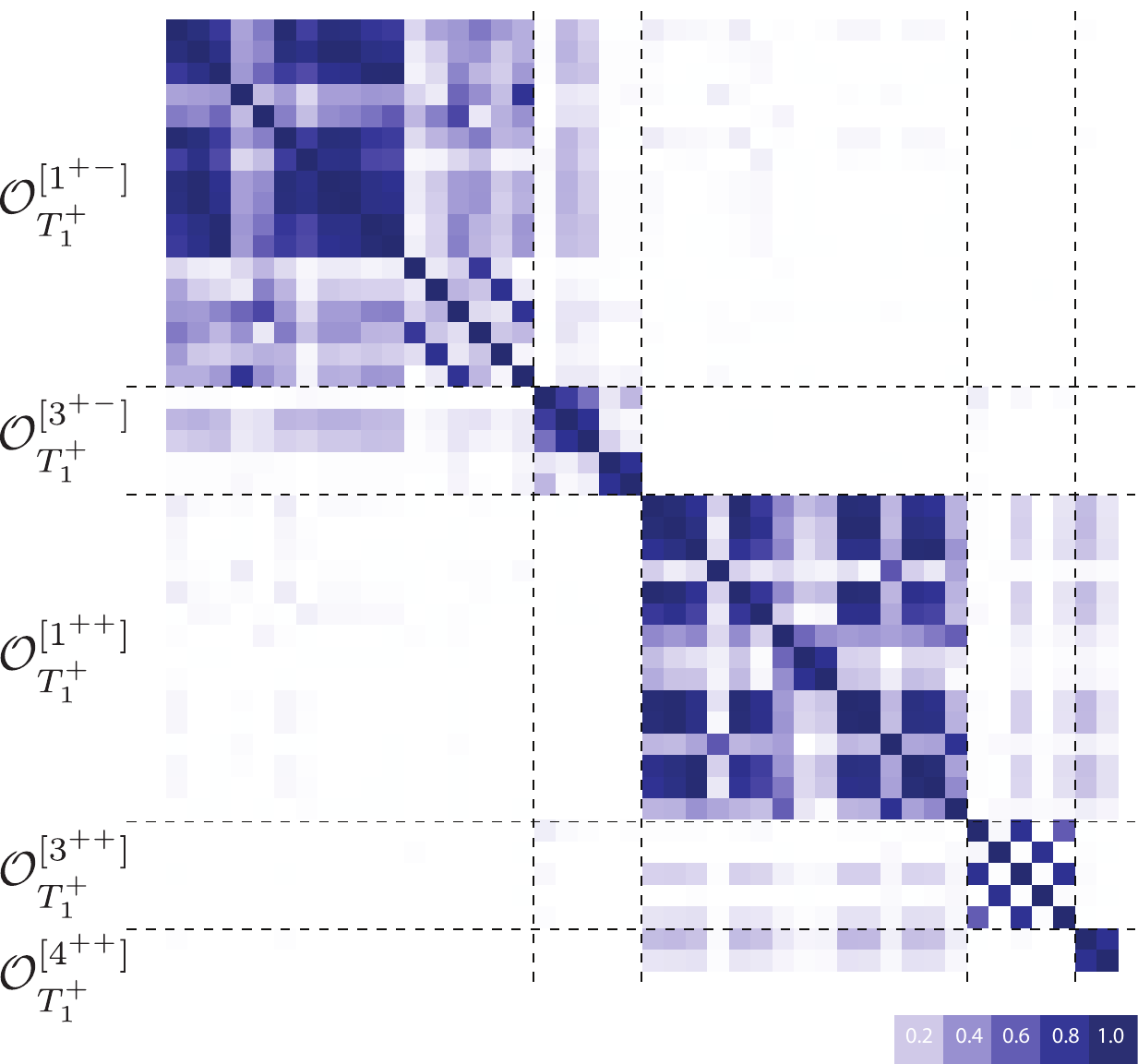}        
\caption{Normalised corrrelation matrix ($C_{ij}/\sqrt{C_{ii} C_{jj}}$) on timeslice 5 in the $T_1^{+}$ kaon irrep on \emph{840} $16^3$ lattice. Operators with $C=-$ first, $C=+$ second.  \label{kaon_matrix_plot}}
\end{figure}

\begin{figure}
 \centering
\includegraphics[width=0.4\textwidth,bb= 0 0 421 716]{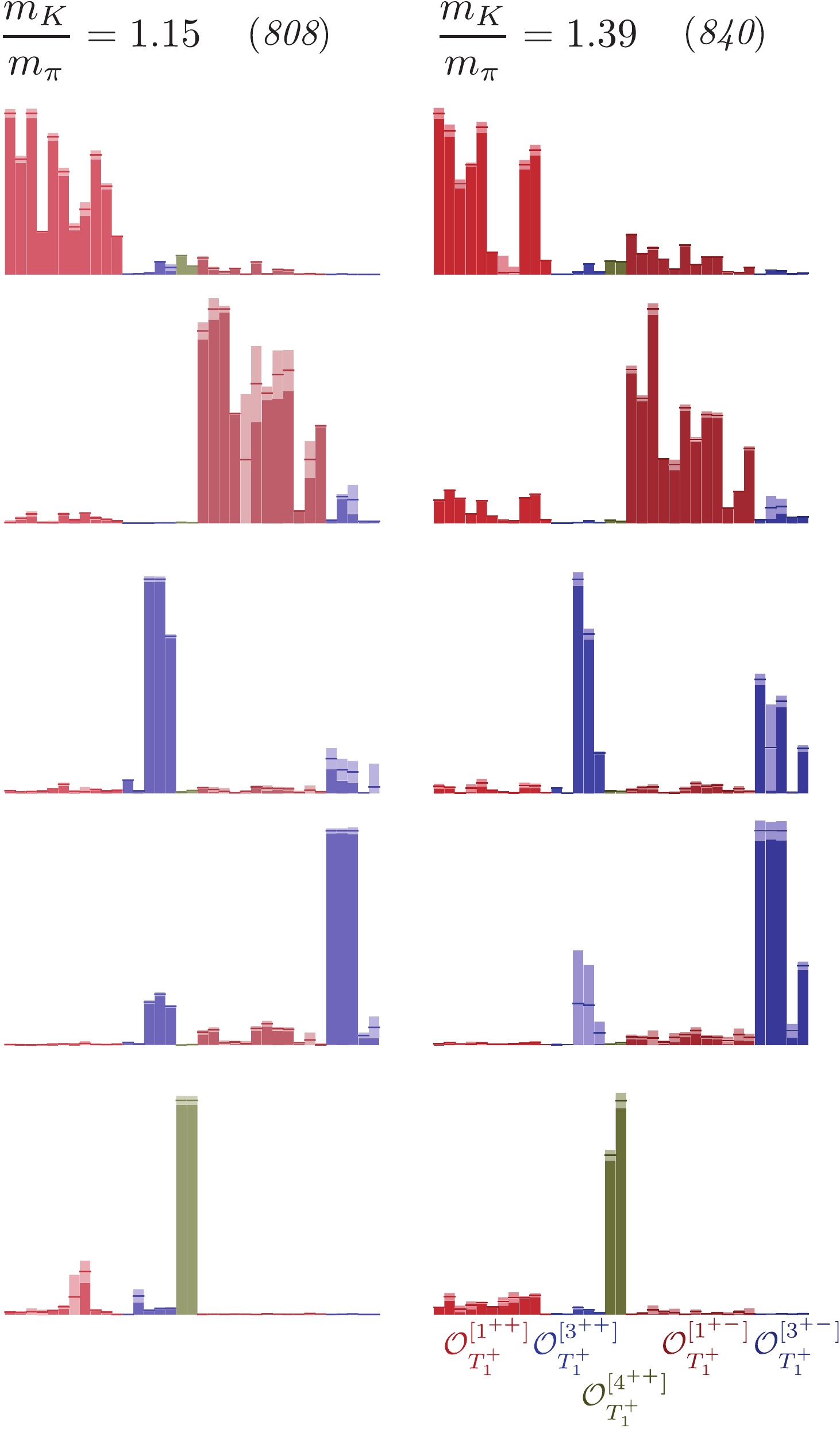}        
\caption{Overlaps, $Z$, of $T_1^+$ kaon operators onto lightest two $J=1$, lightest two $J=3$ and lightest $J=4$ states for \emph{808} $16^3$ (left) and \emph{840} $16^3$ (right) lattices. Operators with $C=\pm$ are grouped to show clear separation. Normalisation as in figure \ref{histogram}. \label{kaon_histogram}}
\end{figure}

\begin{figure*}
 \centering
\includegraphics[width=0.7\textwidth,bb= 0 0 628 459]{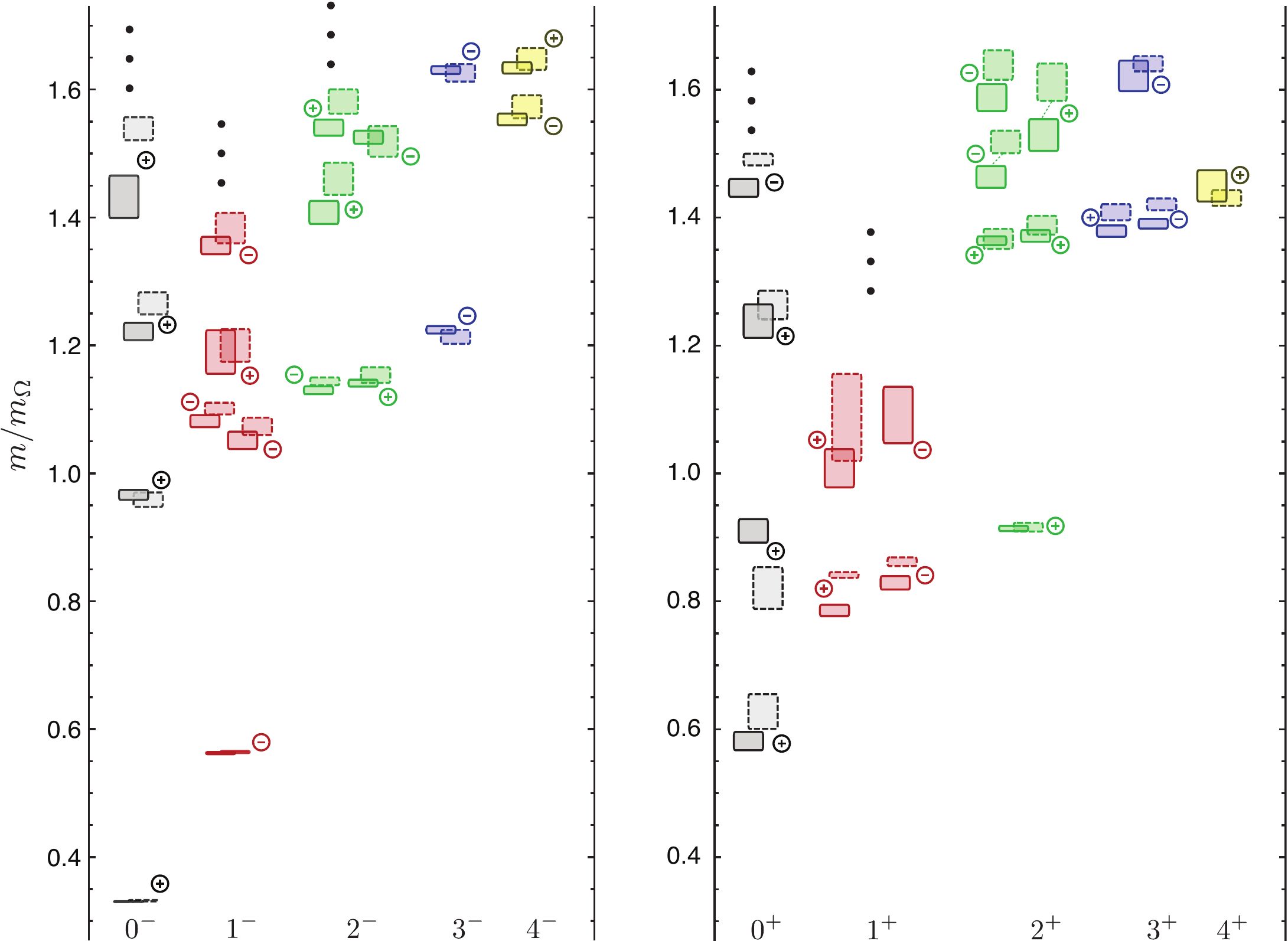}        
\caption{Spin-identified spectrum of kaons from the \emph{840} lattices. $16^3$(solid) and $20^3$(dashed) spectra mostly agree well. The plus and minus symbols indicate the dominance of a $C=\pm$ eigenstate in that state. Ellipses indicate that there are heavier states with a given $J^{PC}$ but that they are not well determined in this calculation.\label{840_kaon} The rather dense spectrum of axial kaons above $m/m_\Omega \sim 1.3$ is suppressed for clarity.}
\end{figure*}

Using a combination of experiment and models\cite{Asner:2000nx, Barnes:2002mu} there are suggestions that resonant kaon states are mixtures of basis states of opposite $C$ with a rather large mixing angle. For example the axial kaons, $K_1(1270), K_1(1400)$, are suggested to be mixtures of basis states $K_{1A}(C=+)$, $K_{1B}(C=-)$ with a mixing angle near $45^\circ$. Clearly this mixing relies upon being far away ($m_K^{\mathrm{phys}}/m_\pi^{\mathrm{phys}} = 3.5$) from the $SU(3)_F$ limit, since in that limit there is effectively restoration of (a generalisation of) $C$-symmetry. All the lattices presented in this paper can be considered to be rather close to the $SU(3)_F$ limit ($ 1 \le m_K/m_\pi \le 1.39$) and we observe little or no mixing. This is suggested at the correlator level (see figure \ref{kaon_matrix_plot} for the \emph{840} $16^3$ correlator), and verified in the $Z$ values of the spectrum extraction (see figure \ref{kaon_histogram} for the \emph{808} $16^3$ lattice (left) and the \emph{840} $16^3$ lattice (right)). While the \emph{840} lattice shows a greater degree of opposite $C$ mixing than the \emph{808}, indicating an increased breaking of $SU(3)_F$ symmetry, the mixing is still very small in absolute terms and states are approximately eigenstates of $C$. The \emph{840} kaon spectrum is shown in figure \ref{840_kaon} with the dominant $C$-eigenstate noted for each state.

\begin{figure*}
 \centering  
\includegraphics[width=0.45\textwidth,bb= 0 0 591 382]{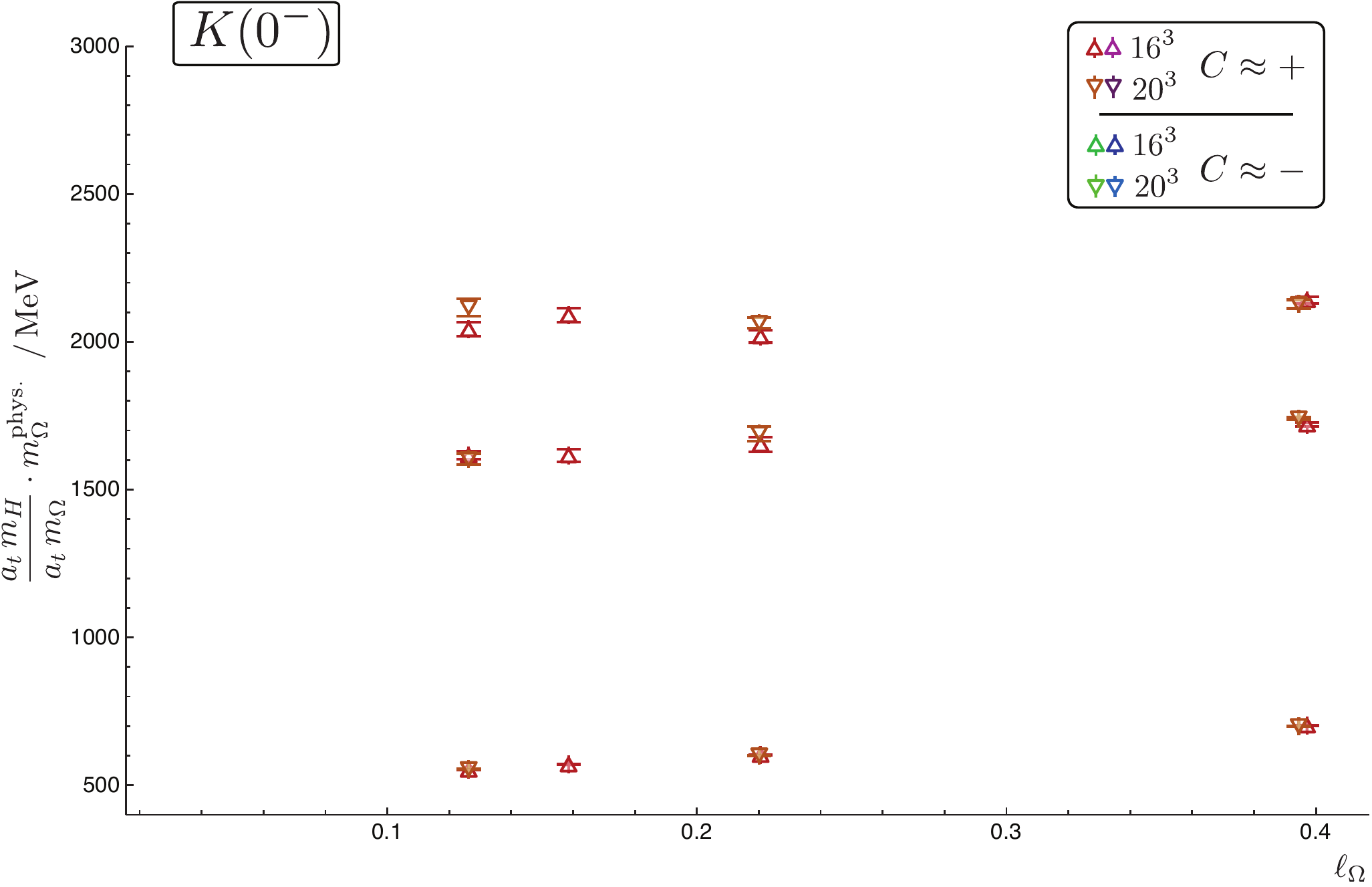}     
\includegraphics[width=0.45\textwidth,bb= 0 0 593 381]{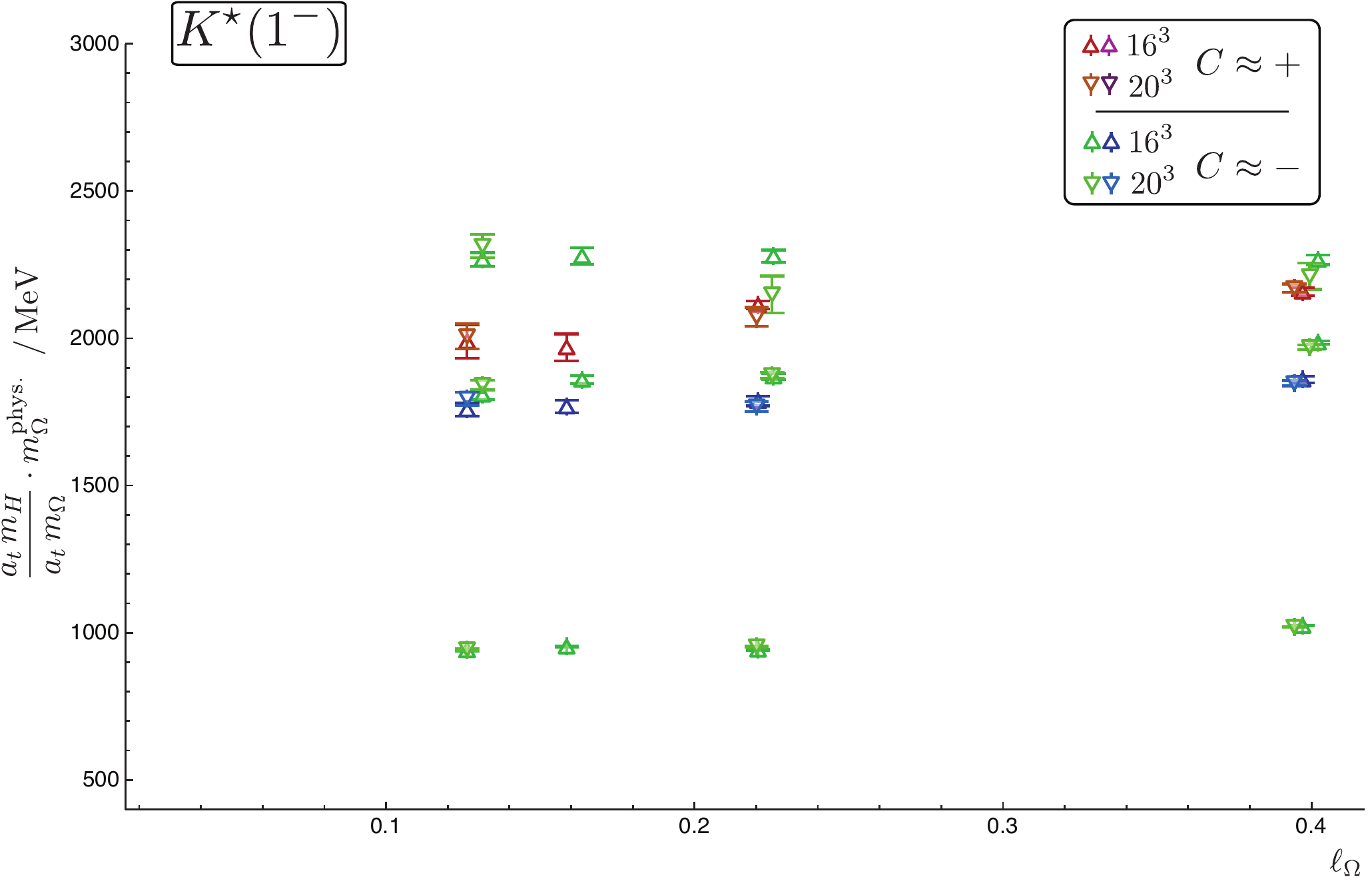}     
\includegraphics[width=0.45\textwidth,bb= 0 0 593 382]{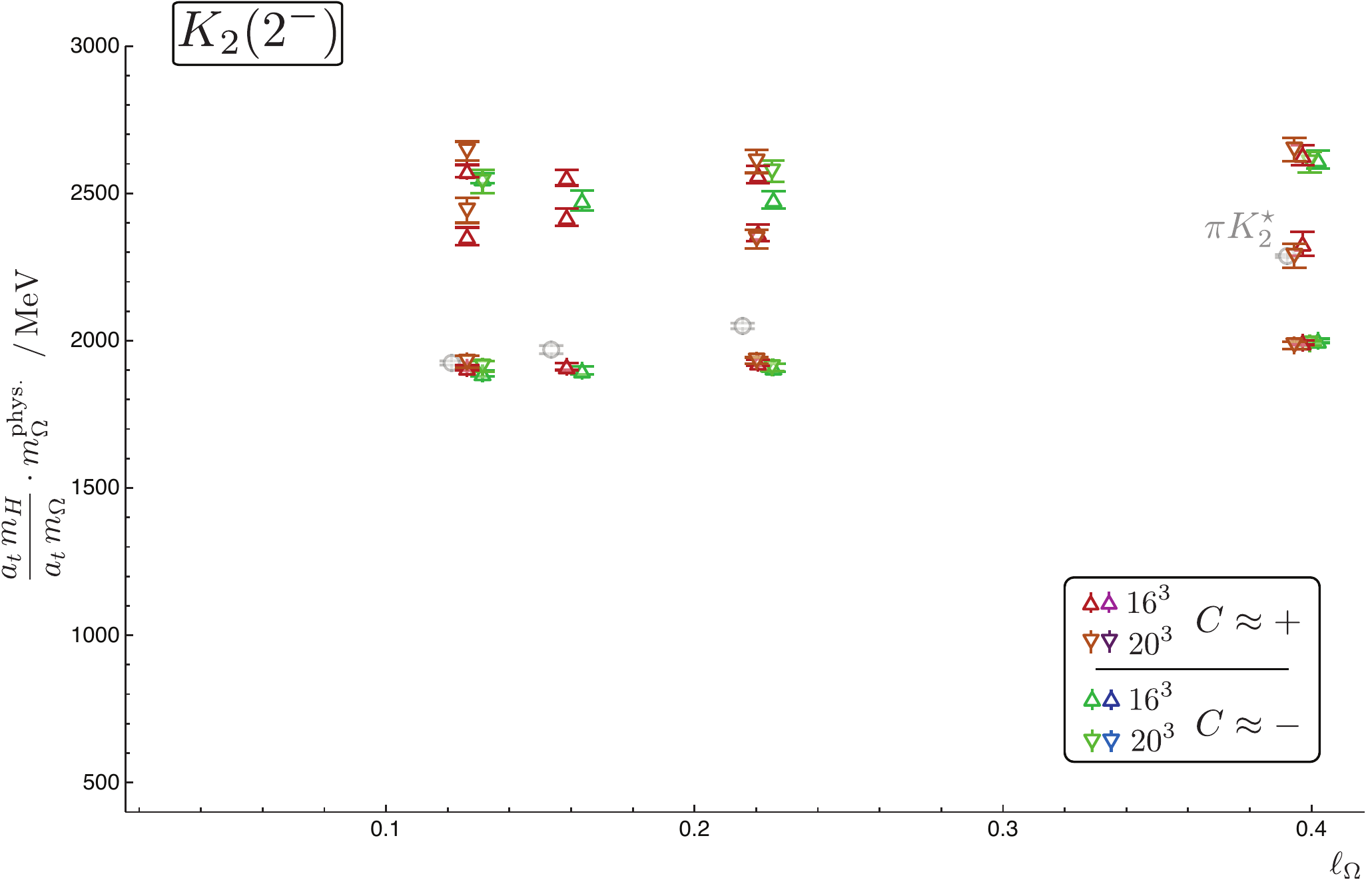}    
\includegraphics[width=0.45\textwidth,bb= 0 0 593 382]{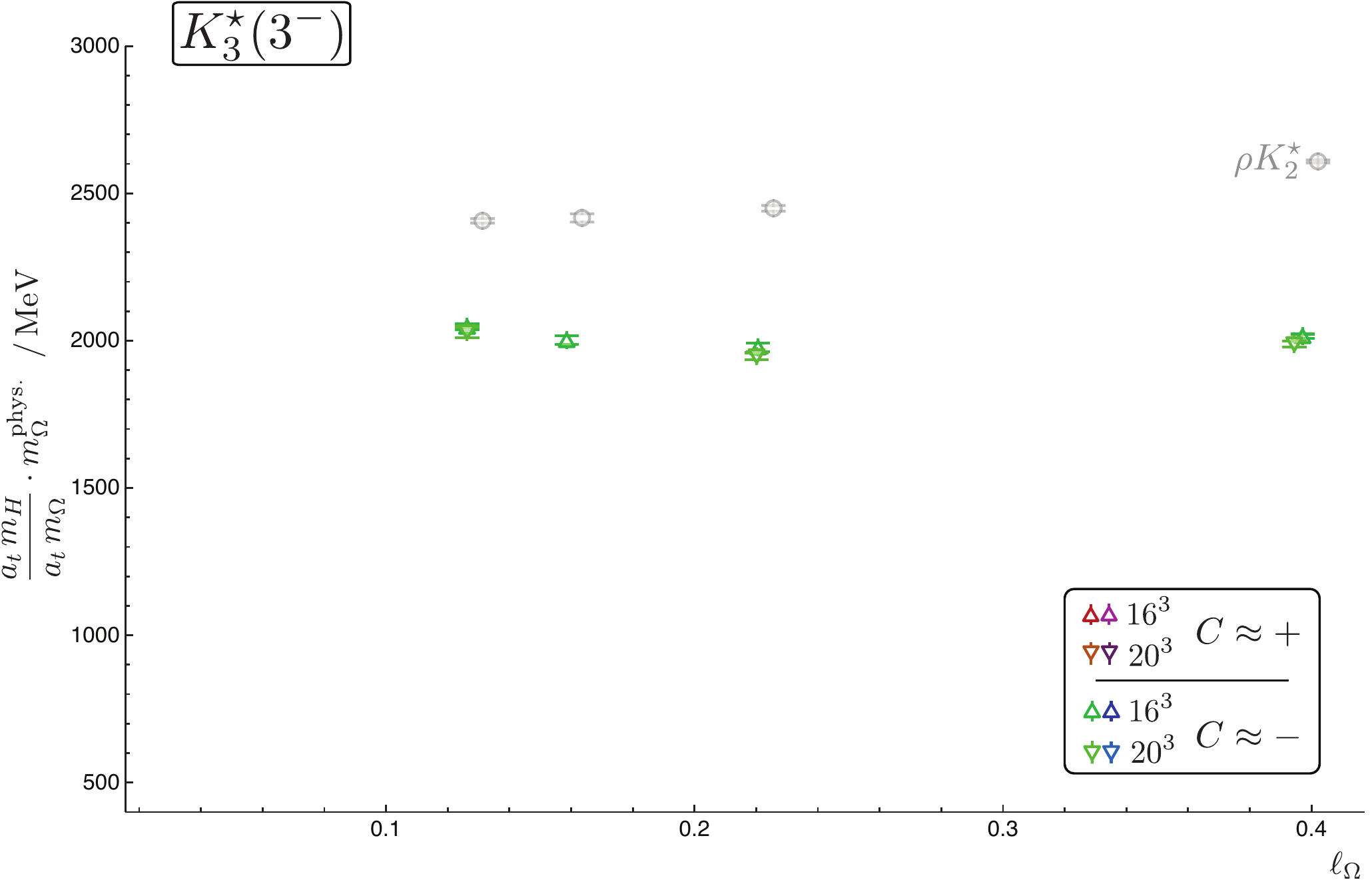}        
\caption{Lightest negative parity kaons. Color coding indicates dominance of a particular charge-conjugation eigenstate. \label{kaons_neg}}
\end{figure*}

\begin{figure*}
 \centering  
\includegraphics[width=0.45\textwidth,bb= 0 0 596 388]{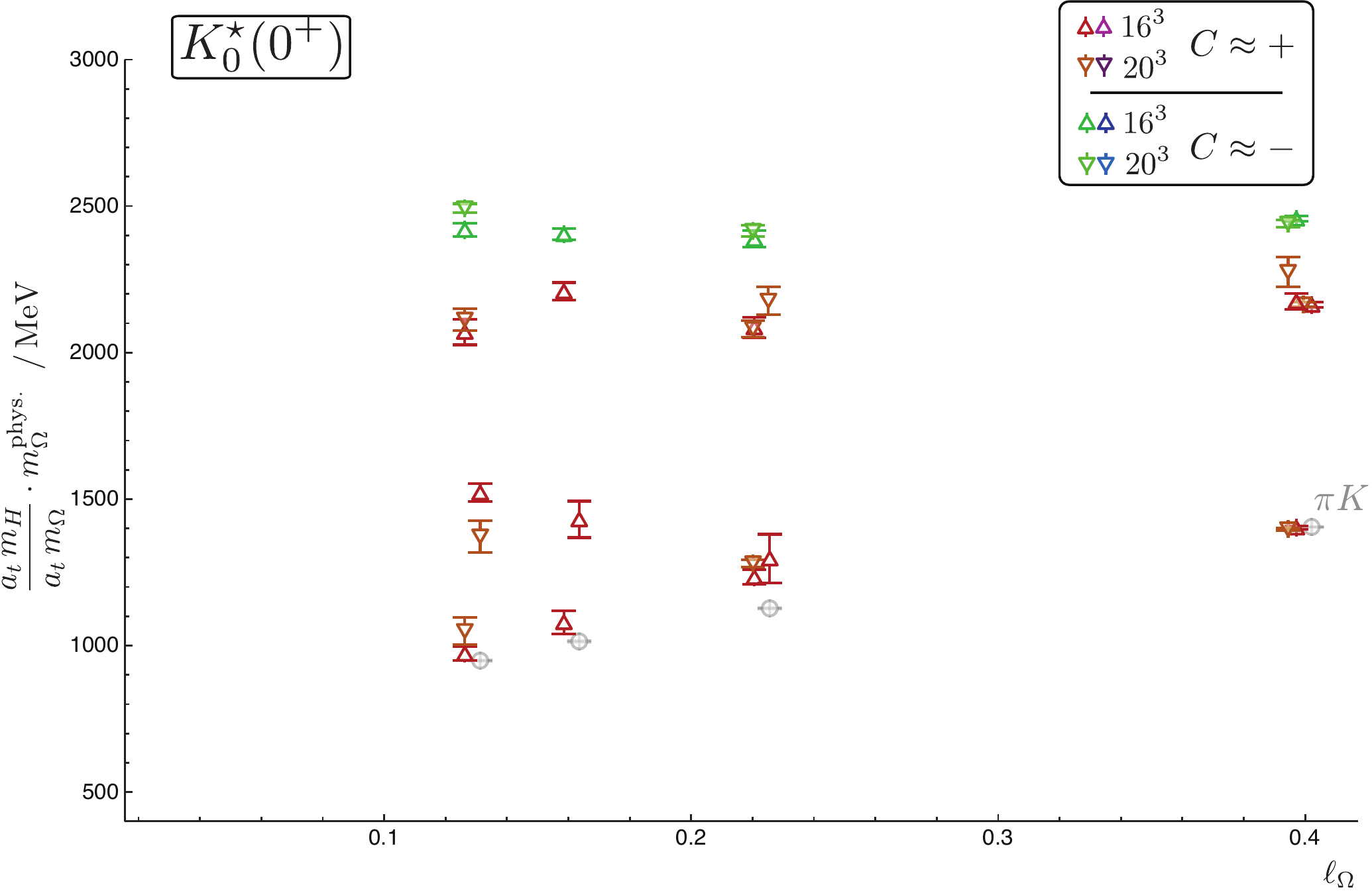}     
\includegraphics[width=0.45\textwidth,bb= 0 0 593 382]{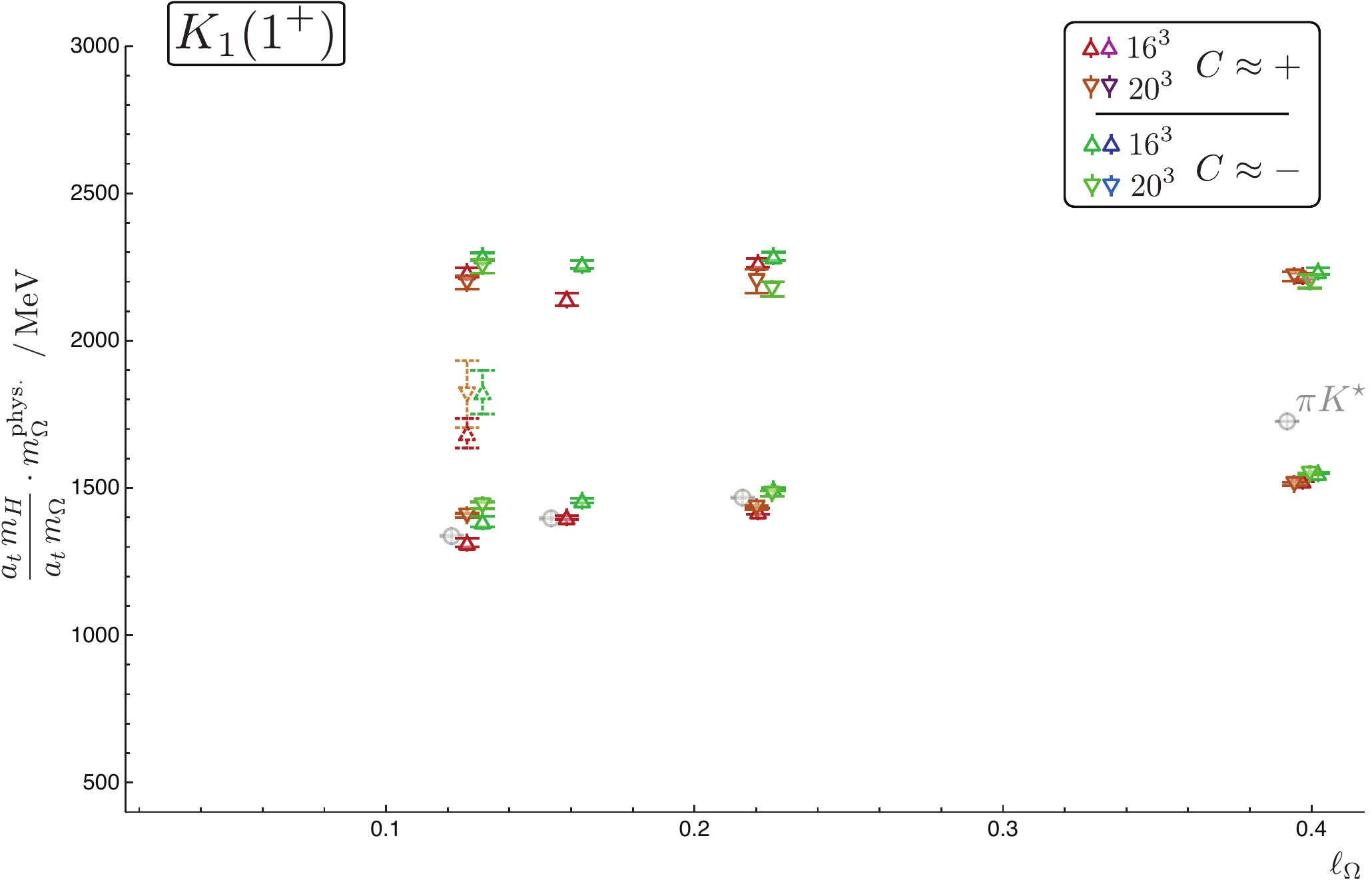}     
\includegraphics[width=0.45\textwidth,bb= 0 0 593 383]{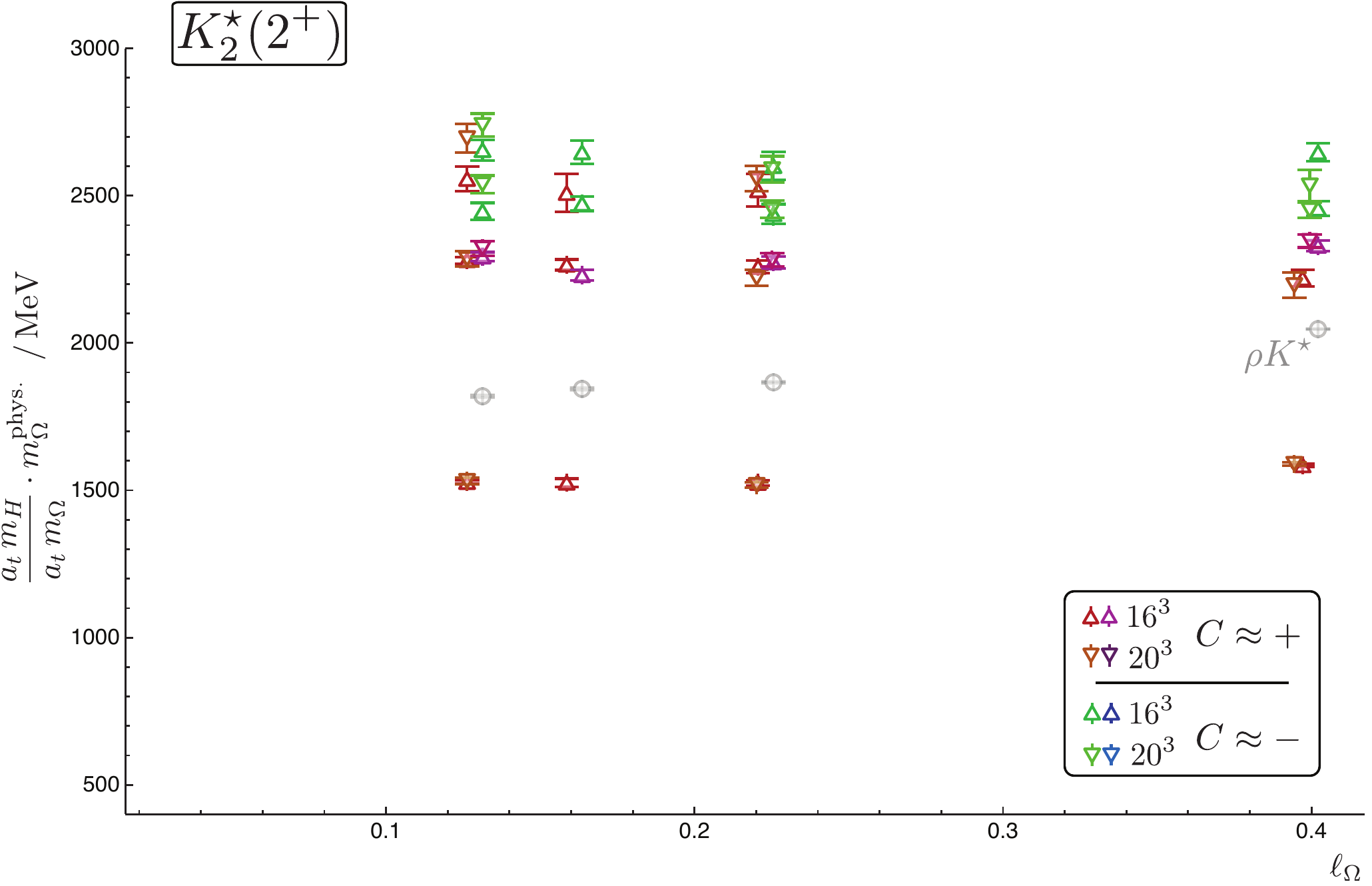}    
\includegraphics[width=0.45\textwidth,bb= 0 0 593 382]{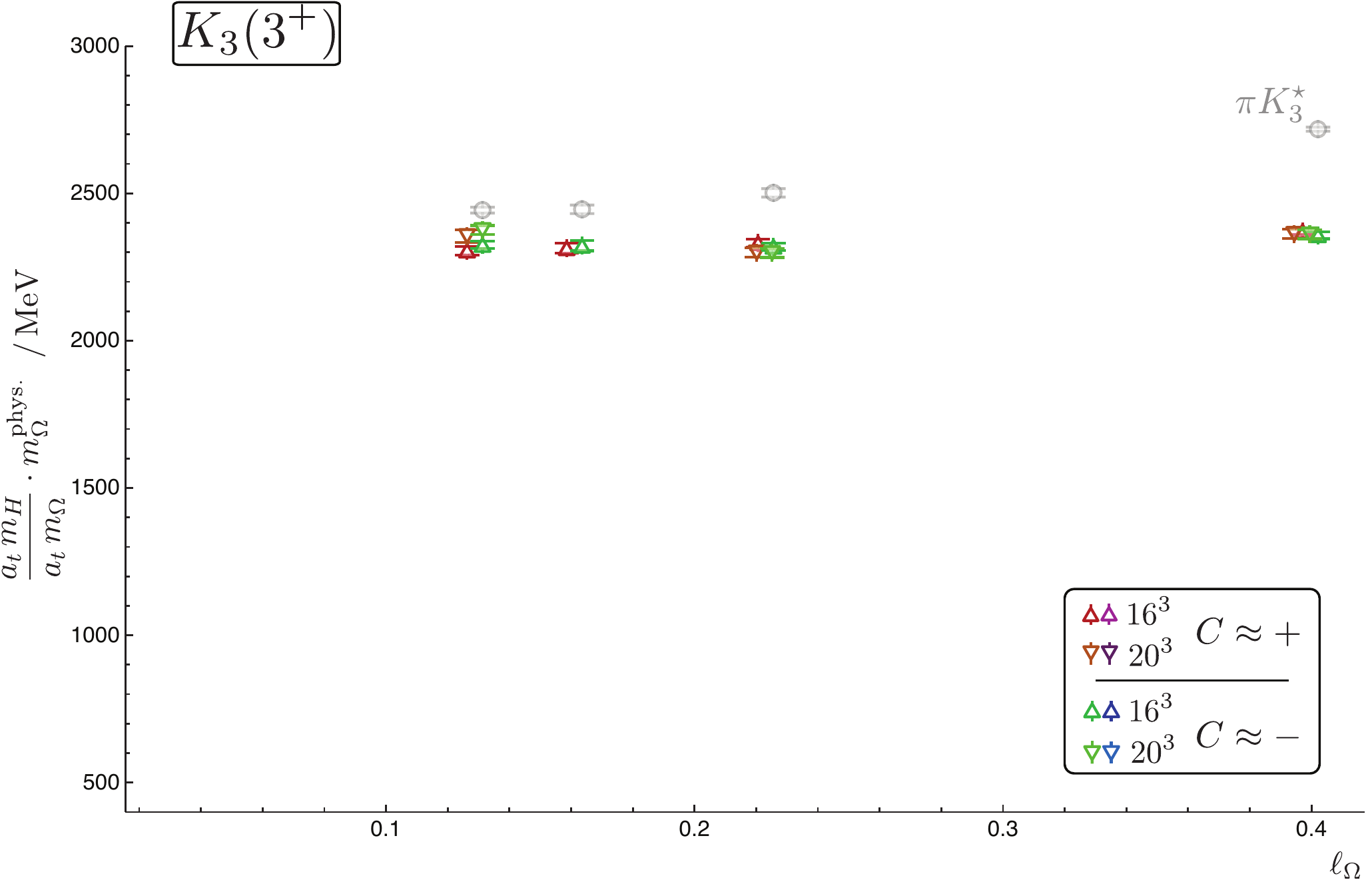}        
\caption{Lightest positive parity kaons. Color coding indicates dominance of a particular charge-conjugation eigenstate. \label{kaons_pos}}
\end{figure*}

\begin{figure*}
 \centering  
\includegraphics[width=0.45\textwidth,bb= 0 0 591 382]{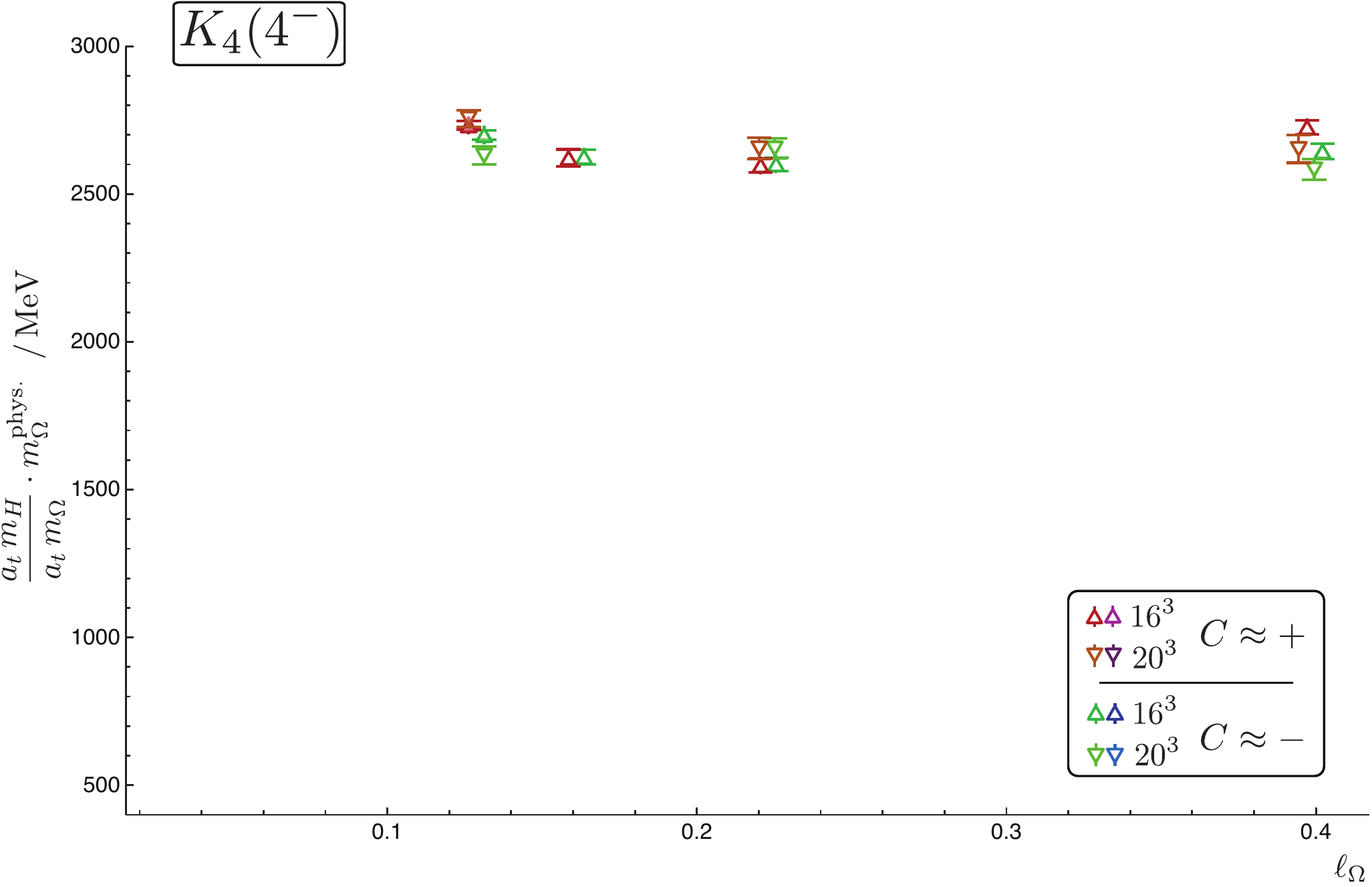}     
\includegraphics[width=0.45\textwidth,bb= 0 0 591 382]{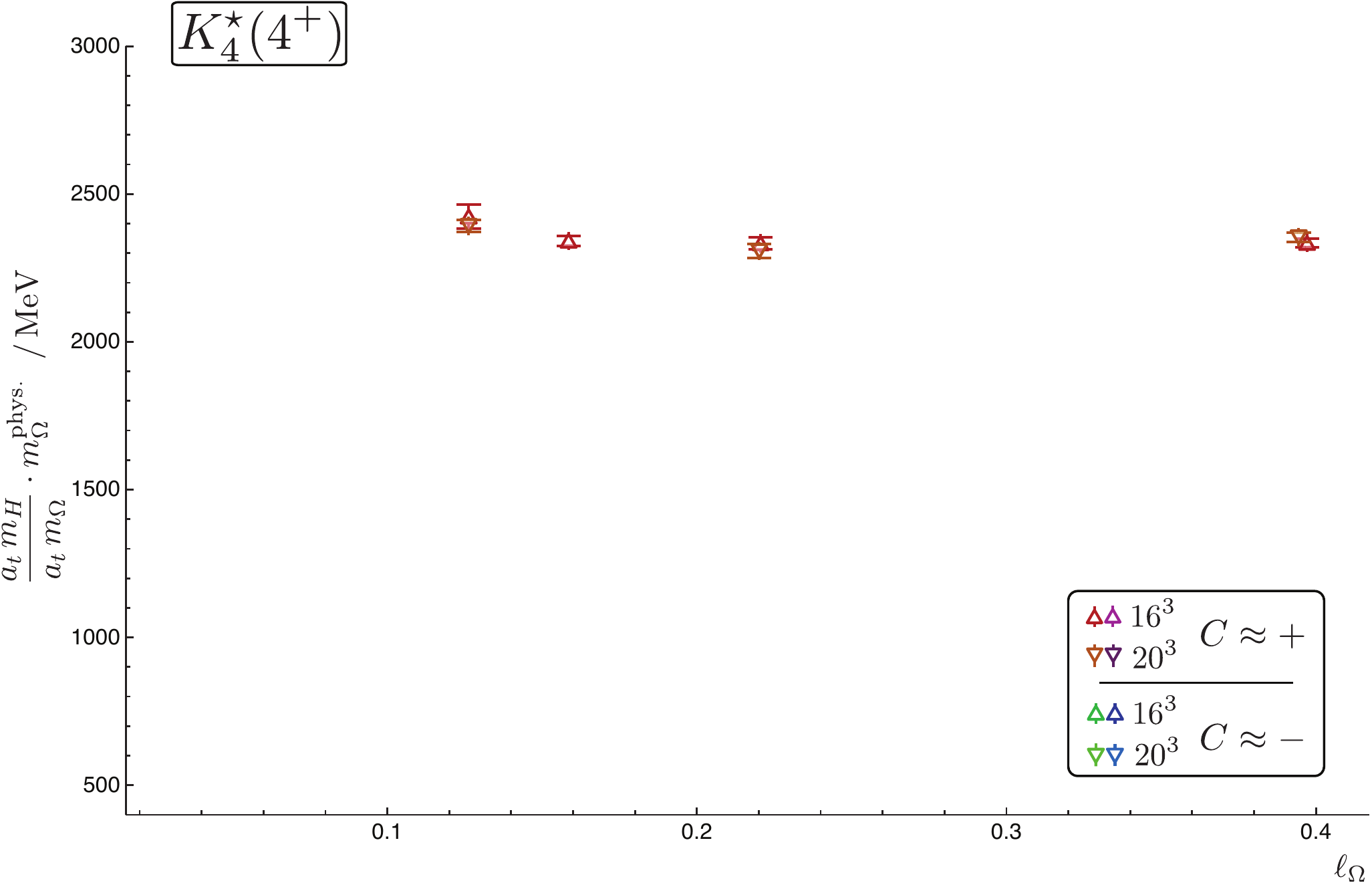}     
\caption{Lightest spin-4 kaons. Color coding indicates dominance of a particular charge-conjugation eigenstate. \label{kaons_4}}
\end{figure*}

The light-quark mass dependence of kaon states is displayed in figures \ref{kaons_neg}, \ref{kaons_pos}, \ref{kaons_4} where color-coding indicates the dominance of $C$-eigenstates within the spectrum.

\subsubsection{``Strangeonium"}
\begin{figure}
 \centering  
\includegraphics[width=0.45\textwidth,bb= 0 0 591 382]{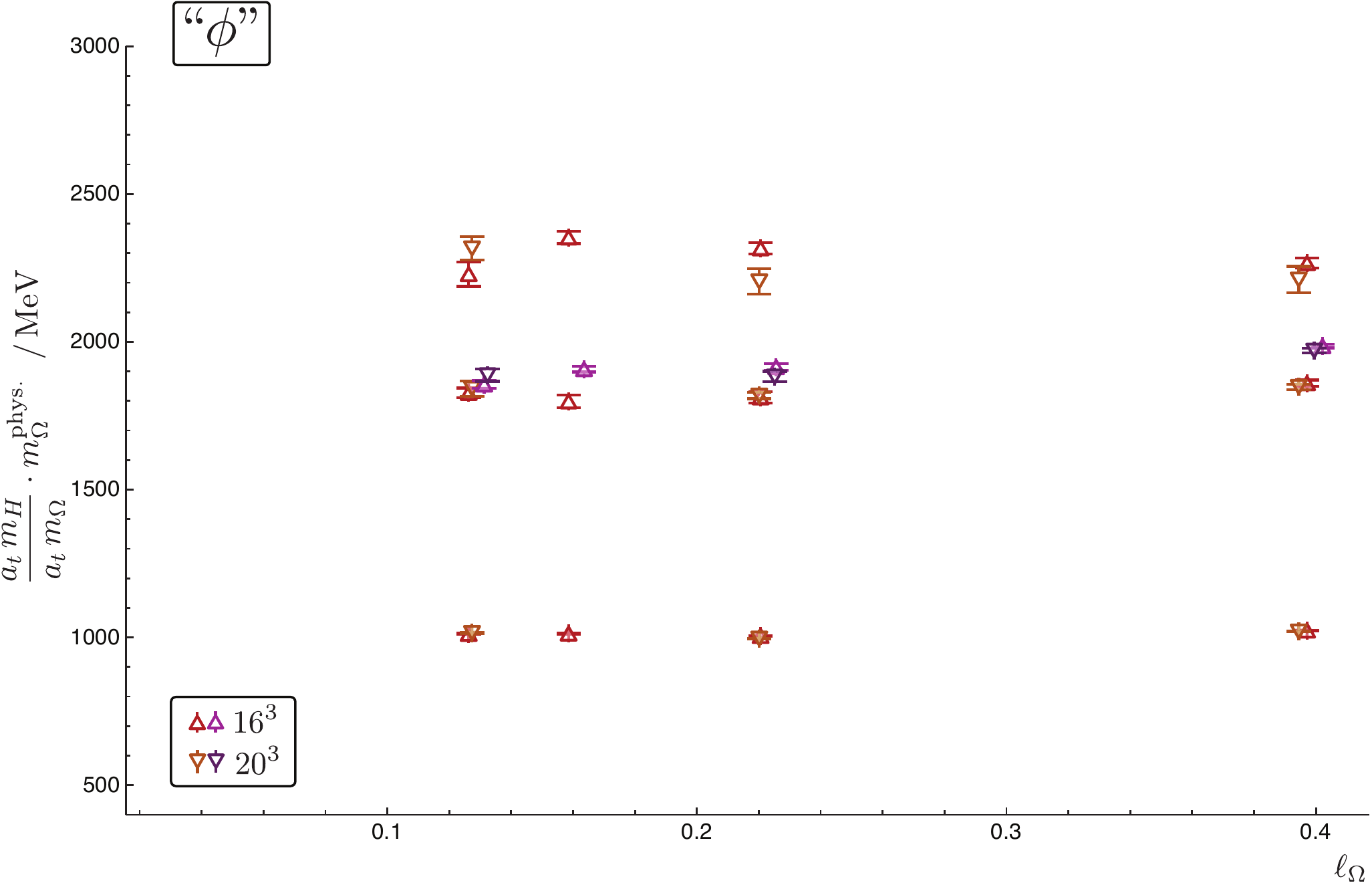}     
\includegraphics[width=0.45\textwidth,bb= 0 0 591 384]{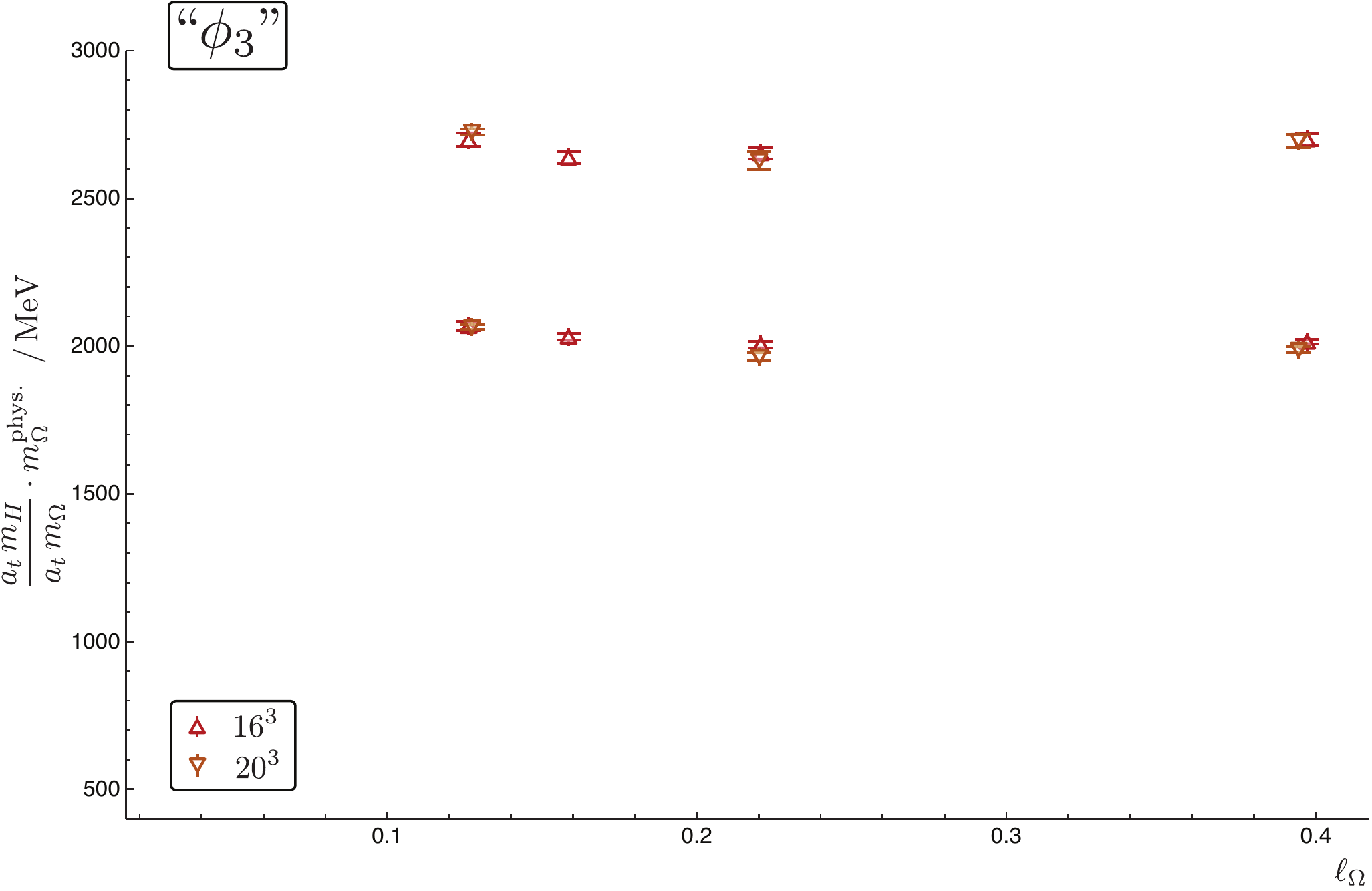}    
\includegraphics[width=0.45\textwidth,bb= 0 0 591 382]{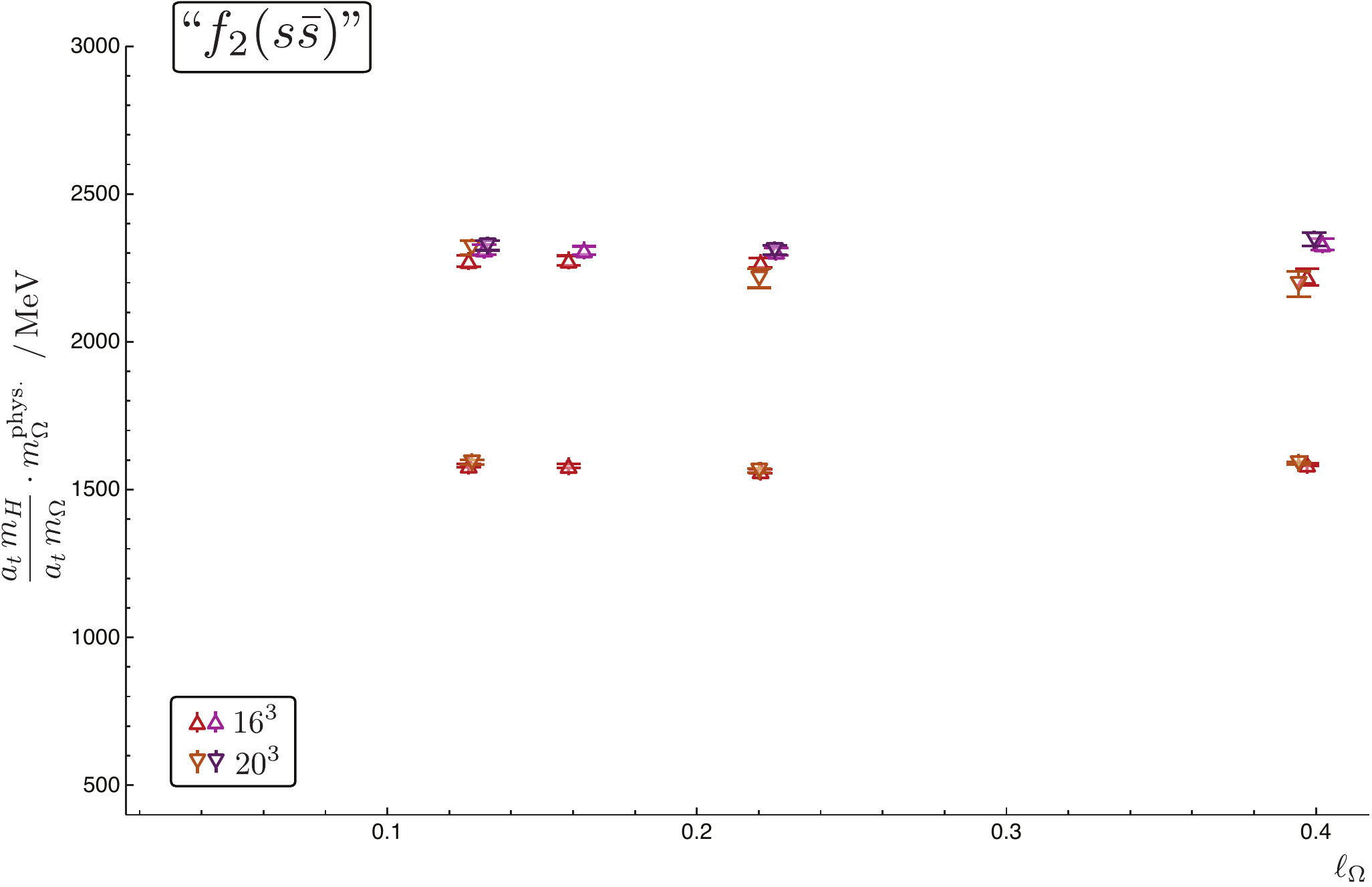}        
\caption{Lightest ``strangeonium" states with $J^{PC} = 1^{--}, 3^{--}, 2^{++}$.\label{ss}}
\end{figure}

In figure \ref{ss} we show only a subset of possible ``strangeonium" $J^{PC}$, those for which there is some phenomenological evidence that the QCD eigenstates are in fact close to being pure $s\bar{s}$. We note that using this particular scale-setting scheme we observe light-quark mass-dependencies that are very flat.

\pagebreak

\section{Multi-meson states}\label{sec:two-meson}

In the previous section we presented the extracted spectra from calculations with four different light quark masses on two different lattice volumes. In each case we were able, using the operator overlaps, to match states across irreps that we believe are subduced from the same continuum spin state. This suggests an interpretation of the spectrum in terms of single-hadron states, while in principle our correlators should receive contributions from all eigenstates of finite-volume QCD having the appropriate quantum numbers. This includes multi-meson states which in finite volume have a discrete spectrum. In fact, in a theory of non-interacting mesons the spectrum is rather simple, being approximately\footnote{we are neglecting small discretisation effects in the dispersion relation.}
\begin{equation}
m\big[A(\vec{p})B(-\vec{p})\big] = \sqrt{m_A^2 + |\vec{p}|^2} + \sqrt{m_B^2 + |\vec{p}|^2}, \label{nonint}
\end{equation}
where only discrete values of the momentum $\vec{p}$ are allowed by the boundary conditions: $a_s \vec{p} = \tfrac{2\pi}{L_s}\big(n_x, n_y, n_z \big)$ ($L_s$ is the spatial lattice extent in lattice units, i.e.\ $16$ or $20$ for the lattices we are using). Clearly this spectrum, with the exception of the states with $\vec{p}=(0,0,0)$, will change considerably under changes in volume.  Ongoing work by the Hadron Spectrum Collaboration is presented in Ref.~\cite{Foley:2010te}. 

Within QCD mesons interact and this interaction has a range of possible forms, for example: the repulsive interaction of two pions in the isospin 2 channel, the somewhat attractive interaction of two pions in isospin 0 that gives rise to the $\sigma$ enhancement and the strong attraction in isospin 1 that corresponds to the relatively narrow $\rho$ resonance. As shown by L\"uscher\cite{Luscher:1991cf}, taking account of hadron interactions, the finite volume energy spectrum will be modified with respect to equation \ref{nonint}.  The modification (at least in the elastic case) can be related to the hadron-hadron scattering phase shift which encodes details of attractive or repulsive interactions and even of resonant behaviour.

A simple schematic framework in which to view the finite-volume eigenstates is in terms of admixtures of idealised non-interacting basis states. For example, at low energy in the $T_1^{--}$ channel, one might consider there to be a space of non-interacting pion-pair states with the various relative momenta allowed in a finite cubic box.  In addition we can allow a space of single-hadron vector bound states like the $\rho$, which we assume to be localised to a region of space somewhat smaller than the size of the lattice box.
The pion-pair state energies vary rapidly with changing box-size while the ``$\rho$" bound-states would be essentially volume-independent for volumes larger than the size of the bound-states. If one supplies a resonant phase shift (such as the $\rho$ in $\pi\pi$ scattering), L\"uscher's formulae give rise to avoided level crossings as a function of lattice size that resemble the behaviour of approximate eigenstates within time-independent quantum mechanical perturbation theory. The finite-volume eigenstates can be viewed then as admixtures of the ``$\rho$" bound-states and the pion-pair states where the degree of mixing is determined by the phase-shift and the volume of the box. 

\begin{figure}
 \centering
\includegraphics[width=0.5\textwidth,bb= 0 0 602 437]{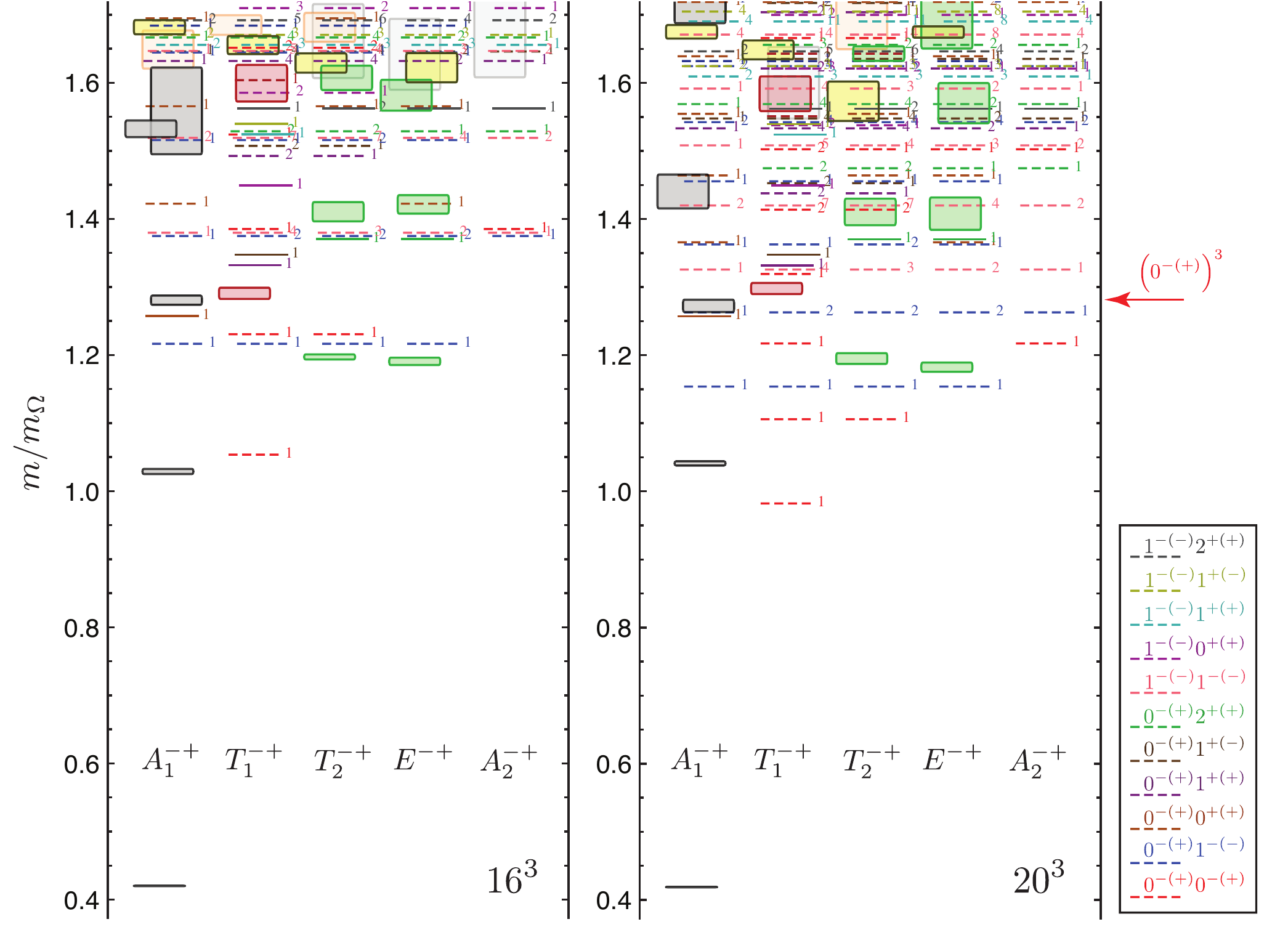}        
\caption{\emph{743} spectrum on $16^3$ and $20^3$ lattices. Boxes show the extracted $\Lambda^{-+}$ spectrum. Lines are the non-interacting two-meson state positions estimated from equation \ref{nonint} and the tables of \cite{Moore:2005dw, Moore:2006ng}(solid lines at $\vec{p}=(000)$), the small numbers indicate the multiplicity. Additional flavour multiplicity not shown. Also shown is the position of the lowest three-meson threshold.\label{twomes_Jmp}}
\end{figure}

\begin{figure}
 \centering
\includegraphics[width=0.5\textwidth,bb= 0 0 582 441]{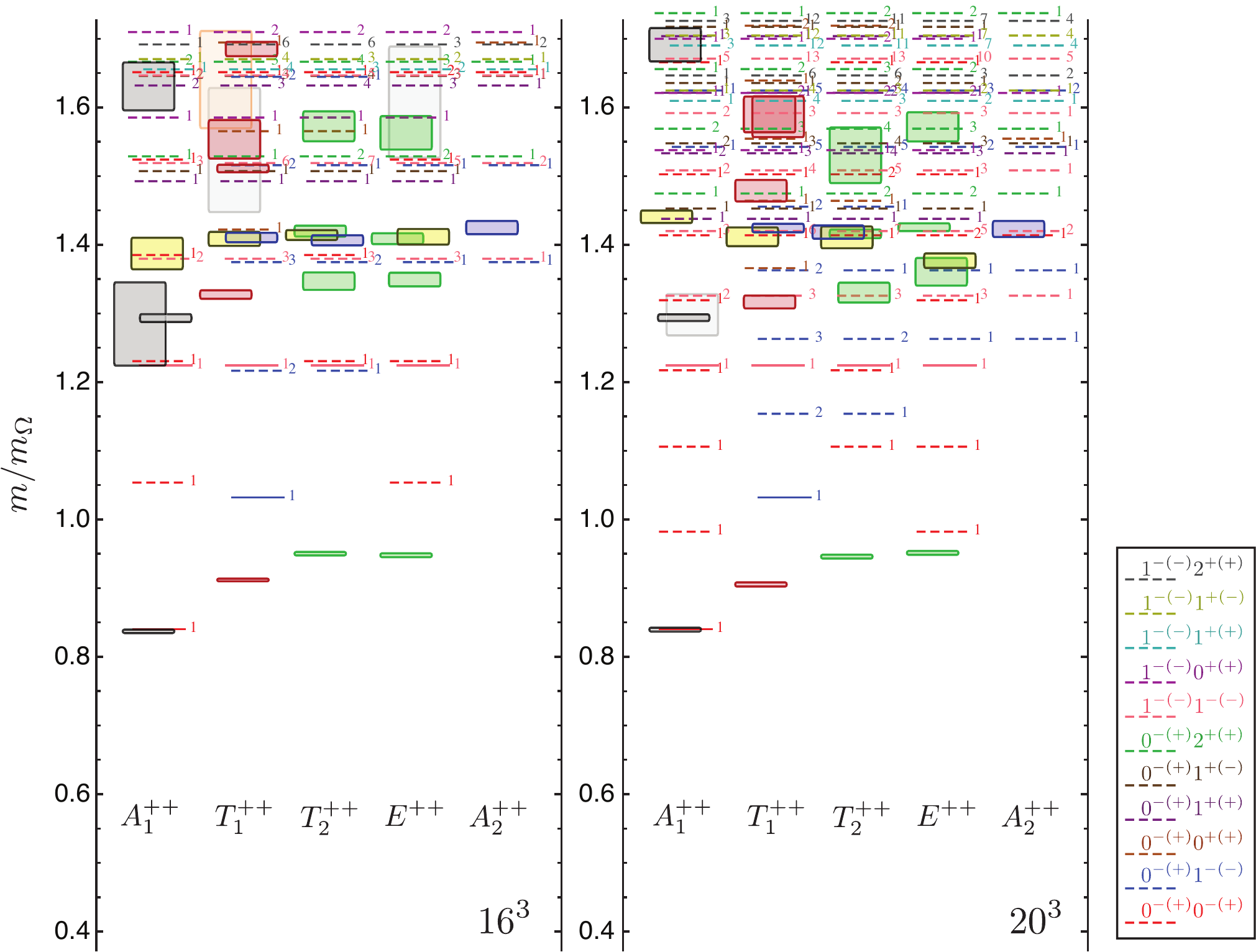}        
\caption{As figure \ref{twomes_Jmp} for $\Lambda^{++}$.
\label{twomes_Jpp}}
\end{figure}

In figures \ref{twomes_Jmp} and \ref{twomes_Jpp} we show the extracted \emph{743} spectrum superimposed with the positions of \emph{non-interacting} meson-meson states for $16^3$ and $20^3$ lattices. The distribution across irreps is determined using the ``in-flight" cubic symmetry group theory tables from \cite{Moore:2005dw, Moore:2006ng}. We show only pairs of $SU(3)_F$ octet states (since we have not determined the masses of the singlets or any other possible multiplets) and do not indicate the multiplicity of flavoured states for each level (which follows from the $SU(3)$ Clebsch-Gordan series for $\mathbf{8} \otimes \mathbf{8} \to \mathbf{8}$).

What is clear from Figures \ref{twomes_Jmp} and \ref{twomes_Jpp} is that the extracted spectrum does not seem to be related in any obvious way to the non-interacting two-particle spectrum.  The distribution of two-particle states across different irreps is determined not by the cubic symmetry of our discretized lattice, but rather by the spatial momenta allowed by the boundary conditions on the cubic volume in which we are performing our calculations.  We have seen that the observed spectrum split across the different irreducible representations conforms to that expected for single-particle states, with only negligible effects from cubic symmetry on the scale $a_s$.  In contrast, the pattern across different irreps expected for multi-particle states would be quite different.  This leads further credence to our assertion that two-particle states are contributing little to the calculated correlators.

How can we explain the lack of two-meson states in the spectrum we observe? 
Clearly, the basis set of operators employed has a very small overlap with 
states that predominantly resemble a composite of two mesons with well-defined 
and opposite momentum. This small overlap means the contribution to the
correlation function coming from the significant number of two-meson states is 
not resolved within the statistical precision of our calculation. 

This effect has been observed before in other dynamical-quark 
Monte Carlo measurements \cite{McNeile:2002fh, Durr:2008zz}
of the energies of states that are above threshold,
although perhaps not to the dramatic degree seen in this study. 
One cause of this supression may come directly from confinement dynamics of 
strongly interacting gluons. The inter-quark gluon flux shows considerable 
reluctance to break in the regime where it is energetically favourable to form 
two well-separated color singlets. This has been seen \cite{Bali:2005fu, 
McNeile:2002az}  as a very small overlap of an operator comprising a gluonic 
string onto the ground-state of this system. 

The overlap of a localised quark-bilinear operator onto a two-meson state will be supressed by $1/\sqrt{V}$, where $V$ is the lattice volume, if the operator creates a resonance with a finite width in the infinite volume limit.  This fall-off is matched by a growth in the density of states with the volume and the resonant state thus maintains a finite width as the mixing with each discrete state falls.  The simulations in this study are carried out in cubic volumes with side-lengths bigger than $2\, \mathrm{fm}$, which might be sufficiently large that the mixing between one of the low-lying two-particle states and a resonance is suppressed sufficiently for it to be undetectable with the quark bilinear operator basis.  

Even if the mixing between localised single-hadron states and two-meson states to form resonance-like finite volume eigenstates is not small, there still remains a practical difficulty associated with using only quark bilinear operators. In this case the state can be produced at the source timeslice through its localised single-hadron component, while the correlator time dependence obtained from $e^{-Ht}$ will indicate the mass of the resonant eigenstate. Consider a hypothetical situation in which a single two-meson state, denoted by $|2\rangle$, mixes arbitrarily strongly with a single 
localised single-hadron state, $|1\rangle$, with all other states being sufficiently distant in energy as to be negligible. There will be two eigenstates
\begin{align}
	\big| \mathfrak{a} \big\rangle &= \cos \theta \big|1\big\rangle + \sin \theta \big|2\big\rangle  \nonumber\\
	\big| \mathfrak{b} \big\rangle &= -\sin \theta \big|1\big\rangle + \cos \theta \big|2\big\rangle, \nonumber
\end{align}
with masses $m_\mathfrak{a}, m_\mathfrak{b}$. At the source (and sink) only the 
localised single-hadron component of each state overlaps with the operators in our basis and hence the overlaps, $Z^{\mathfrak{a},\mathfrak{b}}_i \equiv \langle \mathfrak{a},\mathfrak{b} | {\cal O}_i | 0 \rangle$, will differ only by an overall multiplicative constant, $Z^{\mathfrak{a}}_i = \cos \theta Z^{|1\rangle}_i, \;Z^{\mathfrak{b}}_i = -\sin \theta Z^{|1\rangle}_i$. As such the eigenvectors $v^{\mathfrak{a}}, v^{\mathfrak{b}}$ point in the same direction and cannot be made orthogonal. Thus the time dependence of both states will appear in the \emph{same} principal correlator as
\begin{equation}
	\lambda(t) \sim A_\mathfrak{a} e^{-m_\mathfrak{a}(t-t_0)} + A_\mathfrak{b} e^{-m_\mathfrak{b}(t-t_0)} + \ldots   \nonumber
\end{equation}

Since $m_\mathfrak{a}$ and $m_\mathfrak{b}$ most likely do not differ significantly (on the scale of $a_t^{-1}$) it will prove very difficult to extract a clear signal of two-exponential behavior from the principal correlator. This is precisely why the variational method's orthogonality condition on near degenerate states is so useful, but we see that it cannot work here and we are left trying to extract two nearby states from a $\chi^2$ fit to time-dependence. Typically this is not possible and reasonable looking fits to data are obtained with just one low-mass exponential.

If this is what is occurring in parts of our extracted spectrum, that we are extracting states which are admixtures of ``single-particle" and multi-meson states, but that we are not extracting the orthogonal mixtures, how should we interpret the mass values we are extracting? One conservative approach would be to suggest that our mass values are accurate only up to the hadronic width of the state extracted, since this width is correlated with mixing with multi-meson states via the scattering phase-shift. 

We would like to explicitly observe resonant behaviour of states in
our calculations and as such we must countenance the inclusion in our
operator basis of operators with larger numbers of fermion fields in
order to obtain healthy overlap with multi-hadron states. This can be
done while respecting the lattice symmetries using the tables in
\cite{Moore:2005dw, Moore:2006ng}. By using single-meson operators
subduced into the `in flight' little-group irreps from operators of
definite continuum helicity, it may prove possible to utilise
something similar to the spin-identification carried out in this
paper.  These constructions are underway, and distillation, with the
possible use of a stochastic estimator\cite{Morningstar:2010ae},
affords an efficient numerical means of implementing
them\cite{Foley:2010te}.

\section{Summary}\label{sec:summary}

\begin{figure*}
 \centering
\includegraphics[width=0.7\textwidth,bb= 0 0 673 402]{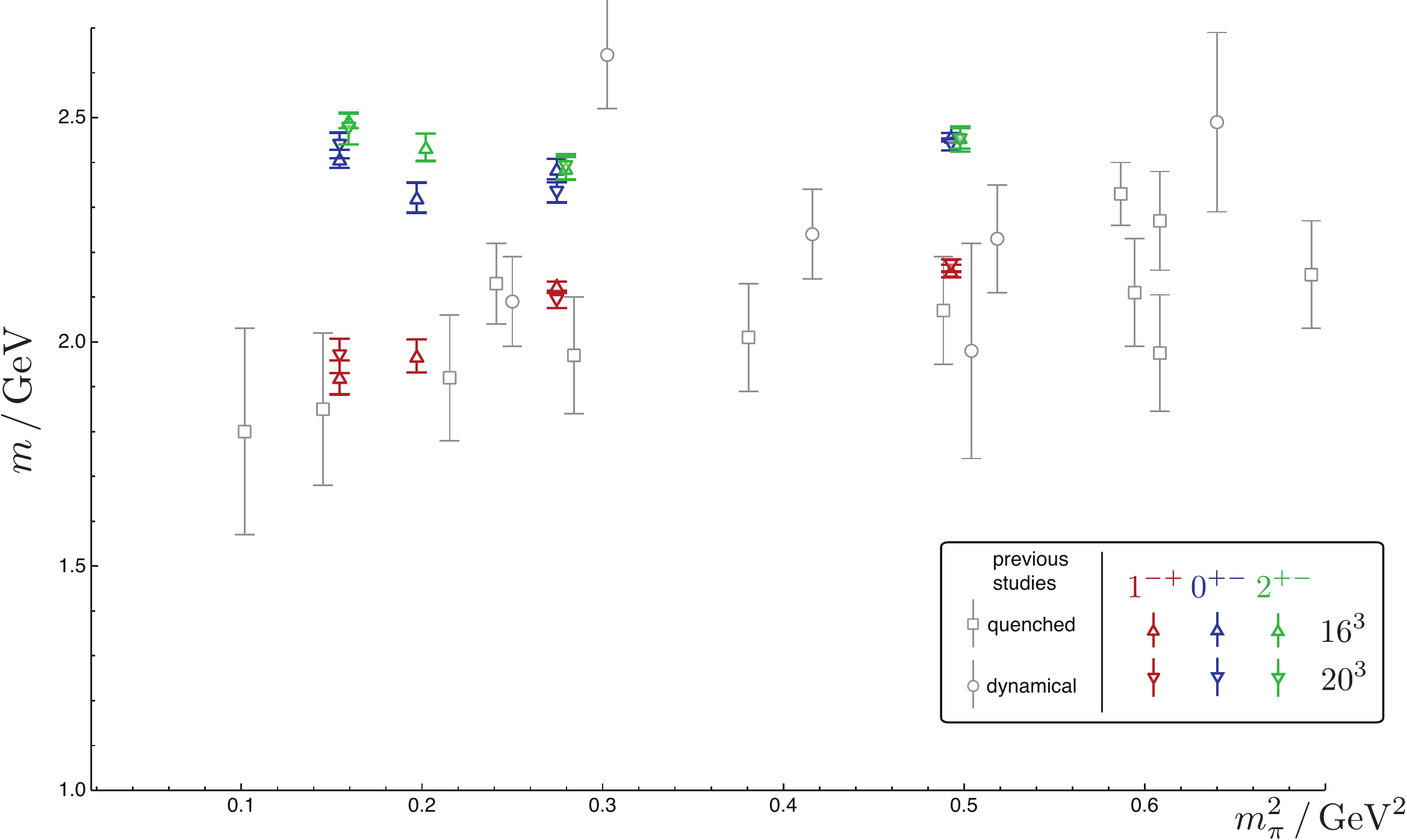}        
\caption{Summary of extracted isovector exotic states.  For comparison $1^{-+}$ results from Refs.\ \cite{Bernard:2003jd, Bernard:1997ib, Lacock:1998be, Lacock:1996ny, McNeile:2006bz, Hedditch:2005zf} are also plotted. \label{exotics}}
\end{figure*}

We have described in some detail our method for extracting a large number of excited states using dynamical anisotropic lattices, distillation technology and a variational analysis with an extensive basis of carefully constructed operators.  We have demonstrated the stability of the spectra with respect to changing the specific set of meson operators used, the number of distillation vectors and the details of the variational analysis.  Our method of spin identification based on operator overlaps has enabled us to confidently assign continuum spin to many states. 

We have successfully applied these techniques for two volumes on multiple mass sets: one with three degenerate flavours of quark (\emph{743}) and three with lighter $u$ and $d$ quarks giving mild breaking of $SU(3)_F$ and pion masses down to $\sim 400$ MeV.  In all cases we see mostly no significant volume dependence.  On all mass sets and volumes we are able to reliably extract a large number of excited states with all $PC$ combinations, states with high spin, up to and including spin four ($4^{++}$, $4^{-+}$ and $4^{--}$), and states with exotic quantum numbers ($0^{+-}$, $1^{-+}$ and $2^{+-}$).    The exotic states are particularly interesting and their presence points to the influence of explicit gluonic degrees of freedom.  In Fig.\ \ref{exotics} we summarise our results on exotic states and compare with previous lattice QCD results from Refs.\ \cite{Bernard:2003jd, Bernard:1997ib, Lacock:1998be, Lacock:1996ny, McNeile:2006bz, Hedditch:2005zf}.  

The extracted spectra show features of the $n^{2S+1}L_J$ state assignment of bound-state quark models, along with states (both exotic and non-exotic) which do not seem to lie within that classification.  A detailed model-dependent interpretation of these spectra is called for, comparing overlaps with quark model expectations and determining the degree of mixing of non-exotic hybrids and quark model states.  This work is ongoing.

We have presented kaon spectra and observe little or no mixing between the two charge conjugation eigenstates ($C=+$ and $C=-$); the resulting spectrum largely corresponds to the superposition of the $C=-$ and $C=+$ isovector spectra modulo the mass shift due to the light quark - strange quark mass difference.  In the $SU(3)_F$ limit there can be no such mixing and on the mass sets considered we are still rather close to this limit ($1 \le m_K/m_{\pi} \le 1.39$), so it is therefore not surprising that the mixing is small. Of particular interest at lighter quark masses, closer to the physical $m_K/m_\pi = 3.5$, will be the mixing between the axial kaons, $K_1(1270)$ and $K_1(1400)$, which, using a combination of experiment and models\cite{Asner:2000nx,Barnes:2002mu} is expected to be large.

We have argued that we see little evidence for two-particle states in our spectra and that to study such states we need to construct operators with a larger number of fermion fields. Such constructions are in progress and we believe that the addition of these operators will lead to a denser spectrum of states which can be interpreted in terms of resonances via techniques like L\"uscher's and its inelastic extensions. 

With the excited state spectra extracted herein, we argue that it does not make sense to attempt chiral extrapolation given that we cannot form a clear field-theoretic interpretion of the extracted energy levels. Once we have a handle on the two-meson levels, we can apply the techniques mentioned above to extract something like a phase shift, or more generally elements of the $S$-matrix, at discrete energy values. The phase shift may show resonant behaviour, which can be fitted with, in the simplest case, a Breit-Wigner form. The mass and width parameters of this Breit-Wigner are quantities which should be more amenable to chiral extrapolation.

A further avenue of study is the computation of disconnected two-point correlators giving access to isoscalar mesons; here we are interested in determining how QCD decides to mix light and strange. In addition, methods similar to those detailed in this paper are being applied in the baryon sector where the lattice irrep spectrum suffers from an even greater degree of degeneracy. An important aim of the Hadron Spectrum Collaboration is the calculation of light meson photocouplings which are relevant for, amongst other things, the GlueX experiment at the JLab 12 GeV upgrade where light mesons will be studied in photoproduction, with particular interest in exotics.

\begin{acknowledgments}
We thank our colleagues within the Hadron Spectrum Collaboration. Particular thanks go to B\'alint Jo\'o for his tireless efforts supplying us with gauge-field configurations, Jie Chen for his work on databases and Steve Wallace for providing us with determinations of the $\Omega$-baryon masses.

The Chroma software suite~\cite{Edwards:2004sx} was used to perform this work 
on clusters at Jefferson Laboratory and Fermilab using time awarded under the USQCD 
Initiative. 

Authored by Jefferson Science Associates, LLC under U.S. DOE Contract No.
DE-AC05-06OR23177. 
The U.S. Government retains a non-exclusive, paid-up,
irrevocable, world-wide license to publish or reproduce this manuscript for
U.S. Government purposes. MP is supported by Science Foundation Ireland under
research grant 07/RFP/PHYF168.

\end{acknowledgments}

\bibliography{bibliography}

\appendix


\section{Subduction Coefficients}\label{App:SubductionCoeffs}

Here we give a derivation of the subduction coefficients before listing their 
explicit values for all integer spins up to $J=4$.  An alternative derivation using the group 
theoretic projection formula is also described.

The continuum spin $J$ is reducible under the group of lattice rotations (the octahedral 
group or equivalently the cubic group).  We use ``subduction'' coefficients to project 
the continuum based operators onto their suitable octahedral group based versions via
\begin{equation}
{\cal O}^{[J]}_{\Lambda,\lambda} = \sum_M {\cal S}^{J,M}_{\Lambda,\lambda} {\cal O}^{J,M}  ,
\end{equation}
where ${\cal O}^{J,M}$ are the continuum operators with some definite total spin $J$ 
and spin component $M$.  For each $J\rightarrow\Lambda$ there is a matrix in the values 
of $M$ and the rows of the irrep, $\lambda$, that performs this mapping, 
i.e. the subduction coefficients, $\mathcal{S}^{J,M}_{\Lambda,\lambda}$.  

As shown in Table \ref{Table:Subduce}, the $J=0$ continuum spin subduces 
only onto the one-dimensional $A_1$ irrep and so trivially we have $\mathcal{S}^{0,0}_{A_1,1} = 1$.  
The simplest non-trivial subduction is that of continuum $J=1$ into the $T_1$ irrep and 
this is given by
\begin{equation}
(J=1)\rightarrow T_1 : 
\begin{tabular}{c|ccc}
$\begin{smallmatrix} & M \\ \lambda & \end{smallmatrix}$ & 1 & 0 & -1 \\
\hline
1 & 1 & 0 & 0\\
2 & 0 & 1 & 0\\
3 & 0 & 0 & 1
\end{tabular}
\end{equation}
where the $T_1$ is a faithful representation of $J=1$ (The basis used here follows Ref.~\cite{Basak:2005ir}).  
\begin{widetext}
The subduction coefficients for higher spins can be built up using continuum and octahedral 
group Clebsch-Gordan coefficients using
\begin{equation}
S^{J,M}_{\Lambda,\lambda} = N \sum_{\lambda_1,\lambda_2} \sum_{m_1,m_2} 
C\begin{pmatrix} 
\Lambda & \Lambda_1 & \Lambda_2 \\
\lambda & \lambda_1 & \lambda_2
\end{pmatrix}
S^{J_1,m_1}_{\Lambda_1,\lambda_1} S^{J_2,m_2}_{\Lambda_2,\lambda_2}
\langle J_1,m_1;J_2,m_2|J,M \rangle\quad .
\label{Equ:AppSubductionIteration}
\end{equation}
Here $\langle J_1, M_1; J_2, M_2 | J, M \rangle$ is the usual $SO(3)$ Clebsch-Gordan coefficient for $J_1 \otimes J_2 \rightarrow J$ and $C\Bigl( \begin{array}{ccc}\Lambda & \Lambda_1 & \Lambda_2 \\ \lambda & \lambda_1 & \lambda_2\end{array} \Bigr)$ is the octahedral group Clebsch-Gordan coefficient for $\Lambda_1 \otimes \Lambda_2 \rightarrow \Lambda$.   $N$ is a normalisation factor, fixed by the requirement that the subduction coefficients form an orthogonal matrix, $\sum_M {\cal S}^{J,M}_{\Lambda, \lambda} {\cal S}^{J,M*}_{\Lambda', \lambda'} = \delta_{\Lambda, \Lambda'} \delta_{\lambda, \lambda'}$.

This iteration formula can be constructed by noting that, for appropriately normalised states, $\mathcal{S}^{J,M}_{\Lambda, \lambda}=\left<\Lambda,\lambda|J,M\right>$ and so 
\begin{equation}
\mathcal{S}^{J,M}_{\Lambda, \lambda} = N \sum_{\lambda_1,\lambda_2} \sum_{m_1,m_2} \left<\Lambda,\lambda|\Lambda_1,\lambda_1; \Lambda_2, \lambda_2\right>\left<\Lambda_1,\lambda_1; \Lambda_2, \lambda_2|J_1,m_1; J_2,m_2\right> \left<J_1,m_1; J_2,m_2|J,M\right>  .
\end{equation}
Substituting for the continuum and octahedral group Clebsch-Gordan coefficients and $\left<\Lambda_1,\lambda_1;\Lambda_2,\lambda_2|J_1,m_1; J_2,m_2\right> = \left<\Lambda_1,\lambda_1|J_1,m_1\right> \left<\Lambda_2,\lambda_2|J_2,m_2\right> = \mathcal{S}^{J_1,m_1}_{\Lambda_1, \lambda_1} \mathcal{S}^{J_2,m_2}_{\Lambda_2, \lambda_2}$ gives the result in Eq.\ \ref{Equ:AppSubductionIteration}.

\hfill

For $J=2$ to $J=4$ the subduction coefficients are shown below:
\begin{equation*}
\begin{tabular}{c|ccccc}
\multicolumn{6}{c}{$(J=2)\rightarrow T_2$}\\
$\begin{smallmatrix} & M \\ \lambda & \end{smallmatrix}$ & 2 & 1 & 0 & -1 & -2\\
\hline
1 & 0 &  1 & 0 & 0 & 0\\
2 &  $\frac{1}{\sqrt{2}}$ & 0 & 0 & 0 & -$\frac{1}{\sqrt{2}}$\\
3 & 0 & 0 & 0 &  1 & 0\\
\end{tabular}
\hspace{1cm}
\begin{tabular}{c|ccccc}
\multicolumn{6}{c}{$(J=2)\rightarrow E$}\\
$\begin{smallmatrix} & M \\ \lambda & \end{smallmatrix}$ & 2 & 1 & 0 & -1 & -2\\
\hline
1 & 0 & 0 &  1 & 0 & 0\\
2 &  $\frac{1}{\sqrt{2}}$ & 0 & 0 & 0 &  $\frac{1}{\sqrt{2}}$\\
\end{tabular}
\end{equation*}

\begin{equation*}
\begin{tabular}{c|ccccccc}
\multicolumn{8}{c}{$(J=3)\rightarrow T_1$}\\
$\begin{smallmatrix} & M \\ \lambda & \end{smallmatrix}$ & 3 & 2 & 1 & 0 & -1 & -2 & -3\\
\hline
1 & 0 & 0 &  $\sqrt{\frac{3}{8}}$ & 0 & 0 & 0 &  $\sqrt{\frac{5}{8}}$\\
2 & 0 & 0 & 0 & -1 & 0 & 0 & 0\\
3 &  $\sqrt{\frac{5}{8}}$ & 0 & 0 & 0 &  $\sqrt{\frac{3}{8}}$ & 0 & 0\\
\end{tabular}
\hspace{.5cm}
\begin{tabular}{c|ccccccc}
\multicolumn{8}{c}{$(J=3)\rightarrow T_2$}\\
$\begin{smallmatrix} & M \\ \lambda & \end{smallmatrix}$ & 3 & 2 & 1 & 0 & -1 & -2 & -3\\
\hline
1 & 0 & 0 &  $\sqrt{\frac{5}{8}}$ & 0 & 0 & 0 & -$\sqrt{\frac{3}{8}}$\\
2 & 0 & -$\frac{1}{\sqrt{2}}$ & 0 & 0 & 0 & -$\frac{1}{\sqrt{2}}$ & 0\\
3 &  $\sqrt{\frac{3}{8}}$ & 0 & 0 & 0 & -$\sqrt{\frac{5}{8}}$ & 0 & 0\\
\end{tabular}
\hspace{.5cm}
\begin{tabular}{c|ccccccc}
\multicolumn{8}{c}{$(J=3)\rightarrow A_2$}\\
$\begin{smallmatrix} & M \\ \lambda & \end{smallmatrix}$ & 3 & 2 & 1 & 0 & -1 & -2 & -3\\
\hline
1 & 0 &  $\frac{1}{\sqrt{2}}$ & 0 & 0 & 0 & -$\frac{1}{\sqrt{2}}$ & 0\\
\end{tabular}
\end{equation*}

\begin{equation*}
\begin{tabular}{c|ccccccccc}
\multicolumn{10}{c}{$(J=4)\rightarrow A_1$}\\
$\begin{smallmatrix} & M \\ \lambda & \end{smallmatrix}$ & 4 & 3 & 2 & 1 & 0 & -1 & -2 & -3 & -4\\
\hline
1 &  $\sqrt{\frac{5}{24}}$ & 0 & 0 & 0 &  $\sqrt{\frac{7}{12}}$ & 0 & 0 & 0 &  $\sqrt{\frac{5}{24}}$\\
\end{tabular}
\hspace{.5cm}
\begin{tabular}{c|ccccccccc}
\multicolumn{10}{c}{$(J=4)\rightarrow T_1$}\\
$\begin{smallmatrix} & M \\ \lambda & \end{smallmatrix}$ & 4 & 3 & 2 & 1 & 0 & -1 & -2 & -3 & -4\\
\hline
1 & 0 & 0 & 0 & -$\sqrt{\frac{7}{8}}$ & 0 & 0 & 0 & -$\frac{1}{\sqrt{8}}$ & 0\\
2 &  $\frac{1}{\sqrt{2}}$ & 0 & 0 & 0 & 0 & 0 & 0 & 0 & -$\frac{1}{\sqrt{2}}$\\
3 & 0 &  $\frac{1}{\sqrt{8}}$ & 0 & 0 & 0 &  $\sqrt{\frac{7}{8}}$ & 0 & 0 & 0\\
\end{tabular}
\end{equation*}
\begin{equation*}
\begin{tabular}{c|ccccccccc}
\multicolumn{10}{c}{$(J=4)\rightarrow T_2$}\\
$\begin{smallmatrix} & M \\ \lambda & \end{smallmatrix}$& 4 & 3 & 2 & 1 & 0 & -1 & -2 & -3 & -4\\
\hline
1 & 0 & 0 & 0 & -$\frac{1}{\sqrt{8}}$ & 0 & 0 & 0 &  $\sqrt{\frac{7}{8}}$ & 0\\
2 & 0 & 0 &  $\frac{1}{\sqrt{2}}$ & 0 & 0 & 0 & -$\frac{1}{\sqrt{2}}$ & 0 & 0\\
3 & 0 &  $\sqrt{\frac{7}{8}}$ & 0 & 0 & 0 & -$\frac{1}{\sqrt{8}}$ & 0 & 0 & 0\\
\end{tabular}
\hspace{.5cm}
\begin{tabular}{c|ccccccccc}
\multicolumn{10}{c}{$(J=4)\rightarrow E$}\\
$\begin{smallmatrix} & M \\ \lambda & \end{smallmatrix}$& 4 & 3 & 2 & 1 & 0 & -1 & -2 & -3 & -4\\
\hline
1 &  $\sqrt{\frac{7}{24}}$ & 0 & 0 & 0 & -$\sqrt{\frac{5}{12}}$ & 0 & 0 & 0 &  $\sqrt{\frac{7}{24}}$\\
2 & 0 & 0 &  $\frac{1}{\sqrt{2}}$ & 0 & 0 & 0 &  $\frac{1}{\sqrt{2}}$ & 0 & 0\\
\end{tabular}
\end{equation*}

\hfill

An alternative method for constructing the subduction coefficients is by using the group theoretic projection formula:
\begin{equation}
\mathcal{O}^{[J]}_{\Lambda, \lambda} = \frac{d_\Lambda}{g_G} \sum_{R \in G} \Gamma^{\Lambda}_{\lambda, \mu}(R) \sum_{M'} R_{M M'} \mathcal{O}^{J,M'}  ,
\label{Equ:ProjectionFormula}
\end{equation}
where $G$ is the octahdedral group, $g_G$ is the order of the group $G$ (i.e. 24), $d_{\Lambda}$ is the dimension of octahedral group irrep $\Lambda$, $R_{M M'}$ is an element of $G$ acting on $O^{J,M'}$ and $\Gamma^{\Lambda}_{\lambda, \mu}(R)$ is the representation of $R$ in $\Lambda$.  Here the operators $\mathcal{O}^{J,M'}$ have definite spin and so $R_{M M'} = D^{(J)}_{M' M}(R)$ (a Wigner-$D$ matrix).  The $\Gamma^{\Lambda}_{\lambda, \mu}(R)$ are found, for example, by considering a basis for the irreps in terms of spherical harmonics\cite{Basak:2005ir} and then using the transformation properties of spherical harmonics under rotations.  Once all possible operators $\mathcal{O}^{[J]}_{\Lambda,\lambda}$ have been found (considering all $M$ and $\lambda$), the linearly independent combinations are constructed.  These combinations then give the subduction coefficients which are identical, up to possible phases and choice of basis, to those obtained using the method described above.

\end{widetext}


\end{document}